\documentclass[hyper,a4paper,11pt,oneside]{JHEP3}

\usepackage{graphicx}
\usepackage{dcolumn}
\usepackage{bm}
\usepackage{epsfig,multicol}
\usepackage[title]{appendix}
\newcommand\fverb{\setbox\fverbbox=\hbox\bgroup\verb}
\newcommand\fverbdo{\egroup\medskip\noindent%
            \fbox{\unhbox\fverbbox}\ }
\newcommand\fverbit{\egroup\item[\fbox{\unhbox\fverbbox}]}
\newbox\fverbbox

\newcommand{\M}{{\cal {M}}}
\newcommand{\B}{{\cal {B}}}

\newcommand{\HH}{{\cal {H}}}

\newcommand{\CV}{{\cal {V}}}

 \renewcommand{\figurename}{Figura}
 \renewcommand{\contentsname}{\'{\i}ndice}
 
 \renewcommand{\tablename}{Tabla}
 \renewcommand{\contentsname}{\'{\i}ndice}

\title{Gravitational collapse and entropy of Black Holes with magnetic sources.}
\author{A. Ulacia Rey$^{\diamondsuit,\spadesuit}$ \\
$^{\diamondsuit}$Departamento de F\'{\i}sica Te\'{o}rica.
\\Instituto de Cibern\'etica, Matem\'atica y F\'{\i}sica, ICIMAF.\\
Calle E No-309 Vedado, cp-10400. La Habana, Cuba.\\\email{alain@icmf.inf.cu}\\\\
$^{\spadesuit}${Departamento de Gravitaci\'{o}n y Teor\'{\i}as de
Campo.\\ Instituto de Ciencias Nucleares, ICN, \\
Universidad Aut\'{o}noma de M\'{e}xico UNAM, Mexico,DF.
04510.}\\\email{alain.ulacia@nucleares.unam.mx}}

\abstract{This thesis is divided in two parts, each one addressing problems that can be relevant in the study of compact objects.
The first part deals with the study of a magnetized and self-gravitating gas of degenerated fermions (electrons and neutrons) as
sources of a Bianchi-I space-time. We solve numerically the Einstein-Maxwell field equations for a large set of initial conditions
of the dynamical variables. The collapsing singularity is isotropic for the neutron gas and can be anisotropic for the electron gas.
This result is consistent with the fact that electrons exhibit a stronger coupling with the magnetic field, which is the source of
anisotropy in the dynamical variables. In the second part we calculate the entropy of extremal black holes in 4 and 5 dimensions,
using the entropy function formalism of Sen and taking into account higher order derivative terms that come from the complete set of
Riemann invariants. The resulting entropies show the deviations from the well know Bekenstein-Hawking area law.}

\begin{document}
\newpage

\section*{}
\vspace{3in}
\hspace{4in}
{\it A la memoria de mis abuelos}

\newpage
\section*{Agradecimientos}

\vspace{2.5in} \noindent{
 \it 
A mis tutores, Roberto A. Sussman y A. P\'erez Mart\'inez por ayudarme a crecer.\\
A mi familia por todo el  amor y comprensi\'on infinita.\\
A mi esposa por su inagotable paciencia, su apoyo incondicional y sobre
todo por su sincero amor.\\
A todos los que me han brindado su experiencia y con los que he tenido fruct\'{\i}feras
discusiones t\'ecnicas, en espacial a Jos\'e F. Morales.\\
Al grupo de F\'{\i}sica Te\'orica por brindarme la posibilidad de compartir con excelentes
profesionales y magn\'{\i}ficas personas, de los cuales estar\'e agradecido toda una vida.\\
\\
Este trabajo ha sido posible gracias al apoyo del ICIMAF, de la UNAM, de la beca CLAF--ICTP y del INFN.}

\newpage

\section*{Introducci\'on}
Los objetos compactos: Enanas Blancas (EBs), Estrellas Neutr\'onicas (ENs) y Agujeros Negros (ANs) son objetos muy densos que invaden el Universo y son el \'ultimo estad\'io por el que transita una estrella antes de apagarse. Que una estrella se convierta en uno u otro tipo de objeto compacto depende decisivamente de su masa. Por ejemplo para estrellas con una masa entre $9\sim 10M_{\odot}$, que han agotado su combustible nuclear evolucionan hacia una EB. Sin embargo, estrellas con masa mayor que $10M_{\odot}$ forman ANs.

En particular las EBs y ENs son laboratorios astrof\'isicos naturales en los cuales se puede probar la veracidad de las teor\'ias f\'isicas m\'as actuales ya que en ellas viven fen\'omenos en los que est\'an presentes part\'iculas fuertemente interactuando, en presencia de grandes campos gravitacionales y electromagn\'eticos.

 El modelo m\'as aceptado para la descripci\'on de part\'iculas en el micro mundo es el Modelo Est\'andar (ME). Por otra parte, la Teor\'ia General de la Relatividad (TGR) describe fielmente los fen\'omenos que ocurren a una mayor escala. Por tanto, un modelo f\'isico que pretenda describir estrellas de este tipo, debe incluir la gravedad y las interacciones entre part\'iculas para que sea v\'alido estrictamente hablando. Sin embargo en nuestro cuadro f\'isico del Universo, todav\'ia no tenemos una teor\'ia unificada y experimentalmente demostrada que pueda juntar las cuatro fuerzas fundamentales de la naturaleza (fuerte, d\'ebil, electromagn\'etica y gravitatoria) de modo que se puedan realizar predicciones te\'oricas medibles experimentalmente. De las fuerzas mencionadas, la \'unica que se niega a ser unificada es la gravitatoria, aunque en estos \'ultimos tiempos se viene trabajando fuertemente en teor\'ias prometedoras (como las ``teor\'ias de cuerdas'') \cite{SmolinLee}. Las Teor\'ias de Cuerdas no tienen por ahora muchas esperanzas de ser probadas experimentalmente debido a que se requiere para verificarlas, de experimentos que produzcan energ\'ias superiores a las que se espera, se generen en el LHC\footnote{ LHC significa ``Large Hadron Collider'' o sea el Gran Colisionador de Hadrones.}. El LHC fue dise\~nado y puesto en marcha para encontrar el bos\'on de Higgs que es un pilar fundamental del ME \cite{John_Ellis_Nature}. Sin embargo, existen casos dentro de la misma teor\'ia de cuerdas en que el espacio de configuraci\'on es tan grande que las cantidades de energ\'ia para su detecci\'on no es tan alta, si este fuera el caso real, entonces con el LHC se podr\'ian ver indicios de estos objetos.

 Por otro lado, desde hace varias d\'ecadas se le da gran importancia al estudio de campos magn\'eticos intensos en el universo y en particular en objetos compactos. En los \'ultimos tiempos, con la aparici\'on de nuevas y m\'as precisas observaciones, se avanza en las mediciones astrof\'isicas y se introducen nuevas cotas para estos campos magn\'eticos. As\'i en los a\~nos 80 campos magn\'eticos de $10^{9}\,Tesla\,(T)$ \footnote{$1\,Tesla=10^{4}Gauss$ conversi\'on para la inducci\'on del campo magn\'etico.} eran considerados m\'aximos. Hoy en d\'ia, seg\'un observaciones m\'as refinadas se estiman que existen campos 1000 veces mayores que el conocido campo l\'imite de Schwinger $4,4\times10^{9}\,T$ \cite{Kim:2003qp}.

Las EBs son estrellas con una masa de $1,2M_{\odot}$, (o sea $<\,1,44M_{\odot}$), densidades entre $10^{9}-10^{10}kg/m^{3}$, radios de $10^{6}\,m$ y campos magn\'eticos del orden de $10^{4}\,T$.

 Sin embargo, una Estrella Neutr\'onica tiene una masa M entre $1.5$ y $2$ masas solares $M_{\odot}$, radios entre $10-12\,km$ y una densidad central de $10^{17}\,kg/m^{3}$. Es por tanto una de las formas m\'as densas de materia encontradas en el Universo observable. Aunque los neutrones dominan la componente bari\'onica de la Estrella de Neutrones, existen tambi\'en algunos protones (y suficientes electrones y muones) que garantizan la neutralidad de carga en la estrella. Adem\'as, engendran un s\'uper campo magn\'etico de unos $10^{10}\,T$ \cite{Lattimer1}.

La primera EB encontrada, fue la compa\~nera de la brillante estrella Sirio. Sirio y su compa\~nera est\'an en una \'orbita mutua, una alrededor de la otra, y esto permiti\'o que se determinaran las masas de cada una. A partir del brillo y temperatura de la compa\~nera, podemos determinar su tama\~no, que es de unos $10^{7}\,m$ de di\'ametro, menos que el de la Tierra \cite{Mediciones_Sirio}.

Las EBs son intr\'insecamente muy poco brillantes, y son por tanto dif\'iciles de detectar. A\'un as\'i, ellas son el estado final de todas las estrellas de masa mediana, y por ello podr\'iamos esperar hallar muchas EBs en el Universo.

Los astr\'onomos han logrado encontrar muchas, usando t\'ecnicas que dependen ya sea de que son compa\~neras de otras estrellas, o de que son estrellas calientes con grandes movimientos relativos respecto a otras estrellas (indicando que ellas est\'an mucho m\'as cercanas que otras estrellas de la Secuencias Principal con la misma temperatura), y a partir de su emisi\'on de radiaci\'on de alta energ\'ia, tal como luz ultravioleta. La presencia de EBs en sistemas binarios ha sido muy importante para entender violentos estallidos en sistemas estelares. Supernovas del Tipo I, Novas, y estrellas variables catacl\'ismicas, son todos casos en los que la compa\~nera de una EB ha alcanzado un punto en su evoluci\'on en el que est\'a aumentando en di\'ametro y perdiendo masa hacia la EB.

La deposici\'on de material en un disco de acreci\'on alrededor de la EB, o en la superficie de la EB, determinar\'a la naturaleza de cualquier estallido. La estrella aparentemente m\'as brillante en el cielo, Sirio, fue observada por Bessel en 1844 mostrando un bamboleo en su movimiento a trav\'es del cielo. Bessel atribuy\'o esto a la presencia de una compa\~nera, pero no se observ\'o alguna hasta que Alvan Clark, mientras probaba un nuevo telescopio, vio la tenue estrella compa\~nera. En 1925, el espectro de la estrella compa\~nera confirm\'o que era una estrella con aproximadamente la misma temperatura que Sirio A.

 Ella tiene un per\'iodo de 50 a\~nos, con una m\'axima separaci\'on en el cielo de 7,6 arcosegundos. La diferencia de luminosidad entre Sirio A y B llega a un factor de m\'as de $8\,000$. La soluci\'on de su movimiento orbital arroj\'o para A y B, masas de 2,3 y 1 veces la masa del Sol. Sirio A tiene un radio de cerca de $10^{9}\,m$, mientras que Sirio B tiene un radio de solo $10^{7}\,m$.

En 1926 R. H. Fowler junto con Paul Dirac trabajaron en aplicar la mec\'anica estad\'istica para explicar la estabilidad de una estrella EB y en 1930 Chandrasekhar descubre la masa m\'axima de $1.44\,M_{\odot}$ para estas estrellas \cite{Chandrasekhar1,Chandrasekhar2}.

Para las estrellas neutr\'onicas los resultados te\'oricos antecedieron a los observacionales. En 1932 Landau habl\'o por primera vez de la EN. En el a\~no 1934 se predijo por primera vez, bas\'andose en c\'alculos te\'oricos, la existencia de una ENs y a partir de esa fecha se estudiaron muchos modelos para describirlas, pero no es hasta el a\~no 1967 en que se verifica observacionalmente por azar, la existencia de un objeto con las caracter\'isticas de una EN \cite{PremioNobel1974}.

Las ENs no se hab\'ian observado antes debido a que no emiten en el visible, por lo cual fue necesario que pasaran m\'as de 30 a\~nos y que la tecnolog\'ia permitiera la construcci\'on de radiotelescopios capaces de detectar las primeras ENs. El descubrimiento de la primera EN provoc\'o un aumento en el n\'umero de trabajos te\'oricos encaminados a describir los procesos f\'isicos que ocurren en el interior de estos objetos \cite{Hartle1,Hartle2,Gusakov:2002hh,Lattimer:1990zz,Gusakov:2010ce}.

Desde hace veinte a\~nos los estudios de las Estrellas de Neutrones se nutren de cuantiosas observaciones que se obtienen gracias a observatorios que detectan emisiones de rayos X y radiaci\'on gamma colocados en sat\'elites. Entre los observatorios m\'as conocidos est\'an el Telescopio Espacial Hubble, el Observatorio de Rayos X Chandra y el Observatorio de Rayos Gama Compton.

Actualmente tambi\'en se habla de la posible existencia de Estrellas de Quark, o Estrellas Extra\~nas. Estas deben aparecer para s\'uper altas densidades, donde la materia presenta caracter\'isticas todav\'ia m\'as ex\'oticas, tales como bariones con extra\~neza, mesones condensados (piones, kaones), o quarks no confinados \cite{Haensel:2009wa,Postnikov:2010yn,Felipe:2008cm}.

Estos fermiones en forma de hadrones o quark no confinados se espera que tambi\'en exhiban propiedades de superfluidez o superconductividad.

Una EN normal, tiene materia hadr\'onica en su exterior y tanto la presi\'on como la densidad de materia bari\'onica se anula en la frontera exterior (en el interior puede contener cualquier combinaci\'on de part\'iculas ex\'oticas permitidas por la f\'isica de interacciones fuertes). Sin embargo, una Estrella de quarks, puede tener tambi\'en una superficie de quark desnuda cuya presi\'on se desvanece en la frontera y una s\'uper alta densidad de materia, o una fina capa de materia normal soportada por las fuerzas Coulombianas sobre la superficie de materia de quark. Tales objetos est\'an constituidos por la materia en el \'ultimo estado fundamental cuya energ\'ia es menor que la del n\'ucleo de Hierro. Esta materia comprimida a una densidad suficientemente alta prodr\'ia espont\'aneamente convertirse en materia de quark desconfinada.

Hasta el presente ning\'un experimento ha encontrado quark no confinados, aunque se supone que en las Estrellas Extra\~nas los quarks est\'an libres. A diferencia de una estrella normal, la estrella de materia extra\~na, si tiene \'este estado auto-ligado de energ\'ia, no requerir\'ia de gravedad para unirse.

Los ANs son los remanentes estelares m\'as controversiales que habitan el Universo, ellos aparecen tras la explosi\'on de una Supernova cuya estrella progenitora es muy masiva ($M\,>\,10\,M_{\odot}$) o como resultado de la coalescencia de un sistema binario de dos ENs. El colapso gravitacional no puede ser evitado por la presi\'on del gas degenerado de part\'iculas que constituye la estrella y por ello colapsa apareciendo una singularidad en el espacio-tiempo.

Se presupone que existan en el Universo infinidad de ellos pero observacionalmente no han sido detectados hasta hoy. Sin embargo la existencia de ellos ayudar\'ia a comprender observaciones astron\'omicas, emisi\'on de rayos X por estrellas binaria y galaxias activas. Actualmente, muy pocos astr\'onomos ponen en duda la existencia de los ANs, aunque la evidencia de su existencia es indirecta, a partir del comportamiento de otros objetos cercanos a los ANs, tales como estrellas brillantes.

En 1963, Roy Kerr \cite{Kerr} demostr\'o que en un espacio-tiempo de cuatro dimensiones todos los ANs, deb\'ian tener una geometr\'ia cuasi-esf\'ericas determinada por tres par\'ametros: su masa $M$, su carga el\'ectrica $Q$ y su momento angular $J$.

Con posterioridad, la teor\'ia de los ANs se nutri\'o de los trabajos de Hawking, Ellis y Penrose de los a\~nos 70 que demostraron varios teoremas importantes sobre la ocurrencia y geometr\'ia de los ANs \cite{HawkingTLSSoST}.

Como la presencia de campos magn\'eticos intensos es relevante en los objetos compactos, su inclusi\'on en las Ecuaciones de Estado (EE) se convierte en un importante tema a tener en cuenta para explicar las observaciones, adem\'as de considerar que tenemos un sistema en presencia de gravedad.

El gas magnetizado de electrones, neutrones y quarks ha sido estudiado previamente en las Ref. \cite{Aurora1,Aurora2,Aurora3}.

Entre muchos resultados, el m\'as importante es que se descubre una anisotrop\'ia entre las presiones paralela y perpendicular al campo magn\'etico, siendo la presi\'on paralela $(p_{\|} > p_{\bot})$ mayor. En el caso extremo de campos magn\'eticos muy intensos ($B=10^{10}T$) la presi\'on $p_{\|}$ puede hacerse cero y entonces el sistema tiende a colapsar anisotr\'opicamente. Esto trae como consecuencia una l\'ogica conclusi\'on cualitativa sobre la forma que va a tener dicho sistema: \'el deber\'a tomar la forma de un objeto alargado en la direcci\'on del campo magn\'etico y a medida que este campo sea m\'as intenso pues m\'as alargado se har\'a el objeto.

Es importante se\~nalar que los estudios citados anteriormente se realizaron sin incluir el efecto de la gravedad. Ahora, si queremos describir sistemas con alta densidad de materia, los cuales puedan producir su propia gravedad (sistemas auto-gravitantes) en una proporci\'on tan elevada, que no podamos ignorar su efecto, entonces podr\'iamos referirnos a la posibilidad de que este sistema pueda describir un objeto estelar. Debido a que los sistemas con este tipo de anisotrop\'ia, en general no son estables en el universo, entonces un objeto de este tipo debe colapsar \cite{AAS} o realizar una transici\'on a un estado estable \cite{Thorne1965}.

Incluir la presencia de gravedad, ahora nos permite obtener informaci\'on sobre el papel que ella juega cuando aparecen estos tipos de anisotrop\'ias en las presiones, esta es unas de las principales motivaciones de este trabajo y nos orienta a responder las siguientes preguntas:
 \begin{itemize}
  \item \textquestiondown Aparecer\'a o no el colapso cuando la gravedad esta presente en un sistema magnetizado de Fermi?
  \item En caso que colapsaran, \textquestiondown Como ser\'ia el tipo de singularidad que aparecer\'ia en estos sistemas?
  \item \textquestiondown C\'omo es en general la din\'amica de un gas magnetizado de Fermi autogravitante? 
 \end{itemize}

 Con el prop\'osito de responder estas interrogantes, el primer objetivo de esta tesis est\'a dirigido a estudiar una fuente de materia densa y magnetizada de fermiones (electrones y neutrones), y con el uso de una m\'etrica no estacionaria obtener la evoluci\'on en el tiempo del sistema.

Escogeremos un espacio-tiempo Bianchi I que ha sido utilizado con \'exito para la descripci\'on de modelos cosmol\'ogicos \cite{WainwrightEllis}. Esta vez sin embargo estudiaremos un escenario astrof\'isico, consideraremos un gas degenerado de fermiones, es decir, la temperatura se toma igual a cero. La utilidad de estos espacio-tiempos se debe a que nos permiten tomar en cuenta las anisotrop\'ias debidas al campo magn\'etico. Una extensi\'on de este formalismo que tome en cuenta la temperatura podr\'ia describir una situaci\'on cosmol\'ogica \cite{Coley:2008gh}.

Por supuesto la formulaci\'on de un modelo para una estrella EB o una ENs magnetizada requiere del uso de una m\'etrica m\'as complicada, que se traduce en un problema num\'erico m\'as engorroso. Nuestro modelo aunque no muy realista puede dar informaci\'on cualitativa muy interesante sobre los procesos din\'amicos que ocurren en el n\'ucleo de tales objetos astrof\'isicos, en particular para una EB o una ENs. Tal estudio cualitativo podr\'ia dar luz para entender estos sistemas tan complejos.

Por otro lado a pesar de la descripci\'on cl\'asica de los ANs y sus ``pelos'' o caracter\'isticas principales derivadas de la descripci\'on de su $M,\,Q,\,J$ (Masa, Carga y Momento Angular de Rotaci\'on) \cite{MTW,RMW_GR}, descripciones de estos entes a partir de modernas teor\'ias de cuerdas, le atribuyen situaciones intermedias, donde otras magnitudes pueden ser consideradas, como por ejemplo campos escalares, o cargas magn\'eticas. Por otra parte, el estudio de la entrop\'ia y la perdida de informaci\'on de los ANs es un tema abierto. Ha sido probado te\'oricamente que la entrop\'ia de un AN es proporcional al \'area de la garganta del AN, lo cual ha mantenido a los Cosm\'ologos y F\'isicos Te\'oricos a la expectativa y en busca de nuevas interpretaciones.

A pesar de que los objetos astrof\'isicos deben de ser neutros el\'ectricamente existen soluciones matem\'aticas que caracterizan a los ANs como objetos cargados el\'ectricamente. Son soluciones conocidas por el nombre de Reissner-Nordstrom (RN) y tiene el l\'imite extremal cuando la masa se iguala a la carga el\'etrica. La conjetura de Penrose objeta a todo AN con carga el\'ectrica mayor que la masa, ya que significar\'ia una densidad de energ\'ia negativa y por lo que conocemos, la densidad de energ\'ia de todo sistema f\'isico conocido siempre es positiva.

La teor\'ia de cuerdas surgida con la pretensi\'on de unificar las interacciones fundamentales de la naturaleza, ha contribuido a que se desarrollen numerosas t\'ecnicas que con \'exito logran ser aplicadas a otras \'areas de la F\'isica. Uno de los \'exitos mayores lo ha logrado con la gravitaci\'on. En particular el m\'etodo de Sen o formalismo entr\'opico, fue dise\~nado originalmente como mecanismo entr\'opico en teor\'ias de Supergravedad \cite{Renata1,Renata2} y extendido con \'exito al estudio de la entrop\'ia de ANs con diferentes topolog\'ias. Este formalismo simplifica los procedimientos de c\'alculo de entrop\'ia conocidos.

Por tanto el segundo objetivo de esta tesis ser\'a entonces utilizar este mecanismo de Sen para estudiar ANs con constante cosmol\'ogica y campo electromagn\'etico en cuatro y cinco dimensiones teniendo en cuenta los invariantes de Riemann como conjunto completo introducido por Carminati-McLenagha \cite{JCarminati,Zakhary_McIntosh}.

Esta tesis consta de dos partes que se asocian con los dos objetivos que nos hemos propuesto con este trabajo. La primera parte aborda el estudio de las fuentes magnetizadas autogravitantes de electrones y neutrones. En la segunda parte nos centramos en el c\'alculo de la entrop\'ia de ANs estudiada a partir del m\'etodo de Sen.

Aqu\'i se estudian ANs en cuatro y cinco dimensiones que tienen campo electromagn\'etico y constante cosmol\'ogica. Este tipo de objeto tiene un inter\'es fundamental en la F\'isica de Part\'iculas donde algunas teor\'ias de campo predicen la aparici\'on de soluciones de este tipo debido a que la curvatura se modifica por la energ\'ia y no por la masa del objeto. De igual manera conocer estas soluciones, podr\'ia ser interesante para la Cosmolog\'ia y la Astrof\'isica.

La primera parte de las tesis consta de 4 cap\'itulos. El primero es introductorio y los cap\'itulos 2,3,4, contienen la contribuci\'on original del autor a esta tem\'atica.

En el cap\'itulo introductorio describimos las ecuaciones de la TGR, dos de sus soluciones exactas, se explica que son las singularidades y la din\'amica de un fluido covariante.

El segundo cap\'itulo se dedica a estudiar el gas de electrones magnetizado y degenerado autogravitante, se escriben las ecuaciones de estado, se estudia la din\'amica y se discuten los resultados num\'ericos obtenidos.

El tercer cap\'itulo se dedica a estudiar el gas de neutrones magnetizado y degenerado autogravitante, se escriben las ecuaciones de estado, se estudia la din\'amica y se discuten los resultados num\'ericos obtenidos.

El cuarto cap\'itulo se dedica a estudiar cualitativamente las condiciones que frenar\'ian el colapso. Esto se hace perturbando el espacio-tiempo Bianchi I y analizando la ecuaci\'on de Raychaudhuri \cite{RMW_GR}.

La segunda parte de la tesis se divide en dos cap\'itulos. El primero es introductorio y en \'el se exponen los resultados de la entrop\'ia de ANs a partir de la TGR y se explica el mecanismo de Sen. El segundo cap\'itulo contiene el aporte original del autor a esta tem\'atica. O sea, la aplicaci\'on del mecanismo de Sen al estudio de la entrop\'ia de ANs, obteniendo soluciones en 4 y 5 dimensiones teniendo en cuenta el conjunto de invariantes de Riemann. Se discuten adem\'as los resultados obtenidos.

Finalmente nos dedicamos a enumerar las conclusiones de la tesis y las perspectivas del trabajo futuro. 
\newpage
\part{GASES AUTO-GRAVITANTES MAGNETIZADOS}
\newpage
%
%

%
\section[RELATIVIDAD GENERAL]{RELATIVIDAD GENERAL}
\label{Capitulo_1_RG}
\subsection{Fuentes de campo en Relatividad Especial}
 En esta secci\'on obtenemos en forma heur\'istica las ecuaciones de Einstein, las cuales son las ecuaciones de campo de la TGR. Como punto de partida consideramos la Teor\'ia de Relatividad Especial (TRE), la cual se basa en suponer un espacio-tiempo seudo-euclideano de 4-dimensiones de Minkowski, cuya m\'etrica en coordenadas cartesianas $x^a=(x,y,z,t)$ es, 
\begin{equation} ds^2 = \eta_{ab} dx^a dx^b = -dt^2+ dx^2+ dy^2+dz^2,\label{mink}\end{equation}
donde $\eta_{ab}=\hbox{diag}[-1,1,1,1]$. Las transformaciones de Lorentz, representadas por las matrices de Lorentz $\Lambda^{a}_{\,\,\,b}$, describen las transformaciones de coordenadas entre sistemas de referencia inerciales $x^a$ y $\tilde x^b$ arbitrarios en el espacio de Minkowski. La forma expl\'icita de dichas transformaciones es,
\begin{equation}
  x^{a}=\Lambda^{a}_{\,\,\,b}\,\tilde{x}^{b}, \qquad  \tilde{x}^{a}=\tilde{\Lambda}^{a}_{\,\,\,b}\,x^{b}\,,   
 \end{equation}
donde las matrices satisfacen  $\Lambda^{a}_{\,\,b}\tilde{\Lambda}^{b}_{\,\,c}=\delta^{a}_{\,\,\,c}$. Las transformaciones de Lorentz describen rotaciones arbitrarias (ver forma expl\'icita en p\'ag 67 de \cite{MTW}).
Toda teor\'ia f\'isica debe ser invariante bajo transformaciones de Lorentz al menos en una regi\'on local del espacio-tiempo. Esta invariancia asegura el cumplimiento de las leyes de conservaci\'on locales (de la energ\'ia, momento, etc) que debe cumplir toda teor\'ia f\'isica en sistemas de referencia inerciales arbitrarios.
 Las transformaciones de Lorentz entre dos sistemas de referencia para la 4-velocidad $u^{a}$, el 4-momento $p^{a}$ y el tensor de Maxwell $F^{a b}$ se expresan en t\'erminos de estas matrices de la forma,
\begin{equation}
  u^{a} = \Lambda^{a}_{\,\,\,b}\tilde{u}^{b},\qquad
  p^{a} = \Lambda^{a}_{\,\,\,b}\tilde{p}^{b},\qquad
  F^{a b}= \Lambda^{a}_{\,\,\,c}{\Lambda}^{b}_{\,\,\,d}\tilde{F}^{c d}.\label{Ftrans}
\end{equation}

El tensor de energ\'ia-momento $T^{a b}$, es un mapeo bilineal sim\'etrico que se transforma como un tensor de segundo orden,
\begin{equation}
  T^{a b} = \Lambda^{a}_{\,\,\,c}\Lambda^{b}_{\,\,\,d}\,\tilde{T}^{c d}. 
\end{equation}

Este tensor cuantifica el flujo local de 4-momento $p^a$ de una fuente dada sobre superficies $x^b =$ constante, y por lo tanto proporciona toda la informaci\'on din\'amica sobre las fuentes, las cuales pueden ser part\'iculas masivas, neutrinos, fotones, de ondas electromagn\'eticas, etc. Dada una clase arbitraria de observadores caracterizados por un campo vectorial de 4-velocidad $u^a$, la forma matem\'atica m\'as general de $ T^{a b}$ es (ver p\'ag 7, \cite{Barrow:2006ch}),
\begin{equation}
  T_{a b}=\rho\,u_{a}\,u_{b}+p\,h_{a b}+2\,q_{(a}u_{b)}+\pi_{ab},\label{Tabgen}
\end{equation}
donde $h_{a b}=g_{a b}+u_{a}u_{b}$ es el tensor de proyecci\'on ortogonal a la hipersuperficie ortogonal a $u^{a}$.

 Como toda fuente de campo puede ser matem\'aticamente reducida a la forma (\ref{Tabgen}), se utiliza el concepto formal de ``fluido'' para describir gen\'ericamente a las fuentes.
Las componentes de $ T^{a b}$ en el marco de referencia dado por $u^a$ proporcionan las cantidades din\'amicas del fluido (para cualquier fuente) que los observadores asociados a esta 4-velocidad detectan,
\begin{itemize}
   \item $\rho=T_{a b}u^{a}u^{b}$, es la densidad de energ\'ia del sistema.
   \item $p=T_{a b}h^{a b}/3$ es, la presi\'on is\'otropica del sistema. 
   \item $q_{a}=-h_{a}^{\,\,\,b}\,T_{b c}u^{c}$ es el flujo de energ\'ia en la direcci\'on $x^{a}$. Este tensor es ortogonal a la 4-velocidad: $q_{a}u^{a}=0$.
   \item $\pi_{a b}=h_{\langle a}^{\,\,\,c}h_{b\rangle}^{\,\,\,d}T_{c d}$ \footnote{Los s\'imbolos de corchetes angulares $\langle\,\rangle$ indican simetrizaci\'on, traza nula y proyecci\'on ortogonal mediante $h_{a b}$ para el par de \'indices correspondiente. Adem\'as los par\'entesis en (\ref{Tabgen}) indican simetr\'ia.} describe la presi\'on anis\'otropica y es tambi\'en ortogonal a la 4-velocidad: $\pi_{a b}u^{a}=0$. 
\end{itemize}

Un caso caso particular importante es el llamado ``fluido perfecto'', el cual se obtiene de (\ref{Tabgen}) cuando $q_{a}$ y $\pi_{a b}$ se anulan. 
Dada una fuente cualquiera caracterizada por un tensor de energ\'ia-momento $T^{ab}$, las ecuaciones de conservaci\'on,
\begin{equation}
 T^{a b}_{\,\,\,\,,b}=0\,, \label{Tabcons}
\end{equation}
contienen informaci\'on din\'amica importante. Por ejemplo, proyectando (\ref{Tabcons}) en direcci\'on paralela y ortogonal a $u^a$ obtenemos,
\begin{eqnarray}
\hbox{Ecuaci\'on de Euler:}\qquad  u_{a}T^{a b}\,_{,b} = 0,\label{ecEu}\\
\hbox{Ecuaciones de Navier-Stokes:}\qquad h_{a b}T^{b c}\,_{,c}=0.\label{ecNS} 
\end{eqnarray}

Para el caso del campo electromagn\'etico  el tensor energ\'ia-momento tiene la forma (ver p\'ag 89, \cite{MTW}),
\begin{equation}
  T^{a b}=\frac{1}{4\pi}(F^{a c}F_{c}^{\,\,\,b}-\frac{1}{4}\eta^{a b}\,F_{i k}F^{i k}),\label{TabMax}
\end{equation}
 y es tambi\'en reducible a la forma (\ref{Tabgen}). Las ecuaciones de Maxwell se obtienen aplicando la ley de conservaci\'on (\ref{Tabcons}) a (\ref{TabMax}), 
\begin{eqnarray}
  J^{a}=F^{a b}\,_{,b}\,\,,\\
   F_{a b,c}+F_{b c,a}+F_{c a,b}=0,
\end{eqnarray}
donde el s\'imbolo ${}_{,a}$ indica derivaci\'on parcial con respecto a $x^a$. La conservaci\'on de la carga viene dada por la proyecci\'on (\ref{ecEu}) que conduce a (\ref{TabMax}), a la forma de la ecuaci\'on de continuidad $J^a\,_{,a}=0$, donde $J^a=(\rho, \vec{J})$ es el 4-vector de la corriente electromagn\'etica (con $\rho=u_aJ^a$ la densidad de carga y $\vec{J}$ la corriente de carga). 

\subsection{Sistemas de referencia no-inerciales}

Los sistemas de referencia no-inerciales corresponden a clases de observadores acelerados (por ejemplo, sujetos a rotaci\'on). Las l\'ineas de universo de dichos observadores (desde un marco inercial) son curvil\'ineas, por lo que pueden ser utilizadas para definir un sistema de coordenadas curvil\'ineas relacionado con las coordenadas cartesianas mediante la transformaci\'on general de coordenadas,
\begin{equation} y^{a'}=y^{a'}(x^b),\label{transcur}\end{equation}
de modo que un campo vectorial $V^a$ se transforma como,
\begin{equation}V^a = J^a\,_{b'} V^{b'},\label{Vtrans}\end{equation}
donde, 
\begin{equation} J^a\,_{b'} \equiv \frac{\partial x^a}{\partial y^{b'}},\label{jac} \end{equation}
es la matriz jacobiana de la transformaci\'on de coordenadas.  Es importante notar que $J^a\,_{b'}$ generaliza las matrices de Lorentz $\Lambda^a\,_b$ a coordenadas generales, de modo que la transformaci\'on (\ref{Vtrans}) se reduce a las formas (\ref{Ftrans}) cuando involucra a dos sistemas de referencia inerciales. La m\'etrica del espacio-tiempo de Minkowski en las coordenadas $y^{a'}$ se obtiene, a partir de la forma est\'andar (\ref{mink}), mediante la transformaci\'on de un tensor de segundo orden,
\begin{equation} g_{a'b'} = J_{a'}\,^c\,J_{b'}\,^d\, \eta_{cd},\label{gab}\end{equation}
donde $J_{a'}\,^c$ es la matriz inversa de $J_c\,^{a'}$. N\'otese que las componentes de la m\'etrica $g_{a'\,b'}$ son, en general, funciones de las coordenadas curvil\'ineas $y^{c'}$. Es importante que las leyes de la f\'isica puedan ser expresadas tambi\'en en forma covariante entre observadores no-inerciales, de modo que un 4-vector en un sistema de referencia sea tambi\'en 4-vector en cualquier otro (y por ende, para tensores arbitrarios y escalares que surgen de sus contracciones).

Consideremos una part\'icula libre en el espacio de Minkowski con 4-velocidad $u^a = dx^a/d\tau$, donde donde $\tau$ es el tiempo propio. El 4-momento asociado a dicha part\'icula es,
\begin{equation} p^a   = m\,\gamma\,u^a=m\,\gamma\,(1,v^{\mu}),\label{partlib}\end{equation}
donde $\mu=1,2,3$, $\gamma=1/ \sqrt{1-(v/c)^{2}},\, m$ es la masa en reposo y $v^2=\delta_{\mu\nu}v^\mu v^\nu$ con $\delta_{\mu\nu}=\hbox{diag}[1,1,1]$. Es f\'acil mostrar que la 4-velocidad y el 4-momento de la part\'icula libre, desde sistemas de referencia no-inerciales, se transforman de acuerdo a la ley (\ref{Vtrans}),   
\begin{equation} u^{a} =\frac{d x^a}{d\tau} = J^a\,_{b'} \frac{dy^{b'}}{d\tau} = J^a\,_{b'} u^{b'}.\label{u_trans}\end{equation}

Al tratarse de una part\'icula libre, su 4-aceleraci\'on debe ser nula, lo cual es evidente si consideramos el 4-momento (\ref{partlib}) desde un sistema de referencia inercial en los que $v^\mu$ es un 3-vector constante,    
\begin{equation}\dot p^a = \frac{dp^a}{d\tau}=m\,\gamma\, \dot u^a=0.\label{udot1}\end{equation}

Es evidente que $\dot u^a$ en (\ref{udot1}) se transforma como 4-vector bajo transformaciones de Lorentz entre sistemas inerciales arbitrarios (al igual que $u^a$ y $p^a$), de modo que si $\dot u^a$ se anula en un sistema de referencia inercial, se deber\'a anular en todos. Sin embargo, la 4-aceleraci\'on no se transforma como un 4-vector entre sistemas de referencia no-inerciales, lo cual es evidente al derivar (\ref{u_trans}) con respecto a $\tau$ y comparar con (\ref{Vtrans}),
\begin{equation} \dot u^a = \frac{d u^a}{d\tau} = u^a\,_{,b}u^b
= J^a\,_{b',c'}\frac{dy^{b'}}{d\tau}\frac{dy^{c'}}{d\tau}+J^a\,_{b'} \frac{d u^{b'}}{d\tau}=J^a\,_{b',c'}u^{b'}u^{c'}+J^a\,_{b'}  \dot u^{b'}.\label{a_trans}\end{equation}

Este hecho es sumamente importante, ya que se espera que una part\'icula libre carezca de aceleraci\'on (ya que no est\'a sujeta a interacci\'on alguna). Por lo tanto, para poder generalizar la invariancia de Lorentz en forma adecuada a sistemas de referencia arbitrarios (por ejemplo, no-inerciales), debemos introducir un operador diferencial que generalice a la derivada parcial ordinaria, de modo que obtengamos una definici\'on alterna de aceleraci\'on que se transforme como (\ref{Vtrans}). Dicho operador es la llamada ``derivada covariante'', la cual denotamos con el s\'imbolo ${}_{;}$ (punto y coma). La derivada covariante de un 4-vector $V^{a}$ toma la forma,
\begin{equation}   V^{a}\,_{;b} = V^{a}\,_{,b}+\Gamma^{a}\,_{b\,c}\,V^{c},\label{derivcov}\end{equation} 
donde $\Gamma^{a}_{b c}$ son los s\'imbolos de Christoffel,
\begin{equation} \Gamma^a\,_{bc}=\frac{1}{2}g^{ad}\left(g_{bd,c}+g_{cd,b}-g_{bc,d}\right),\end{equation}
y $g_{ab}$ es la m\'etrica (\ref{gab}) asociada a las coordenadas curvil\'ineas $y^{a'}$. Los s\'imbolos de Christoffel no se transforman como tensores. N\'otese que la derivada covariante se reduce a la derivada ordinaria cuando consideramos sistemas inerciales, ya que que estos sistemas est\'an asociados a la m\'etrica est\'andar de Minkowski (\ref{mink}) en la que las derivadas de los componentes son cero y por tanto $\Gamma^a\,_{bc}=0$.

Tomando en cuenta (\ref{derivcov}), es natural generalizar la 4-aceleraci\'on dada en (\ref{a_trans}) por,
\begin{equation} \dot u^a = u^a\,_{;b}u^b = (u^{a}\,_{,b}+\Gamma^{a}\,_{b\,c}\,u^{c})u^{b},\label{udot}\end{equation}
la cual podemos demostrar f\'acilmente (aunque es algebraicamente engorroso) que se transforma como (\ref{Vtrans}).

Por lo tanto, con la nueva definici\'on (\ref{udot}) podemos categorizar a una part\'icula libre por $\dot u^a =0$ en sistemas de referencia arbitrarios. Como consecuencia de este hecho, la derivada covariante $ V^{a}\,_{;b}$ se transforma como un tensor de segundo orden,
\begin{equation} V^{a}\,_{;b} = J^a\,_{c'}J^{d'}\,_b\, V^{c'}\,_{;d'},\end{equation}
lo cual es sumamente relevante, ya que las leyes de la f\'isica se expresan mediante tensores y derivadas de estos. 

Esto generaliza la invariancia de Lorentz a sistemas de referencia no-inerciales, simplemente sustituyendo en las ecuaciones tensoriales la derivada parcial ordinaria por la derivada covariante: {\it{coma}} pasa a ser {\it{punto y coma}}. Es importante enfatizar que hemos introducido el uso de coordenadas curvil\'ineas (\ref{transcur}) para describir observadores acelerados (no-inerciales) en el espacio-tiempo seudo-euclideano de Minkowski. Es necesario recalcar que las coordenadas curvil\'ineas no implican que el espacio-tiempo es curvo, ya que siempre es posible encontrar una transformaci\'on de coordenadas que reduzca la m\'etrica (\ref{gab}) a la forma estandar de Minkowski (\ref{mink}).

Una vez asumida la validez de la invariancia de Lorentz, contenida en la TRE, se pensaba hacia principios del siglo XX que el espacio de Minkowski era el escenario natural para formular las ecuaciones de campo de toda teor\'ia f\'isica (ya sea en sistemas de referencia inerciales o no-inerciales). Como ejemplo importante, se consider\'o dentro de este esquema a la Teor\'ia Electromagn\'etica, cuyas ecuaciones de campo (Maxwell) toman la siguiente forma en un sistema de referencia arbitrario,
 \begin{equation}
   J^{a}=F^{a b}_{\,\,\,\,;b}\,, \qquad F_{ab;c}+F_{bc;a}+F_{ca;b}=0.
 \end{equation}

El siguiente paso ser\'ia elaborar una teor\'ia de gravitaci\'on en el espacio-tiempo de Minkowski, que generalizara la teor\'ia newtoniana y que cumpliera con la invariancia de Lorentz. Sin embargo, los esfuerzos de Einstein por elaborar dicha teor\'ia fracasaron (ver cap\'itulo 7, \cite{MTW}). 

\subsection{Teor\'ia de la Relatividad General}

Es evidente que un campo gravitacional local puede ser emulado o ``cancelado'' mediante observadores no-inerciales (como, por ejemplo, un ascensor en caida libre). Sin embargo, existen serias dificultades para la descripci\'on no-local de campos gravitacionales mediante los sistemas de coordenadas curvil\'ineas asociados a observadores no-inerciales. A modo de ejemplo, podemos citar el caso de movimientos acelerados en trayectorias hiperb\'olicas (ver p\'ag 166, \cite{MTW}), donde se puede demostrar que en general existe una ruptura de la comunicaci\'on entre los observadores fundamentales para distancias mayores que el inverso de la aceleraci\'on (ver p\'ag 170, \cite{MTW}).

Dadas las inconsistencias en el intento de incorporar la gravedad en el marco te\'orico de la TRE (espacio-tiempo seudo-euclideano de Minkowski), Einstein plante\'o lo que ser\'a considerado como una de las propuestas te\'oricas m\'as innovadoras y revolucionarias en la historia de la ciencia: {\it{La gravitaci\'on es la manifestaci\'on din\'amica de la curvatura del espacio-tiempo}}. Esta sugerencia no invalida a la TRE, ya que todo espacio curvo es localmente plano, por lo que el espacio-tiempo curvo en el cual se manifiesta la gravedad se reduce localmente al espacio seudo-euclideano (sin gravedad) de la TRE. Esta propiedad es consistente con el poder ``anular'' localmente a todo campo gravitacional mediante observadores no inerciales. De hecho, Einstein llam\'o a esta propiedad el ``Principio de Equivalencia", seg\'un el cual, la din\'amica de todo sistema, a escala local, puede ser descrita en ausencia de gravedad por interacciones que suceden en el espacio-tiempo de Minkowski y entre observadores inerciales.

Bajo este esquema, la gravedad a nivel local se manifiesta (y es equivalente) a aceleraciones detectadas por observadores no-inerciales. Sin embargo, en escalas m\'as extensas es imposible ignorar la gravedad (curvatura del espacio-tiempo), la cual necesariamente influye en las ecuaciones de campo de toda teor\'ia al sustituir derivadas ordinarias por covariantes. Este hecho es muy relevante, ya que diversos campos que aparentemente no ``producen'' gravedad (como el campo electromagn\'etico) son efectivamente fuentes de curvatura del espacio-tiempo, y por lo tanto, fuentes de gravitaci\'on.

La nueva teor\'ia de la gravitaci\'on basada en asumirla como una manifestaci\'on de la curvatura del espacio-tiempo es conocida como la ``Teor\'ia General de la Relatividad'' (TGR).
Existen muchas evidencias experimentales que han validado a la TGR: las trayectorias curvas de los rayos de luz, el corrimiento hacia el rojo de un fot\'on bajo la influencia del campo gravitatorio, la precesi\'on en la \'orbita de Mercurio, etc \cite{MTW}. Todos estos efectos son predicciones de la TGR, y son imposibles de predecir o justificar a partir de la TRE.

En base a los razonamientos anteriores, y tomando en cuenta que todo flujo de materia-energ\'ia es una fuente de curvatura del espacio-tiempo, las ecuaciones de campo de la TGR deben tener como fuente al tensor de energ\'ia-momento $T^{a b}$, el cual es un tensor de segundo orden. Por lo tanto, la curvatura del espacio-tiempo que ser\'a asociada a estos flujos de materia-energ\'ia debe ser expresada por un tensor de curvatura que sea del mismo orden. Si llamamos $G^{a b}$ a este tensor, podemos proponer las siguientes ecuaciones de campo,
\begin{equation}
 G^{a b}=\kappa T^{a b},\label{eineqs}
\end{equation}
donde $\kappa$ es una constante de acoplamiento que ser\'a derivada m\'as adelante. El tensor $G^{ab}$ debe cumplir con las siguientes propiedades:
\begin{itemize}
 \item Debe estar relacionado mediante contracciones con el tensor de Riemanian, $R^{a}\,_{b c d}$, el cual es el tensor fundamental de curvatura.
 \item Debe ser una combinaci\'on lineal del tensor de Ricci (que resulta de contraer el tensor de Riemann) y de la m\'etrica $g_{a b}$.
 \item Debe satisfacer la condici\'on de conservaci\'on $G^{a b}\,_{;b}=0$, ya que esta condici\'on es satisfecha por $T^{a b}$.
\end{itemize}

Es f\'acil demostrar que el tensor $G_{a b}$ que cumple con estas propiedades es llamado tensor de Einstein, el cual tiene la forma,
\begin{equation}
  G_{a b}=R_{a b}-\frac{1}{2}g_{a b}R+\Lambda g_{a b},\label{eintens}
\end{equation}
donde $\Lambda$ es la constante cosmol\'ogica y la constante de acoplamiento toma la forma $\kappa=8\pi G/c^4$, la cual se obtiene recurriendo al l\'imite cl\'asico newtoniano al asumir que las velocidades asociadas a los flujos de materia-energ\'ia son mucho menores que la de la luz. Las ecuaciones de campo (\ref{eineqs}) con $G^{ab}$ dado por el tensor de Einstein (\ref{eintens}) son conocidas como ``Ecuaciones de Einstein'', y son las ecuaciones de campo de la TGR.

Los  tensores que aparecen en las ecuaciones de Einstein son sim\'etricos, de modo que en 4 dimensiones tienen 10 componentes independientes. Dada la libertad de elecci\'on de las cuatro coordenadas del espacio-tiempo, las ecuaciones independientes se reducen a 6. Las ecuaciones de Einstein son un sistema de ecuaciones diferenciales parciales no-lineales de alta complejidad, por lo que es dif\'icil encontrar soluciones exactas de ellas. 

\section*{SOLUCIONES EXACTAS}

 En la siguientes secciones presentamos dos clases de soluciones exactas de las ecuaciones de Einstein. Asimismo, se discute el concepto de {\it singularidad de curvatura}, y se derivan los par\'ametros cinem\'aticos que describen localmente la evoluci\'on de elementos de un fluido arbitrario. 
 \subsection[Soluci\'on de Schwarzschild]{Soluci\'on de Schwarzschild} 
\label{Sol_Schwarzschild}

La soluci\'on de Schwarzschild fue la primera soluci\'on exacta de las ecuaciones de Einstein (\ref{eineqs}). Esta es una soluci\'on de vac\'io ($T^{ab}=0$) que asume simetr\'ia esf\'erica, por lo que es adecuada para describir el exterior de un objeto esf\'erico. Es posible demostrar que dicha soluci\'on es adem\'as estatica (existen coordenadas en que la m\'etrica no depende del tiempo) y \'unica (Teorema de Birkhoff). La soluci\'on de Schwarzschild en su m\'axima extensi\'on anal\'itica permite describir un AN esf\'erico (por lo tanto, sin momento angular) y sin carga el\'ectrica.

Suponiendo coordenadas esf\'ericas $x^a =(t,r,\theta,\phi)$, la m\'etrica de Schwarzschild tiene la forma,

\begin{eqnarray}\label{SED_2}
\mu_{,\tau}&=&\frac{1}{\Gamma_{U,\mu}}\biggl[(2\HH-S_3)(\Gamma_M-2\Gamma_{U,\beta})\beta-3\HH(\Gamma_P+\Gamma_U)
\biggr],
\\
S_{3,\tau}&=&2\beta \Gamma_M-3S_3\HH,
\\
\HH{,\tau}&=&\beta \Gamma_M+\frac{3}{2}(\Gamma_U-\Gamma_P)-3\HH^2,
\\
\beta_{,\tau}&=&2\beta(S_3-2\HH),
\end{eqnarray}
\begin{equation}
  ds^{2}=-\left(1-\frac{2M}{r}\right)dt^2+\frac{dr^2}{\left(1-2M/r\right)}+r^2\,(d\theta^2+\sin^{2}\theta\,d\phi^2),\label{schw}
\end{equation}
donde $M$ es la masa del objeto esf\'erico en unidades de longitud (ya que suponemos $G/c^2=1$).  N\'otese que a distancias lejanas de la fuente ($r\gg M$) la m\'etrica (\ref{schw}) tiende al espacio plano de Minkowski (m\'etrica (\ref{mink})).

Es evidente que la m\'etrica de Schwarzschild (\ref{schw}) se torna problem\'atica cuando $r\to 2M$, ya que $g_{tt}\to 0$ y $g_{rr}\to \infty$. Como $M$ es una longitud, se le llama el ``radio de Schwarszschild''. \footnote{El radio de Schwarzschild corresponde al radio que deber\'ia tener un objeto esf\'erico para que la velocidad de escape (newtoniana) fuera igual a la de la luz.} Sin embargo, es f\'acil mostrar que el comportamiento problem\'atico de (\ref{schw}) no corresponde a condiciones f\'isicas singulares,  ya que todos los escalares de curvatura son finitos en $r=2M$, como por ejemplo, 
\begin{equation}
  I=R_{a b c d}\,R^{a b c d}=\frac{48\,M^2}{r^6}.\label{esccurv}
\end{equation}

Adem\'as, toda clase de observadores en caida libre alcanzan el valor $r=2M$, y llegan a $r=0$ en un tiempo propio finito. El problema con $r=2M$ se debe a que las coordenadas $(t,r)$ no son adecuadas para la descripci\'on de regiones del espacio-tiempo cercanas a este radio. Debido a que la m\'etrica (\ref{schw}) no es v\'alida para $r\leq 2M$, esta no da una descripci\'on completa del espacio-tiempo. Podr\'iamos omitir por el momento el valor problem\'atico $r=2M$ y extender estas coordenadas al rango $0<r<2M$, de modo que la m\'etrica completa es (\ref{schw}) para $r>2M$ y para $0<r<2M$ tomar la forma,
\begin{equation}
  ds^{2}=-\frac{dr^2}{\left(2M/r-1\right)}+\left(\frac{2M}{r}-1\right)dt^2+r^2\,(d\theta^2+\sin^{2}\theta\,d\phi^2),\qquad 0<r<2M,  \label{schw_2}
\end{equation}
con lo que el car\'acter causal de $(t,r)$ se invierte: para $r>2M$, la coordenada $t$ es de tipo tiempo y $r$ de tipo espacio, con lo cual todo observador puede seguir una l\'inea de universo dada por $r =$ constante, $t=$ arbitrario (observador est\'atico). Sin embargo, para $0<r<2M$, $r$ pasa a ser una coordenada tipo tiempo, por lo que ning\'un observador puede seguir la l\'inea de universo $r=$ constante (la cual queda fuera del cono de luz). En otras palabras, para un observador en esta regi\'on $r$ debe disminuir hasta $r=0$. Evidentemente, el radio de Schwarzschild marca este cambio en la causalidad de $r$ y $t$.

Debido a que la extensi\'on de las coordenadas discutida en el p\'arrafo anterior es singular (matem\'aticamente hablando) en $r=2M$, se utilizan otras coordenadas para entender el comportamiento del espacio-tiempo alrededor de este valor. Por ejemplo, Kruskal y Szekeres propusieron el sistema de coordenadas dado por las relaciones $(t,r,\theta,\phi) \to (T,R,\theta,\phi)$ dadas por,
\begin{eqnarray}
  \left(\frac{r}{2M}-1\right)\exp\left(\frac{r}{2M}\right) =\,R^2-T^2, \label{Sz1}\\
  \left(\frac{ct}{2M}\right)=\ln\left(\frac{T+R}{R-T}\right)=2\,\tanh^{-1}(T/R),\label{Sz2}
\end{eqnarray}
las cuales transforman la m\'etrica (\ref{schw}) a la conocida forma de Kruskal-Szekeres,
\begin{equation}
 ds^2\,=\, \frac{4\,(2M)^3}{r}\,\exp\left(\frac{r}{2M}\right)(-dT^2+dR^2)+r^2\,(d\theta^2+\sin^{2}\theta\,d\phi^2),
\end{equation}
donde ahora $r=r(T,R)$ se debe obtener de la soluci\'on del sistema de ecuaciones (\ref{Sz1}) y (\ref{Sz2}).

En esta forma de la m\'etrica, las coordenadas $(T,R)$ est\'an bien definidas para todo valor real. Las geod\'esicas nulas son l\'ineas rectas a $\pm$ 45 grados de inclinaci\'on, por lo que podemos examinar la estructura causal de este espacio-tiempo. Inmediatamente podemos identificar cuatro regiones de inter\'es,

\begin{itemize}
 \item Regi\'on I o regi\'on exterior de la soluci\'on de Schwarzschild, caracterizada por $R > |T|$; que corresponde a la m\'etrica original (\ref{schw}), y que describe al campo gravitatorio externo (vac\'io) de un objeto esf\'erico. V\'ease en el diagrama del Ap\'endice \ref{Ap_AN}, Figura \ref{KruskalDiagram_1}. 
 \item Regi\'on II o regi\'on de AN de la soluci\'on de Schwarzschild, caracterizada por $\sqrt{1+R^2}>T>|R|$. La direcci\'on de las curvas $r=constante$ son ahora tipo tiempo, por lo que todo observador dentro del cono de luz debe evolucionar hacia $r=0$. A esta regi\'on se le denomina ``Agujero Negro'' (AN), ya que los observadores dentro de ella no pueden transmitir causalmente informaci\'on alguna a observadores en la regi\'on I (sus se\~nales luminosas no llegan a $r>2M$).  
 \item Regi\'on III o regi\'on de Agujero Blanco de la soluci\'on de Schwarzschild, caracterizada por $-\sqrt{1+R^2}<T<-|R|$. Es an\'aloga a la regi\'on II, excepto que la direcci\'on causal es hacia $r$ creciente (desde $r=0$ hasta $r=2M$). Observadores en esta regi\'on no pueden ser influenciados por los de las dem\'as regiones, pero ellos pueden afectar causalmente a estas. 
 \item Regi\'on IV o regi\'on exterior paralela, caracterizada por $R<-|T|$. Esta regi\'on es id\'entica a la regi\'on I, pero est\'a desconectada causalmente. 
\end{itemize}

El valor $r=2M$ corresponde a una 2-esfera generada por segmentos de geod\'esicas nulas (v\'ease diagramas del Ap\'endice \ref{Ap_AN}), y se le denomina ``horizonte de Schwarzschild'', ya que encierra a la regi\'on I (el AN).  Como ya se coment\'o, el espacio-tiempo es regular en $r=2M$, sin embargo, los escalares de curvatura, como por ejemplo (\ref{esccurv}), divergen en el l\'imite $r\to 0$, lo cual se puede asociar a fuerzas de marea colosales que destrozar\'ian a todo objeto f\'isico. \footnote{ Las fuerzas de marea se denominan en ingl\'es ``tidal forces''.} Como los escalares de curvatura divergen, no es posible extender el espacio-tiempo a $r=0$, por lo que toda curva causal (geod\'esica temporal o nula) debe terminar en $r=0$.

El valor $r=0$ corresponde entonces a una {\it singularidad de curvatura}, la cual puede ser definida como un conjunto de valores de las coordenadas $x^a=C^a$, tales que todos los escalares del tipo (\ref{esccurv}) divergen conforme $x^a\to C^a$ a lo largo de las geod\'esicas causales en valores finitos del par\'ametro af\'in de dichas curvas. Bajo este criterio,  la soluci\'on de Schwarzschild en su m\'axima extensi\'on anal\'itica tiene una singularidad de curvatura en $r=0$.  

\subsection[Soluci\'on de FLRW]{Soluci\'on de FLRW}

Al asumir la validez del {\it{Principio Cosmol\'ogico}}
\footnote{El Principio Cosmol\'ogico es la hip\'otesis principal de la Cosmolog\'ia moderna, la cual lleva a una interpretaci\'on teoricamente consistente a un n\'umero creciente de evidencias observacionales. El principio implica que el universo a escalas lo suficientemente grandes (del orden de cientos de megap\'arsecs) es isotr\'opico y homogeneo. La isotrop\'ia exige que las cantidades observables sean independientes de la direcci\'on de observaci\'on. El Principio Cosmol\'ogico sugiere que esta propiedad sea v\'alida (a grandes escalas) para todo observador c\'osmico, lo cual implica a su vez la ``homogeneidad'', o sea, que las variables f\'isicas y geom\'etricas sean independientes de la posici\'on de los observadores, dependiendo solo del tiempo c\'osmico (el cual es el tiempo propio de estos). Estas propiedades solo son satisfechas por la m\'etrica Robertson-Walker.},
la geomet\'ia del espacio-tiempo viene a ser descrita por la m\'etrica de Robertson-Walker (RW), 
\begin{equation}\label{Metr_RW}
  \hbox{d}s^2 = -\hbox{d}t^2+\frac{R^2(t)\,\left[\hbox{d}r^2+r^2\left(\hbox{d}\theta^2+
\sin^2 \theta \hbox{d}\varphi^2\right)\right]}{\left(1+\frac{k}{4}r^2\right)^2},  
\end{equation}
donde  $k=0,\pm 1$ representa la 3-curvatura (cero, positiva o negativa) de las secciones espaciales $t=$ constante.

Los espacio-tiempos asociados a esta m\'etrica, para diferentes fuentes del campo, son los llamados modelos Friedman-Lema\^\i tre-Robertson-Walker (FLRW), los cuales son ampliamente utilizados en la Cosmolog\'ia contempor\'anea %
\footnote{Las iniciales FLRW corresponden a Friedman, Lema\^\i tre, Robertson y Walker, que son los f\'isicos y matem\'aticos involucrados en el estudio de estos modelos. El f\'isico ruso Alexander Friedman obtuvo entre 1922-1924 la primera soluci\'on de las ecuaciones de campo (para polvo) con la m\'etrica RW. Esta solucion fue generalizada extensamente en 1927, para diversas fuentes de campo y valores de la constante cosmol\'ogica, por el jesuita belga Georges Lema\^\i tre. Entre 1935 y 1936, los matem\'aticos ingl\'es y norteamericano, Robertson y Walker, estudiaron las propiedades formales de la m\'etrica RW, probando rigurosamente que esta representa a la \'unica geometr\'ia del espacio-tiempo seudo-riemanniano que satisface al Principio Cosmol\'ogico.}.

El tensor de energ\'ia-momento compatibe con la m\'etrica (\ref{Metr_RW}) en una representaci\'on com\'ovil
\footnote{En una representaci\'on com\'ovil, la 4-velocidad asociada a la m\'etrica RW es $u^a=\delta^a\,_0$.}
es necesariamente de la forma del fluido perfecto, 
\begin{equation}
T^a\,_b=\rho\,u^a\,u_b+p\,h^a\,_b=\hbox{diag}[-\rho(t),\,p(t),\,p(t),\,p(t)],  \label{TEM_FP}
\end{equation}
donde $\rho=u_au_bT^{ab}$ y $p=(1/3)h_{ab}T^{ab}$ pueden ser identificados con la densidad de masa-energ\'ia total y la presi\'on
medidas por un observador com\'ovil (la m\'etrica RW en esta representaci\'on no admite presiones anisotr\'opicas ni flujos de energ\'ia).
Las ecuaciones de Einstein son,
\footnote{Consideramos $\Lambda$, la constante cosmol\'ogica, como una forma adicional en la densidad de energ\'ia cumpliendo con el
v\'inculo $p_{\Lambda}+\rho_{\Lambda}=0$, lo que se conoce como ``energ\'ia del vac\'io'' o ``vac\'io cu\'antico'' y puede ser incorporado
en las cantidades totales $\rho$ y $p$ reemplazando $\rho+\Lambda$ y $p-\Lambda$.} 
%
   \begin{eqnarray}
     8\pi\rho &=& \frac{3(\dot R^2+k)}{R^2}=-G^t\,_t\,,  \label{EE_FLRW_1}\\
     8\pi p &=& -\frac{\dot R^2+k}{R^2}-\frac{2\ddot R}{R}=G^r\,_r=G^\theta\,_\theta=G^\varphi\,_\varphi. \label{EE_FLRW_2}
   \end{eqnarray}
%
Las ecuaciones (\ref{EE_FLRW_1})--(\ref{EE_FLRW_2}) pueden ser complementadas con la ley de conservaci\'on $u_a\nabla_{b}T^{a b}=0$,
\begin{equation}\label{EEB_RW}
  \dot\rho+3(\rho+p)\frac{\dot R}{R}=0\,, 
\end{equation}
 la cual provee una condici\'on de integrabilidad. Por tanto, tenemos dos ecuaciones efectivas (cualesquiera dos de las tres ecuaciones
(\ref{EE_FLRW_1})--(\ref{EEB_RW})) para tres inc\'ognitas $(\rho,p,R)$, por lo que hace falta especificar un v\'inculo extra, o
``ecuaci\'on de estado'', para obtener un sistema de ecuaciones determinado. La ecuaci\'on de estado  no puede ser
proporcionada por la TGR, sino que debe ser sugerida por consideraciones f\'isicas concernientes a la fuente de materia.
Es \'util reescribir las ecuaciones (\ref{EEB_RW}) y (\ref{EE_FLRW_1}) como,
\begin{eqnarray}
 \frac{\hbox{d}}{\hbox{d}t}\left(\rho R^3\right)+p\frac{\hbox{d}}{\hbox{d}t}\left(R^3\right)=0, \label{EEEB_1}\\ 
\dot R^2 = \frac{8\pi}{3}\rho(R)\, R^2 -k.  \label{EEEB_2}
\end{eqnarray}

Dada una relaci\'on entre $p$ y $\rho$, podemos resolver (\ref{EEEB_1}) y obtener una relaci\'on funcional del tipo $\rho=\rho(R)$,
con la cual la ecuaci\'on (\ref{EEEB_2}) se convierte en una ecuaci\'on de evoluci\'on efectiva, conocida como ``ecuaci\'on de Friedmann'' cuya integraci\'on ser\'a de la forma $R=R(t)$.
%
\subsubsection{Fuentes de materia--energ\'\i a}

Una ecuaci\'on de estado de uso sumamente com\'un es la ecuaci\'on barotr\'opica $p=p(\rho)$, y en particular la conocida ``ley gamma'' dada por $p=(\gamma-1)\rho$, donde $\gamma$ es una constante. Ciertos valores espec\'ificos de $\gamma$ se pueden asociar a situaciones f\'isicas, por ejemplo $(\gamma=4/3)$ modela radiaci\'on incoherente, $(\gamma=1)$ modela polvo (gas de presi\'on nula). El caso $\gamma=0$ modela la energ\'ia del vac\'io (constante cosmol\'ogica) y lleva a los modelos de de Sitter. Otros tipos de fuentes de materia, tales como campos escalares o cuerdas c\'osmicas, pueden ser acomodados en el tensor de energ\'ia-momento (\ref{TEM_FP}), aunque en estos casos la 4-velocidad solo tiene sentido formal. Diversas fuentes de materia son tratadas en el Ap\'endice \ref{AP_Fuentes_Materia}.

\subsubsection{Singularidades de curvatura}

De todas las relaciones tipo $\rho=\rho(R)$ obtenidas en los ejemplos del Ap\'endice \ref{AP_Fuentes_Materia}, es evidente que $\rho$ diverge para cuando $R\to 0$ \footnote{El \'unico caso de fuente que logra escapar a este an\'alisis, es el caso de energ\'ia del vac\'io, en donde la densidad de energ\'ia es igual a la constante $\rho^{(vacio)}=\Lambda/8\pi$.}. Esto indica que los modelos FLRW, con las ecuaciones de estado consideradas, inician su evoluci\'on en una singularidad de curvatura escalar, marcada por $R=0$, y caracterizada por la divergencia de todos los escalares construidos a partir del tensor de Riemann y de Ricci. Los modelos FLRW comienzan por este estado de singularidad y se caracterizan por estados de evoluci\'on primordiales a muy altas densidades y temperaturas (en los casos en que la temperatura pueda ser definida). Esta situaci\'on es conocida con el t\'ermino ingl\'es de ``Hot Big Bang scenario'' y la singularidad inicial es llamada singularidad de ``Big Bang''. El factor de escala es una medida de las distancias locales relativas entre observadores fundamentales, de modo que si $R\to 0$ estas distancias empiezan a disminuir y terminan por anularse en el l\'imite. Las l\'ineas de universo de observadores fundamentales convergen y as\'i se caracteriza el Big Bang como un punto c\'austico. 

 
\subsection{Forma covariante de la din\'amica de fluidos y colapso gravitacional}

Podemos definir, para toda congruencia de observadores c\'osmicos asociados a un campo de 4-velocidades $u^{a}(x^b)$, par\'ametros covariantes que determinan el comportamiento cinem\'atico de elementos de volumen de un fluido dados por las mismas l\'ineas de universo de los observadores en las hipersuperficies 3--dimensionales espaciales y ortogonales a $u^a$. Dichos par\'ametros surgen de la descomposici\'on irreducible y covariante de la derivada covariante,
\begin{equation}u_{a;b} = \frac{\theta}{3} h_{ab}+\sigma_{ab}+\omega_{ab}-\dot u_a u_b\,,\end{equation}
donde,
\begin{itemize}
\item $\theta=u^a\,_{;a}$ es el escalar de expansi\'on, el cual describe el cambio isotr\'opico del volumen de los elementos de fluido.
\item $\dot u_a = u_{a;c}u^c$ es la 4--aceleraci\'on, la cual determina que tan diferente es la evoluci\'on de las l\'ineas de universo de los observadores en caida libre, ya que en \'esta \'ultima dichas l\'ineas de universo ser\'\i an geod\'esicas ($\dot u_a =0$). 
\item $\sigma_{a b}=h^c\,_{a}h^d\,_{b}\,u_{(c;d)}-(\theta/3)h_{ab}+\dot u_au_b$ es el tensor de deformaciones o cizalladura, el cual determina la deformaci\'on en las direcciones dadas por sus autovalores (sin cambio de volumen) de los elementos del fluido. La notaci\'on ${}_{(a;b)}$ indica simetrizaci\'on sobre los \'indices $a,b$.
\item $\omega_{ab}=\,h^c\,_{a}h^d\,_{b}\,u_{[c;d]}$ es el tensor de vorticidad, el cual determina la rotaci\'on de elementos del fluido (a volumen constante) en un eje definido por sus autovalores. La notaci\'on ${}_{[a;b]}$ indica anti--simetrizaci\'on sobre los \'\i ndices $a,b$.      
\end{itemize}

Supondremos que la din\'amica de los observadores c\'osmicos viene dada por las ecuaciones de Einstein $G^{ab}+\Lambda\,g^{ab}= 8\pi T^{ab}$, \footnote{Aqu\'i hemos tomado $\kappa=8\pi\,G/c^4$, con $G/c^4=1.$} donde $\Lambda$ es la constante cosmol\'ogica y el tensor de energ\'\i a--momento toma su forma m\'as general posible dada por (\ref{Tabgen}). Es posible demostrar que el tensor de Ricci $R_{ab}$ asociado al espacio--tiempo de los observadores satisface la siguiente ecuaci\'on,
\begin{equation} \dot \Theta+\frac{\Theta^2}{3}=-R_{ab}\,u^a\,u^b+h^b\,_{a}\dot
u^a\,_{;b}+2(\omega^2-\sigma^2)+{\dot{u}}_{a}\,\dot{u}^{a},\label{raych0}\end{equation}
donde  $\omega^2\equiv (1/2)\omega_{ab}\,\omega^{ab}$ y $\sigma^2\equiv
(1/2)\sigma_{ab}\,\sigma^{ab}$. Si consideramos el  tensor de energ\'\i a--momento (\ref{Tabgen}), y las ecuaciones de Einstein que implican la relaci\'on,
\begin{equation} R_{ab}\,u^a\,u^b+\frac{1}{2}\,R^a\,_a=8\pi\,T_{ab}\,u^a\,u^b=8\pi\,\rho-\Lambda.\end{equation}

Dado que $-R^a\,_a=-T^a\,_a=\rho+3p$, podemos transformar (\ref{raych0}) en su forma conocida como ``ecuaci\'on de Raychaudhuri''\footnote{Dicha ecuaci\'on fue descubierta independientemente por Amal Kumar Raychaudhuri y por Lev D Landau. Para ver demostraci\'on a partir de las geod\'esicas cercana revisar Ap\'endice \ref{AP_ECU_RAYCHA}.},
\begin{equation} \frac{\ddot L}{L}=-4\pi(\rho+3p)+\Lambda+2(\omega^2-\sigma^2)+h^b\,_{a}\dot u^a\,_{;b}+\dot u_a\,\dot u^a, \label{raych}\end{equation}
donde hemos definido al factor de escala $L$ como el escalar que satisface, 
\begin{equation} \frac{\dot L}{L} = \frac{\theta}{3}. \label{L}\end{equation}

La ecuaci\'on de Raychaudhuri (\ref{raych}) ofrece una simple validaci\'on general de nuestras expectativas intuitivas de como el efecto din\'amico de la gravitaci\'on como curvatura del espacio-tiempo (dado por $R_{ab}u^au^b$) se traduce tambi\'en en el efecto de una ``fuerza atractiva'' que se manifiesta por la aceleraci\'on $\ddot L/L$ de las distancias caracter\'\i sticas ($L$)  entre dos elementos de fluido definidos por $u^a$. 
\begin{itemize}
  \item Los t\'erminos en (\ref{raych}) que favorecen el colapso para $\tau>\tau_0$ son $\rho+3p>0$ y la deformaci\'on $\sigma^2$. Vale la pena remarcar que a la condici\'on $\rho+3p>0$ se le denomina ``condici\'on de energ\'\i a fuerte'', y es satisfecha por la mayor\'ia de las fuentes de masa--energ\'\i a conocidas.  
  \item Los t\'erminos que en (\ref{raych}) que no favorecen (se oponen) al colapso para $\tau>\tau_0$ son $\rho+3p<0$ (violaci\'on de la condici\'on de energ\'\i a fuerte), vorticidad no nula $\omega^2$, 4--aceleraci\'on (l\'ineas de universo que no son geod\'esicas) y una constante cosmol\'ogica $\Lambda$ no nula y positiva. 
\end{itemize}

El simple an\'alisis cualitativo de la ecuaci\'on de Raychaudhuri que hemos hecho es la base de  los llamados ``teoremas de singularidad'' (ver cap\'itulo 9, \cite{RMW_GR}). En la pr\'actica, los elementos de fluido, cuyos t\'erminos en (\ref{raych}) favorecen el colapso,  necesariamente evolucionan en un tiempo propio, $\tau$, finito hacia un estado singular caracterizado por $\theta\to -\infty$ y volumen espacial nulo (colapso gravitacional). Debido a (\ref{raych0}), los escalares de curvatura (asi como la densidad y presi\'on) tender\'an a infinito para ese tiempo propio finito, lo cual define a una singularidad de curvatura.

En la TGR, existen muchas versiones de los teoremas de singularidad de Penrose-Hawking. 
Muchas versiones establecen que si existen superficies nulas atrapadas y la densidad de energ\'ia es no negativa, entonces all\'i existen geod\'esicas de longitud finita las cuales no pueden ser extendidas. Estos teoremas, estrictamente hablando, prueban que al menos una geod\'esica temporaloidea o nula puede ser finitamente extensible, solamente hacia el pasado, pero existen casos en los cuales las condiciones de estos teoremas se obtienen de tal forma que todos los caminos en el espacio-tiempos hacia el pasado terminan en una singularidad.  Por otro lado la conjetura de Penrose asegura como \'unica singularidad desnuda, la singularidad inicial de Big Bang al principio del origen de nuestro Universo, en otro caso se violar\'ian las condiciones de energ\'ia para ciertos sistemas f\'isicos. Si nos guiamos por esta conjetura, que no ha sido demostrada, entonces s\'olo nos queda pensar en singularidades de colapso donde la singularidad esta escondida dentro del horizonte de eventos del sistema.

\newpage
\section[ESTUDIO DE UN GAS MAGNETIZADO Y AUTO-GRAVITANTE DE ELECTRONES]{\large {ESTUDIO DE UN GAS MAGNETIZADO Y AUTO-GRAVITANTE
DE ELECTRONES}}\label{Cap_de_Electrones}


%
%

\subsection{Introducci\'on}
Indiscutiblemente, los objetos compactos como lo son las EBs pueden ser interpretados como laboratorios astrof\'isicos donde la existencia de altas densidades y campos magn\'eticos fuertes plantean un reto para su estudio. En principio suponer que una secci\'on de volumen de un objeto compacto se encuentre inmerso en un campo magn\'etico fuerte no implica estar muy lejos de la realidad. Adem\'as, cuando decimos campos magn\'eticos fuertes nos referimos a campos muy superiores a los obtenidos en laboratorios terrestres. Es muy interesante estudiar dicho sistema, las l\'ineas del campo magn\'etico y su relaci\'on con la presencia de gravedad. El posible v\'inculo con las condiciones de colapso del sistema en general aumentar\'ia la importancia de dicho estudio.

Por otra parte, gran cantidad de modelos ya sean te\'oricos, fenomenol\'ogicos y n\'umericos han sido propuestos para la descripci\'on de estrellas EBs magnetizadas. Entre los modelos m\'as simples (no magnetizados), est\'a el modelo de Landau (ver \cite{Lattimer1,Lattimer2,Kubica}) que fija la masa en un l\'imite finito para estos objetos. Generalmente un modelo consistente para aplicarlo sobre una estrella EB magnetizada debe requerir considerar c\'odigos num\'ericos hidrodin\'amicos modelando las ecuaciones de Einstein--Maxwell con, al menos, condiciones de sim\'etricas y de frontera.

Es conocido que la teor\'ia de Einstein no produce ecuaciones de estado, por tanto dichas ecuaciones se deben importar desde otra teor\'ia, por ejemplo, la estad\'istica, la teor\'ia de campo, o tambi\'en pueden tener origen fenomenol\'ogico. En el caso especial de las EBs el caso m\'as simple de ecuaci\'on de estado ser\'ia suponer un gas ideal de fermiones magnetizado y a temperatura cero, pues estas estrellas en su etapa final se enfr\'ian. Por tanto, considerar un gas degenerado no resulta ser una mala aproximaci\'on.

Sin embargo, se puede obtener informaci\'on importante sobre la din\'amica local de un gas de Fermi magnetizado y auto-gravitante a partir de la evoluci\'on de esta fuente en un contexto mucho m\'as simple de la geometr\'ia espacio-tiempo. Lidiar con espacios-tiempo m\'as realistas nos har\'ian desembocar en sistemas de ecuaciones de compleja soluci\'on num\'erica. Por tanto, en este cap\'itulo nos proponemos estudiar la din\'amica de un gas magnetizado y auto-gravitante de electrones a temperatura cero. El presente cap\'itulo generaliza el trabajo previo hecho en ~\cite{AAS}.

La ecuaci\'on de estado que consideramos para describir un volumen local de una estrella EB magnetizada es la obtenida en \cite{Canuto1,Canuto2,Canuto3} y posteriormente es desarrollada para fuentes m\'as diversas en \cite{Aurora1,Aurora2,Aurora3,Chakra1,Chakra2}. En estos trabajos se discuten ecuaciones de estado para gases fuertemente magnetizados que contienen presiones anis\'otropas.
  En estos casos las presiones, $p_{\parallel}$ paralela y $p_{\perp}$ perpendicular al campo son diferentes. La presencia del campo magn\'etico provoca estas anisotrop\'ias y la posibilidad de ``colapso'' se sugiere en la direcci\'on perpendicular al campo. Estos resultados nos llevan a pensar en extender tales estudios al marco de la TGR para configuraciones magnetizadas, pregunt\'andonos si se mantendr\'a dicho ``colapso'', y si este podr\'ia dar paso a una singularidad tipo anis\'otropa a lo largo de la direcci\'on del campo magn\'etico. A este tipo de singularidades se les define como singularidades tipo ``l\'inea'' o ``cigarro'', en contraste con las singularidades tipo ``punto'' de sistemas is\'otropos y altamente sim\'etricos\footnote{Ver definiciones de tipos de singularidad en \cite{WainwrightEllis}.}.

En aras de investigar la din\'amica local y el colapso de un gas magnetizado en el contexto de la TGR, nosotros consideramos una de las configuraciones compatibles con presiones anis\'otropas. O sea el espacio tiempo Bianchi-I el cual es descrito en t\'erminos de una m\'etrica de Kasner.

El argumento m\'as claro para usar la geometr\'ia Bianchi es que esta puede proveer una descripci\'on simplificada de la din\'amica local de un elemento de fluido de un gas magnetizado en una configuraci\'on realista, as\'i este volumen desde el punto de vista termodin\'amico, intercambia part\'iculas y energ\'ia con el resto del sistema. Entonces, podemos considerar a este volumen como parte de una distribuci\'on gran can\'onica asociada a todo el gas. Adem\'as, supondremos que el elemento se encuentra lejos de la frontera del sistema donde los efectos de frontera no influyan en la din\'amica de dicho volumen.
O sea, en el caso que sea un objeto compacto, dicho volumen se encontrar\'ia en el centro del sistema, o muy cerca del centro donde el campo magn\'etico se puede suponer que se encuentra apuntando en una direcci\'on preferencial.
Evidentemente, un tratamiento como este es aproximado y no se puede asociar a la descripci\'on del objeto compacto real donde el campo magn\'etico var\'ia con el radio y la aproximada geometr\'ia esf\'erica del objeto tambi\'en ser\'ia importante. No obstante este estudio tiene el valor de acercarnos al estudio din\'amico de un volumen central elemental de dichos sistemas.

Una vez que tomamos el espacio-tiempo Bianchi-I con m\'etrica Kasner como la m\'etrica del campo asociado a un gas magnetizado, el sistema de ecuaciones de Einstein--Maxwell se reduce a un sistema aut\'onomo de cuatro ecuaciones diferenciales. Este sistema representa un espacio de fase de cuatro dimensiones que puede ser estudiado tanto num\'ericamente como cualitativamente, usando las t\'ecnicas ya estandarizadas para sistemas din\'amicos.
Veremos que las magnitudes f\'isicas que caracterizan el espacio de fase ser\'an convenientemente adimensionalizadas. Ellas ser\'an la componente independiente del tensor de deformaciones $S_{\mu \nu}$, el escalar de expansi\'on $\HH$, el campo magn\'etico $\beta$ y el potencial qu\'imico $\mu$ normalizados respectivamente.
Todas las otras cantidades pueden ser expresadas en t\'erminos de estas cuatro cantidades b\'asicas. El examen num\'erico del sistema aclara el tipo de singularidad de colapso que puede tener el volumen en dependencia de las condiciones iniciales.

Las ecuaciones de estado que usaremos aqu\'i son estrictamente v\'alidas para densidades del orden de (al menos) $10^{10}\,kg\,m^{-3}$ y $10^{18}\,kg\,m^{-3}$ esperadas para objetos compactos como las EBs y ENs. Un gas fuertemente magnetizado y altamente degenerado en una EN est\'a en un estado muy cercano a la superfluidez con conductividad infinita (ver p\'ag 291, \cite{Shapiro1}). En estas condiciones el papel de la viscosidad es menor, aunque uno podr\'ia considerar la posibilidad de disipaci\'on o fen\'omenos de transporte, tales como disipaci\'on de la energ\'ia rotacional en ondas electromagn\'eticas o gravitacionales (ver, \cite{Shapiro1,Shapiro2,Shapiro3}). Sin embargo, incluso si la viscosidad no es significativa (al menos para ENs), la raz\'on m\'as importante del por qu\'e la despreciaremos (as\'i como otros efectos disipativos) en nuestros estudios, es por mantener la simplicidad del modelo. O sea, queremos mantener el problema tratable matem\'aticamente. Nosotros pensamos que el caso de equilibrio t\'ermico es suficiente como primera aproximaci\'on para este modelo y dejamos el caso de fen\'omenos de transporte disipativo para trabajos futuros.

\subsection{Ecuaciones de Estado de un gas magnetizado de electrones}

Como ya hemos visto en la introducci\'on, una de las m\'{a}s importantes propiedades de un sistema formado por un gas magnetizado de fermiones, es que exhibe una anisotrop\'{\i}a en las presiones. Esta anisotrop\'{\i}a puede ser directamente asociada con la cuantizaci\'on de los niveles de energ\'{\i}a por la presencia del campo. Cl\'{a}sicamente, el electr\'{o}n orbita en el plano perpendicular al campo magn\'{e}tico formando c\'{\i}rculos, con velocidad angular constante. Su movimiento pueden ser descompuesto en el movimiento de dos osciladores arm\'{o}nicos simples.
Cuando este oscilador arm\'{o}nico simple es cuantizado, sus niveles de energ\'{\i}a vienen dados por la expresi\'{o}n,
\begin{equation}
E(p_{z},n,r)  =\pm mc^2 [1+(\frac{p_{3}}{mc})^2+2\beta
(n+r-1)]^{1/2}. \label{Espectro_Electrones}
\end{equation}

 Este espectro de energ\'ia para los electrones se obtiene resolviendo la ecuaci\'on de Dirac en presencia de campo magn\'etico constante en la direcci\'on $z$. En (\ref{Espectro_Electrones}) $p_{3}$, es el momentum lineal en la direcci\'{o}n $z$, $\beta=\B/\B_{c}$ el cociente entre el campo magn\'{e}tico y el campo magn\'{e}tico cr\'{\i}tico $\B_{c}=m^2c^3/e\hbar=4.414\times 10^{9}T$. Los signos + y - se refieren a electrones y a positrones respectivamente, $r=1,2$ y $n=0,1,2...$, son los n\'{u}meros cu\'{a}nticos de spin y  n\'{u}mero cu\'{a}ntico principal de la \'{o}rbita del electr\'{o}n.

En el l\'{\i}mite de $n$ muy grande, el \'{u}ltimo sumando en la expresi\'{o}n de la energ\'{\i}a corresponde con la suma de los cuadrados de los momentum lineales en las direcciones $x$ y $y$, es decir,
\begin{equation}
2\frac{\B}{\B_{c}}(n+r-1)\longrightarrow
(\frac{p_{x}}{mc})^2+(\frac{p_{y}}{mc})^2,  \label{Limite_de_Campo_cero_Elect}
\end{equation}
 el gas se transforma en un gas de Fermi ordinario. O sea, se puede definir el par\'{a}metro adimensional,
\begin{equation}
\delta= \frac{<p^2_{3}>}{2\beta (mc)^2}\,.
\end{equation}

As\'{\i} las propiedades magn\'{e}ticas del gas son cl\'{a}sicas si $\delta >>1$ o cu\'{a}nticas si $\delta <<1$. Los campos magn\'{e}ticos de inter\'{e}s para el gas de electrones describiendo una EB, est\'an en el rango de $10^{4}\,T$ y pueden estar presentes durante procesos de colapso gravitatorio de objetos estelares.

Los gases magnetizados de Fermi no relativistas han sido estudiados extensivamente en el marco de la f\'{\i}sica
estad\'{\i}stica y sus propiedades m\'{a}s importantes est\'{a}n asociadas con el diamagnetismo de Landau y el paramagnetismo de Pauli.

En adici\'{o}n a esto veremos la forma que toman las ecuaciones de estado para el caso relativista.
El Gran Potencial Termodin\'amico $\Omega$ que se halla con la ecuaci\'on,
\begin{equation}
\Omega=kT \ln\mathcal{Z},
\end{equation}
donde \noindent $k$ es la constante de Boltzman, $T$ la temperatura, $\mathcal{Z}=Tr(\hat{\rho})$ es la funci\'on Gran Potencial Termodin\'amico, $\rho = e^{-(\hat{H}-\mu \hat{N})/kT}$, \, $\hat{H}$ es el Hamiltoniano, $\mu$ es el
potencial qu\'imico y $\hat{N}$ es el operador del n\'umero de part\'iculas.

El tensor de energ\'ia-momentum total, donde en la diagonal principal est\'an las presiones y la densidad de energ\'ia, lo obtendremos haciendo la promediaci\'on macrosc\'opica, $\mathcal{T}_{\mu \nu}=<\hat{\mathcal{T}}^{(micro)}_{\mu \nu}>$ donde $\hat{\mathcal{T}}^{(micro)}_{\mu \nu}=(\partial{\hat{\mathcal{L}}}/\partial{A^{a}_{\mu,\nu}})\,\hat{A}^{a}_{\mu,\nu}+(\partial{\hat{\mathcal{L}}}/\partial{\Psi^{a}_{\mu,\nu}})\,\hat{\Psi}^{a}_{\mu,\nu}-\delta_{\mu \nu}\hat{\mathcal{L}}$, entonces para la funci\'on de partici\'on $\Omega=-(kT)^{-1} ln{<e^{\int^{\beta}_{0}dx_{4}\int d^{3}x \mathcal{L}(x_{4},\vec{x})}>}$.
\footnote{Aqu\'i, los corchetes angulados $<\,>$ significan la integral continual de Feynman en la teor\'ia de Matsubara.}
El gas electr\'on--positr\'on es relevante en el contexto astrof\'isico, y el Gran Potencial Termodin\'amico $\Omega$ para el sector electr\'on--positr\'on tiene dos t\'erminos de la forma,
\begin{equation}
\Omega=\Omega_{se}+\Omega_{Ve},
\end{equation}
donde el primer t\'ermino en el lado derecho, es la contribuci\'on estad\'istica y el segundo la contribuci\'on del
vac\'io \cite{Aurora2}. Expl\'icitamente, tenemos que en la aproximaci\'on de un lazo,
\begin{equation}\label{omega_n}
\Omega_{se}=-\frac{e\B}{4\pi^2\xi}\sum_{\eta=0}^{\infty}\alpha_{n}\int_{-\infty}^\infty
\mathrm{d}p_{3}\mathrm{ln}[f^+(\mu_e,\xi)f^-(\mu_e,\xi)],
\end{equation}
donde $\xi=1/k_BT$, $\alpha_{n}=2-\delta_{0n}$ y $f^{\pm}(\mu_e,\xi)=(1+e^{-(E_e\mp\mu_e)\xi})$
representan respectivamente, las contribuciones para part\'iculas y
antipart\'iculas. El t\'ermino de vac\'io es dado por la
expresi\'on,
\begin{equation}\label{omega_v}
\Omega_{Ve}=\frac{1}{4\pi^2\xi}\sum_{\eta=0}^{\infty}\int_0^{\infty}
p_\bot \mathrm{d}p_\bot \mathrm{d}p_3 E_e,
\end{equation}
el cual es divergente, pero puede ser renormalizado, y para campos de intensidad $\B<10^{9}\,\mathrm{T}$, su contribuci\'on es irrelevante \cite{Aurora2} para campos de inter\'es en los objetos astrof\'isicos actuales, por tanto nosotros no tomaremos en cuenta este t\'ermino. Tampoco incluiremos en $\Omega_{se}$, la contribuci\'on cl\'asica (o de \'arbol) asociada al campo magn\'etico, pues seguimos la corriente de publicaciones en donde este t\'ermino se omite (ver \cite{Aurora3})
\footnote{Otros autores si incluyen este t\'ermino cl\'asico, que finalmente introduce cambios en las direcciones de colapso a nivel newtoniano (ver, \cite{Ferrer:2010wz}).}
, aunque posteriormente en el cap\'itulo \ref{Cap_Perturbado}, si lo tendremos en cuenta.

La ecuaci\'on (\ref{omega_n}) puede ser integrada f\'acilmente para el caso degenerado $(T=0)$. En ese caso
la funci\'on de distribuci\'on de los electrones se convierte en la funci\'{o}n paso unitario y la de los positrones es cero.
El tensor energ\'ia--momentum asociado a un gas de Fermi en presencia de un  campo magn\'etico externo y constante en el tiempo, toma la forma \cite{Aurora3},
\begin{equation}\label{tensor_e-m}
\mathcal{T}^{a}\,_{b}=(T\frac{\partial{\Omega}}{\partial{T}}+\sum{\mu_{n}\frac{\partial{\Omega}}{
\partial{\mu_{n}}}})\delta^{a}\,_{4}\,\delta^{4}\,_{b}+4F^{a c}F_{c b}\frac{\partial{\Omega}}{
\partial{F^{2}}}-\Omega\,\delta^{a}\,_{b},
\end{equation}
 as\'i que en el l\'imite de campo magn\'etico nulo obtenemos el tensor del fluido perfecto,
\,
$\mathcal{T}^{a}\,_{b}=p\delta^{a}\,_{b}-(p+U)\delta^{a}\,_{4}\delta^{4}\,_{b}$.
Las componentes del tensor (\ref{tensor_e-m}) son,
%
\begin{eqnarray}
\mathcal{T}^{3}_{\,\,\,\,3}&=& p_{\|}=-\Omega=p,  \label{componentes de T_a}
\\
\mathcal{T}^{1}_{\,\,\,\,1}&=&\mathcal{T}^{2}_{\,\,\,\,2}=p_{\perp}=-\Omega-\B\M=p-\B\M,  \label{componentes de T_b}
\\
\mathcal{T}^{4}_{\,\,\,\,4}&=&-U=-TS-\mu_{e} N-\Omega,  \label{componentes de T_c}
\end{eqnarray}
%
donde $p_{\|}=p$, $p_{\perp}$ son las presiones, paralela y perpendicular al campo magn\'etico, $U$ la densidad de energ\'ia, $S$ la densidad de entrop\'ia, $N$ la densidad de part\'iculas y $\M$ la magnetizaci\'on.
 Para un objeto compacto, en especial para una estrella EB la temperatura $T<<T_{F}$ \footnote{$T_{F}$ temperatura de Fermi y $T/T_{F}\simeq 10^{-4}$ para una ENs. Por lo tanto, el desorden t\'ermico no es responsable de la presi\'on, ni de la densidad de energ\'ia, ni de la magnetizacion del objeto compacto.}. Es por ello que una descripci\'{o}n de estas estrellas pueden hacerse suponiendo un gas degenerado de part\'{\i}culas.

Todas las variables termodin\'amicas del sistema pueden ser obtenidas a partir del Gran Potencial Termodin\'amico $\Omega$, y las ecuaciones de estado para el l\'imite de campo no nulo y degenerado ($T=0$), tendr\'an las forma expl\'icita,
%
\begin{eqnarray}
p_{\perp}&=& p-\B\M, \qquad
p= \lambda\,\beta\,\Gamma_{p}(\mu_e,\beta),  \label{EE_Electr_a}\\
U&=&\lambda\,\beta\,\Gamma_{U}(\mu_e,\beta), \qquad
\M {\B}=\lambda\,\beta\,\Gamma_{M}(\mu_e,\beta),  \label{EE_Electr_b}
\end{eqnarray}
%
$\lambda=mc^2/(4\pi^{2}\lambda_{c}^3)$ y las funciones $\Gamma_{k}$,
ser\'{a}n, (ver \cite{Aurora1}),
%
\begin{eqnarray}
\Gamma_{p}&=&\frac{a_{_0}}{3}\left(\mu^2_{e}-\frac{5}{2}\right)+\frac{1}{2}\textrm{arcsinh}\left(\frac{a_{_0}}
{\mu_{e}}\right)+\beta\sum^{s}_{n=0}\alpha_{n}(a_{n}-b_{n}-c_{n}),\ \label{KMWAP11a} \\
\Gamma_{\M}&=&\sum^{s}_{n=0}\alpha_{n}(a_{n}-b_{n}-2c_{n}), \label{KMWAP11b} \ \ \ \ \\
\Gamma_{U}&=&a_{_0}\left(\mu^2_{e}-\frac{1}{2}\right)-\frac{1}{2}\textrm{arcsinh}\left(\frac{a_{_0}}{\mu_{e}}\right)+
\beta\sum^{s}_{n=0}\alpha_{n}(a_{n}+b_{n}+c_{n}),\\
\qquad \textrm{donde,}  \nonumber\\
\alpha_{n}&=&2-\delta_{0n}, \ \ n=0,1,...    \label{KMWAP11c} \\
a_{n}&=&{\mu_{e}\sqrt{\mu^2_{e}-1-2n\beta}}, \,\,
b_{n}=\ln\left[\frac{(\mu_{e}+a_{n}/\mu_{e})}{\sqrt{1+2n\beta}}\right],\,\,
c_{n}=2n\beta b_{n}, \label{KMWAP11c} \\
s&=& I\left[\frac{\mu^2_{e}-1}{2\beta}\right],
\end{eqnarray}
%
%
y $I[X]$ denota la parte entera de este argumento $X$.
%

\subsection{Ecuaciones de Einstein--Maxwell}
 La m\'etrica de Kasner es una de las m\'etricas m\'as simples que permiten anisotrop\'ias en las presiones independientemente del origen de las mismas. En el caso de presiones anisotr\'opicas asociadas a una fuente magn\'etica, ella puede ser utilizada. Esta m\'etrica viene dada por,
\begin{equation}
{ds^2} \ = \
-c^2\,dt^2+A^2(t)\,dx^{2}+B^2(t)\,dy^2+C^2(t)\,d{z}^2.\label{KMWAP1}
\end{equation}

 Ella est\'a asociada con un espacio-tiempo Bianchi-I ``no--inclinado''\footnote{Los espacios-tiempo Bianchi inclinados tienen la 4-velocidad ``inclinada'', o sea no son sistemas com\'oviles con la velocidad en la direcci\'on temporal.} \cite{WainwrightEllis}. Para una 4-velocidad com\'ovil $u^{a}=\delta^{a}_{t}$, donde $u^{a}u_{a}=-1$, la 4--aceleraci\'on se anula y el escalar de expansi\'on $\Theta$  y el tensor de deformaciones $\sigma^{a}_{b}$ toma la forma,
\begin{eqnarray}
\sigma^{a}_{b} &=& {\textrm {diag}}[\sigma^{x}_{x},\sigma^{y}_{y},\sigma^{z}_{z},0],\label{KMWAP2a}\\
\Theta &=& \frac{\dot{A}}{A}+\frac{\dot{B}}{B}+\frac{\dot{C}}{C},\label{KMWAP2b}
\end{eqnarray}
donde,
\begin{eqnarray}
\sigma^{x}\,_{x}&=&\frac{2\dot{A}}{3A}-\frac{\dot{B}}{3B}-\frac{\dot{C}}{3C},\
\
\sigma^{y}\,_{y}=\frac{2\dot{B}}{3B}-\frac{\dot{A}}{3A}-\frac{\dot{C}}{3C},\label{KMWAP3a}
\\
\sigma^{z}\,_{z}&=&\frac{2\dot{C}}{3C}-\frac{\dot{A}}{3A}-\frac{\dot{B}}{3B},\
\ \ \ \ ( \sigma^{a}_{a}=0 ).   \label{KMWAP3b}
\end{eqnarray}
%

As\'i, consideramos como fuente de materia para esta m\'etrica el siguiente tensor de energ\'ia-momentum,
\begin{equation}
T^{a}\,_{b}=(U+P)u^{a}u_{b}+P\delta^{a}\,_{b}+\Pi^{a}\,_{b}, \ \
P=p-\frac{2\B\M}{3}, \label{KMWAP4}
\end{equation}
aqu\'i $\Pi^{a}\,_{b}$ es el tensor de presiones anisotr\'opicas, $U$ la densidad de energ\'ia, $P$ es el valor medio de las tres presiones diagonales, $\B$ el campo magn\'etico y $\M$ la magnetizaci\'on, a todos ellos los suponemos funciones del tiempo. Note que la anisotrop\'ia es producida por el campo magn\'etico $\B$. Si este campo se anula, o sea: $\B=0$, entonces el tensor de energ\'ia-momentum se reduce al tensor del fluido perfecto con presi\'on is\'otropa $P = p$. En el caso general $\B\ne 0$ el tensor $\Pi^{a}\,_{b}$ tiene la forma,
\begin{equation}
\Pi^{a}\,_{b}={\textrm {diag}}[\Pi,\Pi,-2\Pi,0], \ \ \ \Pi=-\frac{\B\M}{3},\ \ \
\Pi^{a}\,_{a}=0. \label{KMWAP5}
\end{equation} 

Las ecuaciones de campo de Einstein (ECE) asociada con la m\'etrica de Kasner y con el tensor de energ\'ia-momentum quedan como,
 %
\begin{eqnarray}
-G^{x}_{x}&=&\frac{\dot{B}\dot{C}}{BC}+\frac{\ddot{B}}{B}+\frac{\ddot{C}}{C}=-\kappa(p-\B\M),
\label{KMWAP6a}\\
-G^{y}_{y}&=&\frac{\dot{A}\dot{C}}{AC}+\frac{\ddot{A}}{A}+\frac{\ddot{C}}{C}=-\kappa(p-\B\M),
\label{KMWAP6b}\\
-G^{z}_{z}&=&\frac{\dot{A}\dot{B}}{AB}+\frac{\ddot{A}}{A}+\frac{\ddot{B}}{B}=-\kappa p,\label{KMWAP6c}
\\
-G^{t}_{t}&=&\frac{\dot{A}\dot{B}}{AB}+\frac{\dot{A}\dot{C}}{AC}+\frac{\dot{B}\dot{C}}{BC}=\kappa U,\label{KMWAP6d}
\end{eqnarray}
%
donde el punto denota la derivaci\'on respecto al tiempo propio de observadores fundamentales y $\kappa=8\pi G/c^4$.
A partir de la ecuaci\'on de balance $T^{ab}_{\ \ ;b}=0$ y de las ecuaciones de Maxwell $F^{ab}_{\ ;b}=0$ y $F_{[ab;c]}=0$, tenemos que,
%
%
\begin{eqnarray}
\dot{U}&+&(p+U)\Theta-\B\M(\frac{\dot{A}}{A}+\frac{\dot{B}}{B})=0,
\label{KMWAP7a}\\
\frac{\dot{A}}{A}&+&\frac{\dot{B}}{B}+\frac{1}{2}\frac{\dot{\B}}{\B}=0.
\label{KMWAP7b}
\end{eqnarray}
%

 Nosotros necesitamos construir un sistema auto-consistente de ecuaciones diferenciales de primer orden que puedan ser resueltas
num\'ericamente, as\'i que ser\'ia muy conveniente eliminar primeras y segundas derivadas de las funciones m\'etricas en las ecuaciones
de Einstein--Maxwell y escribirlas en t\'erminos del escalar de expansi\'on y de las componentes del tensor de deformaciones.
Siendo consecuentes con esta idea, combinamos las ecuaciones (\ref{KMWAP2a})-(\ref{KMWAP2b}),(\ref{KMWAP3a})-(\ref{KMWAP3b}),
(\ref{KMWAP6a})-(\ref{KMWAP6d}) y (\ref{KMWAP7a})-(\ref{KMWAP7b}), eliminamos las funciones $A,B,C$ junto con sus derivadas $\dot{A},\ddot{A},\dot{B},\ddot{B},\dot{C},
\ddot{C}$. Despu\'es de algunas manipulaciones algebraicas arribamos al siguiente v\'inculo,
\begin{equation}-(\Sigma^{y})^{2}-(\Sigma^{z})^{2}-\Sigma^{y}\Sigma^{z}+\frac{\Theta^{2}}{3}=\kappa U,\label{constr1}\end{equation}
y a las siguientes ecuaciones diferenciales,
%
\begin{eqnarray}
\dot{U}&+&(U+p)\Theta-\B\M(\frac{2}{3}\Theta-\Sigma^{z})=0, \label{KMWAP8a}
\\
\dot{\Sigma^{y}}&=&-\frac{\kappa}{3} \B\M-\Sigma^{y}\Theta, \label{KMWAP8b}
\\
\dot{\Sigma^{z}}&=&\frac{2\kappa}{3} \B\M-\Sigma^{z}\Theta, \label{KMWAP8c}
\\
\dot{\Theta}&=&\kappa (\B\M-\frac{3}{2}p)-\frac{
\Theta^{2}}{2}-\frac{3}{2}((\Sigma^{y}+\Sigma^{z})^{2}-\Sigma^{y}\Sigma^{z}), \label{KMWAP8d}\\
\dot{\beta}&=&\frac{2}{3}\beta (3\Sigma^{z}-2\Theta),
\quad \textrm{con}\quad \beta\equiv \B/\B_{c}\,,  \label{KMWAP8e}
\end{eqnarray}
%
donde $\Sigma^{z}=\sigma^{z}_{z}$ que la tomamos como la componente independiente del tensor de deformaciones\footnote{Las componentes
$\Sigma^{x}$ y $\Sigma^{y}$ se definen similarmente.}. 

Aunque el tensor de deformaciones se puede determinar por esta \'unica componente, para nosotros ser\'a muy conveniente, desde el punto
de vista de los c\'alculos num\'ericos, utilizar dos componentes de dicho tensor.

Las funciones $U, p$ y $\M$ ahora vendr\'an dadas por las ecuaciones de estado del gas\footnote{Aqu\'i se sobrentiende que $\mu=\mu_e$
potencial qu\'imico de los electrones, para el caso de los neutrones $\mu=\mu_{n}$.},
\begin{equation}
 p= \lambda\,\, \Gamma_{p}(\beta,\mu), \ \ \ \B\M \ = \ \lambda\,
\beta\Gamma_{{\M}}(\beta,\mu), \ \ \ U \ = \ \lambda\,
\Gamma_{_U}(\beta,\mu). \label{KMWAP9} \end{equation}

\subsection{Din\'amica de cantidades covariantes}
\label{DCC_Cap2}
Consideremos ahora las siguientes variables,
\begin{equation}
H=\frac{\Theta}{3}, \qquad\  \frac{d}{d{\tau}}=\frac{1}{H_{0}}\frac{d}{c\ dt},  \label{DV1} 
\end{equation}
y las funciones adimensionales,
\begin{equation}
S^{y}=\frac{\Sigma^{y}}{H_{0}},\,\,\, S^{z}=\frac{\Sigma^{z}}{H_{0}},\,\,\, 
\Omega=\frac{\kappa\lambda\beta}{3H^{2}_{0}},\,\,\, \HH=\frac{H}{H_{0}}, \label{DV2}
\end{equation}
donde $S^{y}$ y $S^{z}$ est\'an relacionados con las componentes $yy$ y $zz$ del tensor de deformaciones, mientras que $\Omega$
est\'a relacionado con el campo magn\'etico. El nuevo tiempo $\tau$ es un tiempo adimensional (o tiempo ``logar\'itmico''). La
cantidad $H(t)$ (debido a (\ref{KMWAP2a})-(\ref{KMWAP2b})) cambia las dimensiones de ${\textrm{cm}}^{-1}$ y el signo de $\tau$ se 
determina por el signo de $H(t)$ (ver Ap\'endice \ref{Apend_Time_Tau}). Sin embargo, nosotros hemos elegido $\Omega=\beta$ por tanto
$\kappa\lambda=3H^{2}_{0}$, para eliminar la presencia de constantes adicionales en el sistema de ecuaciones. Insertando la ecuaci\'on
de estado (\ref{KMWAP9}) y las nuevas definiciones (\ref{DV1}), (\ref{DV2}) en (\ref{constr1}) y (\ref{KMWAP8a})-(\ref{KMWAP8e}), obtenemos
el v\'inculo,
\begin{equation}
 -(S^{y})^{2}-(S^{z})^2-S^{y}S^{z}+3\HH^{2}=3\Gamma_{U},
 \label{DV3d}
\end{equation}
m\'as el sistema,
%
\begin{eqnarray}
S^{y}_{,\tau}&=& -\beta\Gamma_{\M}-3S^{y}\HH, \label{DV3a}
\\
S^{z}_{,\tau}&=& 2\beta\Gamma_{\M}-3S^{z}\HH, \label{DV3b}
\\
\HH_{,\tau}&=&\beta \Gamma_{\M}-\frac{3}{2}(\Gamma_{p}+
\HH^{2}+\frac{(S^{y}+S^{z})^{2}}{3}-\frac{S^{y}S^{z}}{3}),
\label{DV3c}
\\
\beta_{,\tau}&=&2\beta(S^{z}-2\HH),   \label{DV3e}
\\
\mu_{,\tau}&=&\frac{1}{\Gamma_{U,\mu}}\left[(2\HH-S^{z})
(\Gamma_{\M}+2\Gamma_{U,\beta})\beta-3\HH(\Gamma_{p}+\Gamma_{U})\right].
\label{DV3f}
\end{eqnarray}
%

Note que, contrario a los modelos cosmol\'ogicos discutidos en la referencia \cite{MTW} donde $H_{0}=0.59\times
10^{-26}{\textrm{m}^{-1}}$ podr\'ia jugar el rol de la constante de Hubble, en nuestro gas magnetizado de Fermi
tenemos que $H_{0}=0.86\times 10^{-10}{\textrm{m}^{-1}}$, el cual es mucho m\'as peque\~no.
Esto es l\'ogico y consistente porque indica que nuestro modelo simplificado es examinado sobre escalas locales mucho menores
que las escalas c\'osmicas. La escala $1/H_{0}\simeq 1.15\times 10^{10}\,{\textrm{m}}$ es del orden de la distancia de una unidad
astron\'omica.

Los resultados mencionados en las referencias \cite{Aurora1,Aurora2,Aurora3} para el gas de electrones muestran que para un campo magn\'etico
intenso, del orden del campo cr\'itico $B_{c}$, todos los electrones est\'an en el nivel b\'asico de Landau $n=0$, y consecuentemente se
tiene que $p_{\perp} =0$. Consideramos el estudio de como evoluciona un gas en estos casos, en el cual las funciones $\Gamma(\beta,\mu)$ se
simplifican considerablemente, tomando la siguiente forma,
%
\begin{eqnarray}
\Gamma_{p}&=&\frac{a_{_0}}{3}(\mu^2-\frac{5}{2})+\frac{1}{2}\textrm{arcsinh}(\frac{a_{_0}}{\mu})+\beta(a_{_0}-b_{_0}),\ \label{DV4a} \\
\Gamma_{\M}&=&(a_{_0}-b_{_0}), \label{DV4b} \ \ \ \ \\
\Gamma_{U}&=&a_{_0}(\mu^2-\frac{1}{2})-\frac{1}{2}\textrm{arcsinh}(\frac{a_{_0}}{\mu})+\beta(a_{_0}+b_{_0}),     \label{DV4c} 
\,\,\, \hbox{donde}  \\
a_{_0}&=&{\mu\sqrt{\mu^2-1}}, \ \ b_{_0}=\ln (\mu+a_{_0}/\mu),\
c_{_0}=0, \ \alpha_{_0}=1.\nonumber 
\end{eqnarray}
%

As\'i sustituyendo (\ref{DV4a})-(\ref{DV4c}) en (\ref{DV3a})-(\ref{DV3f}) mantenemos el sistema auto--consistente de cinco ecuaciones
diferenciales ordinarias (\ref{DV3a})-(\ref{DV3f}), con las funciones inc\'ognitas $\beta$,\,$\HH$,\,$S^{y}$,\,$S^{z}$ y $\mu$, y el v\'inculo
(\ref{DV3d}), el cual solo admite una soluci\'on num\'erica.

De las ecuaciones de estado (\ref{KMWAP9}) se deduce que el potencial qu\'imico debe satisfacer\, $\mu\geq 1$, lo que es correcto para
sistemas con densidades del orden de $\sim 10^{7}{\textrm{gm/cm}}^{3}$ o mayores. Para EBs o ENs, el potencial qu\'imico toma valores
alrededor de $\sqrt{3}\simeq 1.732$. Tambi\'en $U > 0$, entonces el potencial qu\'imico $\mu\geq 1$ y de (\ref{DV3d}) obtenemos el v\'inculo,
\begin{equation}
-(S^{y})^{2}-(S^{z})^{2}-S^{y}S^{z}+3\HH^{2} = 3 \Gamma_{U}\geq 0,
\label{DV5}
\end{equation}
as\'i nuestro espacio de fase f\'isico de 5-dimensiones, estar\'a restringido por las relaciones,
%
\begin{eqnarray}
3\HH^{2} &\geq& (S^{y})^{2}+(S^{z})^{2}+S^{y}S^{z}, \\
 \mu^2&\geq& 1+2\beta,
\\
 \beta &\geq& \frac{3\Gamma_{p}-\Gamma_{U}}{2\Gamma_{\M}}. \label{DV6}
\end{eqnarray}
%

Para modelos Bianchi-I no-inclinados solo existe una componente independiente del tensor de deformaci\'on, la cual tomamos por $S^{z}$, sin
embargo ser\'a conveniente usar tambi\'en la componente $S^{y}$ para nuestros c\'alculos num\'ericos. Adem\'as, la forma de las componentes
del tensor de deformaciones determina la forma de los coeficientes m\'etricos. De (\ref{KMWAP2a})-(\ref{KMWAP2b}) y (\ref{KMWAP3a})-(\ref{KMWAP3b}), tenemos que,
\begin{equation}
\frac{A_{,\tau}}{A}=(S^{x}+\HH),  \ \
\frac{B_{,\tau}}{B}=(S^{y}+\HH),  \ \
\frac{C_{,\tau}}{C}=(S^{z}+\HH),  \ \
\label{DV7}
\end{equation}
donde $ S^{x}=-S^{y}-S^{z}.$

\subsection{Discusi\'on y resultados}
Usando (\ref{KMWAP2a})-(\ref{KMWAP2b}) y (\ref{DV1}) podemos expresar el volumen local como $\CV\equiv ABC$ en t\'erminos de $\HH$ y el tiempo adimensional
$\tau$ como
\begin{equation}
\CV(\tau)=\CV(0)\,\exp\left(3\,{\int}^{\tau}_{\tau_0}{\HH d{\tau}}\right). \label{NRAD1}
\end{equation}

Claramente se puede notar el por qu\'e al tiempo $\tau$ se le llama tiempo ``logar\'itmico''. Note que el signo de $\HH(\tau)$ implica
expansi\'on del volumen local, si ($\HH(\tau)\,>0$) y colapso si ($\HH(\tau)\,<0$).
\subsubsection{Singularidades}

Las ecuaciones de estado que estamos considerando est\'an asociadas con objetos compactos a muy alta densidad (al menos
$\sim 10^{10}\,kg\,m^{-3}$), el rango de evoluci\'on del modelo para bajas densidades o caso diluido, no es f\'isicamente interesante para
nosotros y no ser\'a tratado. Esto significa que solo consideraremos las fases de colapso del modelo. As\'i solo necesitamos examinar las
condiciones iniciales en las cuales la expansi\'on inicial $\HH_{0}$ es negativa. Nosotros probamos num\'ericamente el modelo usando una
gran cantidad de condiciones iniciales diferentes, cubriendo todo el rango de valores de inter\'es f\'isico para los objetos compactos,
desde EBs hasta ENs. Por ejemplo: $\mu_{0}=2$ corresponde con densidades de $\sim 10^{10}\,kg\,m^{-3}$, mientras que $\beta_{0}=10^{-5}$
representa campos magn\'eticos del unos $10^{4}\,T$. Junto con $\HH_{0}<0$, consideramos en particular: $S^{y}_{0}=0,\pm 1$ y
$S^{z}_{0}=0,\pm 1$, que corresponden con casos de deformaci\'on inicial nula, y deformaciones iniciales positivas y negativas en las
direcciones $y$ o $z$.

As\'i cuando $\HH_{0}<0$ el modelo exhibe un comportamiento en general de colapso $\HH \rightarrow-\infty$, independientemente de los
valores iniciales de las otras funciones (ver ejemplos en la Figura \ref{FIG1}). En todas las configuraciones de colapso la intensidad del
campo magn\'etico diverge a infinito independientemente de las condiciones iniciales. En la Figura \ref{FIG1} mostramos varias soluciones
num\'ericas para diferentes valores del campo magn\'etico inicial. Se puede ver como al aumentar la intensidad del campo (etiquetada por
$\beta1< \beta2< \beta3$ en la figura), la expansi\'on muestra un r\'apido decaimiento a $-\infty$. Consecuentemente, siempre que aumentemos
el campo magn\'etico inicial, el tiempo de colapso decrece.
\FIGURE{\epsfig{file=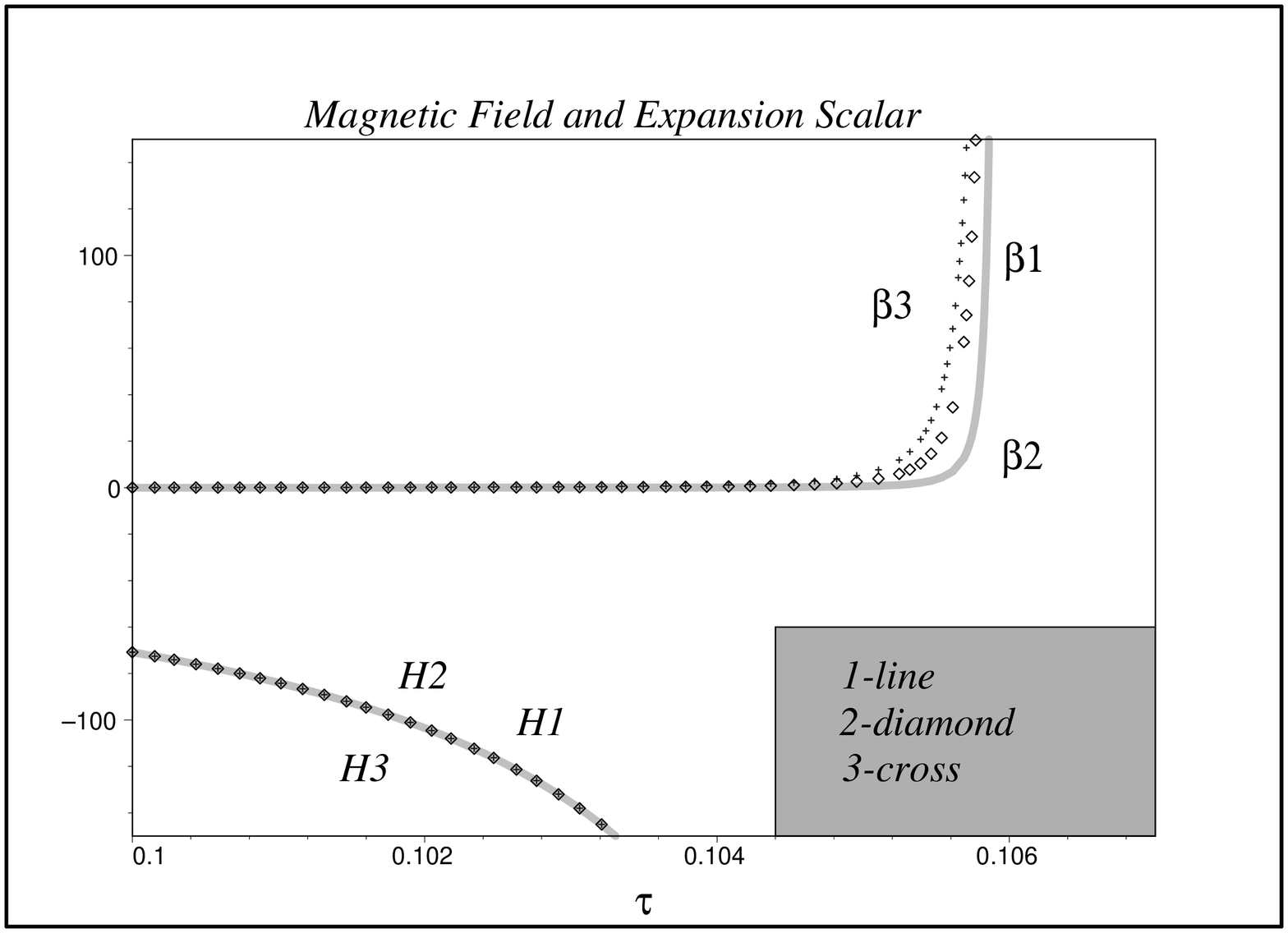,height=14cm,width=14cm,angle=270}
\caption{ Soluciones num\'ericas para el campo magn\'etico adimensional $\beta(\tau)$ y para el escalar de expansi\'on
$\HH(\tau)$. Aqu\'i disponemos de tres condiciones iniciales diferentes, $S^{x}(0)=0,S^{y}(0)=-1, S^{z}(0)=1, \mu(0)=2,$ y
$\beta_{1}(0)=10^{-5},\beta_{2}(0)=5 \times 10^{-5},\beta_{3}(0)=10^{-4}$ respectivamente, para el campo magn\'etico inicial.
Entonces debido a (\ref{DV5}), la expansi\'on inicial debe tomar los valores $\HH_{1}(0)=\HH_{2}(0)=\HH_{3}(0)=-4.82$.
Los resultados num\'ericos para los tiempos de colapso son $\tau_{1}=0.1059, \tau_{2}=0.1058$, y $\tau_{3}=0.1057$
respectivamente.} \label{FIG1}
}
Usamos condiciones iniciales con deformaci\'on inicial no necesariamente en la direcci\'on del campo magn\'etico, as\'i que las
singularidades anis\'otropas del tipo ``cigarro'' emergen a lo largo de cualquiera de estas direcciones, o sea, el colapso puede ser
paralelo o perpendicular al campo magn\'etico.
\FIGURE{\epsfig{file=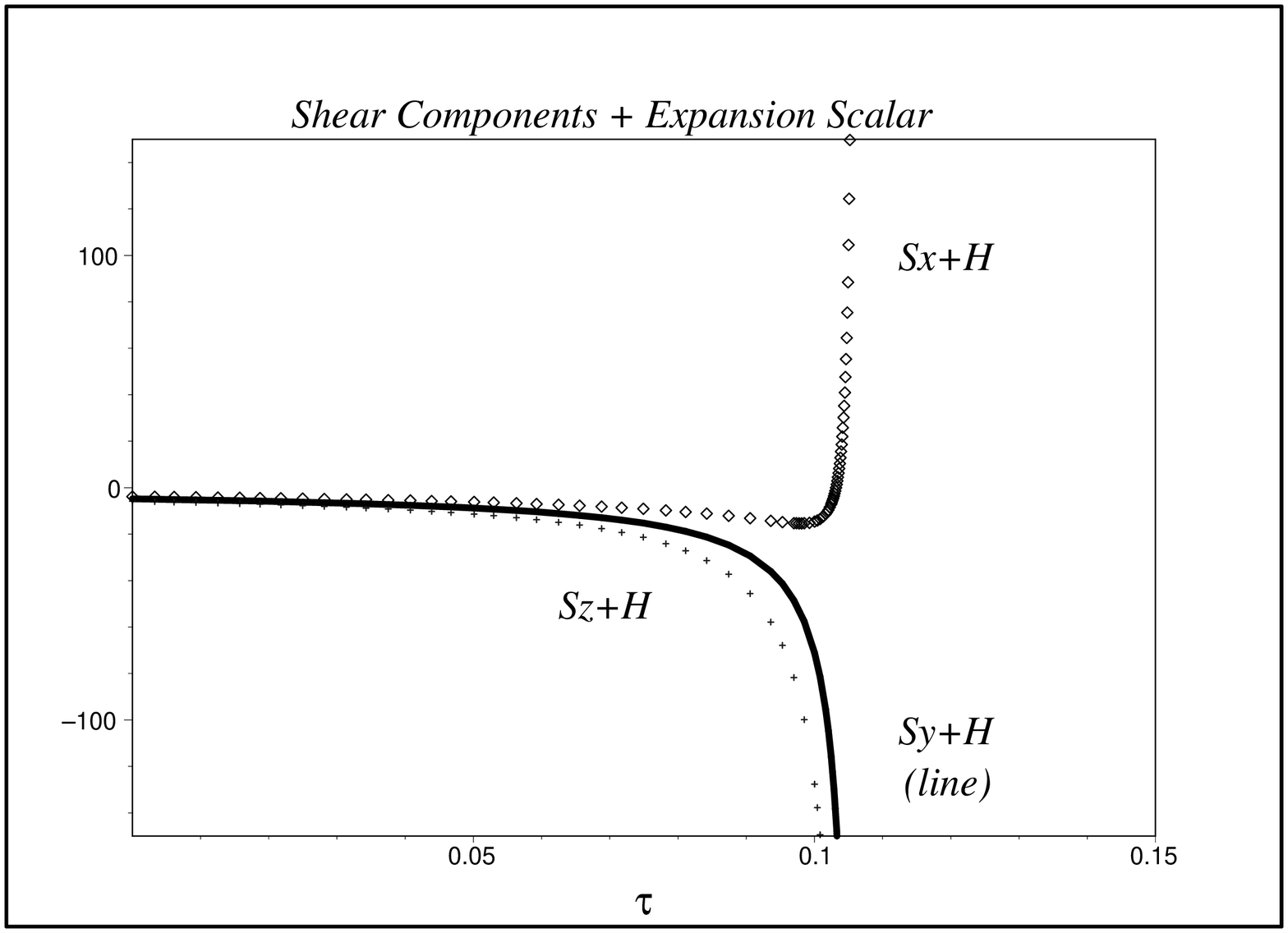,height=14cm,width=14cm,angle=270}
\caption{ Trayectorias de las funciones $S^{i}+\HH$ para $i=x,y,z$. Nosotros tomamos las siguientes condiciones iniciales: 
\,$S^{x}(0)=1,\,S^{y}(0)=0, \,S^{z}(0)=-1, \,\mu(0)=2,\,\beta(0)=0,$ y\, $\HH(0)=-3.43$. Este sistema muestra la deformaci\'on inicial
en la direcci\'on $x$ con el campo magn\'etico apuntando en la direcci\'on $z$. El colapso forma una singularidad tipo ``cigarro'' o
``l\'inea'' en la direcci\'on $x$.} \label{FIG2_EB}
}

Un an\'alisis de todos estos casos revela que los estados de colapso dependen fuertemente de los valores iniciales de la deformaci\'on.
Por ejemplo, si tenemos una deformaci\'on inicial a lo largo de la direcci\'on $x$, se dice que : $S^{x}_{0}\gg S^{y}_{0},S^{z}_{0},
\beta_{0},\mu_{0}$, entonces una singularidad tipo ``cigarro'' emerge a lo largo de la direcci\'on $x$. Esto se muestra en la Figura
\ref{FIG2_EB}, ilustrando (por medio del conjunto de ecuaciones (\ref{DV7})) que la funci\'on m\'etrica $A$ tiende a infinito, mientras
que las otras funciones m\'etricas $B$ y $C$ r\'apidamente decaen a cero. En general, la configuraci\'on inicial del sistema y los valores
iniciales del tensor de deformaciones determinan la direcci\'on (privilegiada) del colapso anis\'otropo.

Las pruebas num\'ericas tambi\'en muestran que existe un valor umbral para el campo magn\'etico inicial y dependiendo de su intensidad,
este tambi\'en puede influir en la direcci\'on que tome la singularidad tipo ``cigarro''. Esto se puede ilustrar apoy\'andonos en la
propia Figura \ref{FIG2_EB}: Si aqu\'i aumentamos el campo magn\'etico a $\beta_{0}=1$ entonces obtendremos una singularidad tipo
``cigarro'' pero en la direcci\'on $z$ para cualquier valor de la deformaci\'on $S_{i}(0)=0,\pm 1$ (donde $i=1,2,3$ direcciones espaciales).
Sin embargo, $\beta_{0}=1 \sim 10^{9}\,T$ es una valor f\'isicamente irreal para estrellas EBs magnetizadas, pero puede ser razonable en
modelos de universos primordiales magnetizados.

Una anisotrop\'ia tipo ``punto'' siempre puede emerger de configuraciones con deformaci\'on inicial nula, o sea con $S^{x}_{0}=S^{y}_{0}
=S^{z}_{0}=0$ y $\beta_{0}=0$. Sin embargo, incluso con deformaci\'on inicial nula existen siempre valores umbrales del campo magn\'etico
para los cuales las singularidades tipo ``punto'', (naturales de configuraciones sin deformaci\'on), pasan a ser singularidades tipo
``cigarro'' extendidas en la direcci\'on del campo magn\'etico.
\subsubsection{Espacio de fase y subespacios cr\'iticos}
\label{EFySubC_Cap2}

Como hemos mencionado previamente, el sistema de ecuaciones de Einstein--Maxwell puede ser escrito como un sistema aut\'onomo asociado a un
espacio de fase de 4-dimensiones, con las variables $(S^{z},\beta,\mu,\HH)$. Note que $\Sigma^{y}$ siempre puede ser encontrada si
determinamos $\Sigma^{z}=\Sigma$, la cual hemos supuesto como la componente independiente del tensor de deformaciones. Considerando las
4-funciones mencionadas, tendremos el siguiente sistema de ecuaciones,
%
\begin{eqnarray}
\dot{U}&=&-(U+p-\frac{2}{3}\B\M)\Theta-\B\M\Sigma, \\
\dot{\Sigma}&=&\frac{2}{3}\kappa\B\M-\Theta\Sigma,\\
\dot{\Theta}&=&\kappa(\B\M+\frac{3}{2}(U-p))-\Theta^2,\\
\dot{\beta}&=&\frac{2}{3}\beta(3\Sigma-2\Theta),
\end{eqnarray}
%
donde $\Sigma=\sigma^{z}$. Si pasamos a trabajar con las funciones adimensionales, definidas en (\ref{DV2}) este sistema se convierte en,
%
\begin{eqnarray}
S^{z}_{,\tau}&=& 2\beta \Gamma_{\M}-3\HH S^{z}, \label{PSaCS3a}\\
\HH_{,\tau}&=& \beta\Gamma_{\M}+\frac{3}{2}(\Gamma_{U}-\Gamma_{p})-3\HH^{2}, \label{PSaCS3b}\\
\beta_{,\tau}&=& 2\beta(S^{z}-2\HH), \label{PSaCS3c}\\
\mu_{,\tau}&=&\frac{1}{\Gamma_{U,\mu}}[(2\HH-
S^{z})(\Gamma_{\M}+2\Gamma_{U,\beta})\beta-3\HH(\Gamma_{p}+\Gamma_{U})].  \label{PSaCS3d}
\end{eqnarray}
%

Note que solo se modifica la ecuaci\'on para $\HH(\tau)$, ya que podemos llegar a la ecuaci\'on (\ref{PSaCS3b}) usando (\ref{DV3c}) y el
v\'inculo (\ref{DV3d}), por consiguiente ambos sistemas son equivalentes.
Entonces, la ecuaci\'on (\ref{DV3a}) es necesaria solamente para calcular los coeficientes m\'etricos restantes.

En la Figura \ref{EF_Electrones} del Ap\'endices \ref{Ap_Esp_Fase}, representamos una secci\'on 3-dimensional del espacio de fase
$(S^{z},\beta,\mu)$ con diferentes curvas para distintas condiciones iniciales. Como se muestra en la figura, el valor inicial de la
expansi\'on $H_{0}$ determina la evoluci\'on global de las soluciones num\'ericas. Todas las curvas que inician en $\tau=0$ con expansi\'on
$H_{0}=\sqrt{\kappa\lambda/3}$, convergen al atractor estable ``$a$''. Pero en el caso que tomemos $H_{0}=-\sqrt{\kappa\lambda/3}$ y
empezando en $\tau=0$, entonces las curvas evolucionar\'an tendiendo a la singularidad anisotr\'opica. Tomando el lado izquierdo del sistema
de ecuaciones en (\ref{PSaCS3a})-(\ref{PSaCS3d}), igual\'andolo a cero y resolvi\'endolo de forma algebraica, encontramos el conjunto de puntos cr\'iticos.
Dentro de este conjunto de puntos cr\'iticos se encuentra el atractor estable ``a'',
\begin{equation}
a=\{S^{z} = 0, \beta = 0, \mu = 1, \HH = 0\}. \label{CS1a}
\end{equation}\label{CS1}
Para el espacio de 4-dimensiones $(S^{z},\beta,\mu,\HH)$ existen cuatro posibles secciones 3-dimensionales. En estas secciones hemos
calculado tambi\'en las soluciones num\'ericas y los resultados son similares.
\subsection{Conclusiones del cap\'itulo}

Hemos presentado un modelo basado en la descripci\'on din\'amica de un volumen local de un gas magnetizado y auto-gravitante de electrones
en el nivel b\'asico de Landau $n=0$. Consideramos las ecuaciones de estado de este sistema. Tambi\'en, hemos trabajado con formas
simplificadas de las ecuaciones de Einstein-Maxwell que fueron escritas asumiendo un espacio-tiempo Bianchi-I con m\'etrica de Kasner.
La fuente de anisotrop\'ia es el propio campo magn\'etico. Este espacio-tiempo simplificado provee un conveniente modelo juguete para
entender el comportamiento local de colapso del sistema que cualitativamente imita las fuentes de materia dentro de un objeto compacto.

La relevancia de este cap\'itulo emerge de nuestros estudios sobre clasificaci\'on de singularidades, las cuales pueden ser de tipo ``punto''
o de tipo ``cigarro''. Las singularidades tipo ``punto'' aparecen bajo condiciones especiales del campo magn\'etico nulo, deformaci\'on nula
o ambas. Singularidades tipo ``cigarro'' pueden ser obtenidas en todas las direcciones, dependiendo de los valores iniciales del tensor de
deformaciones. Sin embargo, para intensidades suficientemente grandes del campo magn\'etico inicial siempre se esperar\'a una singularidad
tipo ``cigarro'' en la direcci\'on del campo magn\'etico.

Este resultado es importante porque el valor del campo magn\'etico determina el tipo de colapso y esto est\'a de acuerdo con la idea previa
no-relativista en la cual se analiza el colapso de estos gases magnetizados dentro de un cuadro Newtoniano\cite{Aurora1},\cite{Aurora2},
\cite{Aurora3}.

Por otro lado, as\'i como es discutido en \cite{SBC} por Collins $\&$ Ellis, los modelos Bianchi como el de tipo-I que estamos considerando,
son globalmente hiperb\'olicos y solo presentan una singularidad, as\'i que las hipersuperficies de tiempo constante (ortogonales a la
4-velocidad) son hipersuperficies globales de Cauchy\footnote{Una superficie de Cauchy es una variedad 3-dimensional espacialoidea, cuyo
futuro y pasado de dependencia describen toda la variedad espacio-tiempo.} y cada punto en el espacio-tiempo puede ser causalmente conectado
con el \'ultimo. La 4-velocidad es una geod\'esica del campo y la singularidad es determinada por un valor espec\'ifico de tiempo constante,
as\'i que esta singularidad no es de tipo temporaloidea\footnote{Seg\'un su car\'acter las singularidades f\'isicas pueden ser: {\bf 1)
Singularidades temporaloideas}, como la que se encuentra en un agujero de Schwarzschild en la que una part\'icula deja de existir por cierto
instante de tiempo; dependiendo de su velocidad, las part\'iculas r\'apidas tardan m\'as en alcanzar la singularidad mientras que las m\'as
lentas desaparecen antes. Este tipo de singularidades son inevitables, ya que tarde o temprano todas las part\'iculas deben atravesar la
hipersuperficie temporal singular. {\bf 2) Singularidades espacialoideas}, como la que se encuentra en agujeros de Reissner-Nordstrom, Kerr
y Kerr-Newman. Al ser hipersuperficies espaciales una part\'icula puede escapar de ellas y por tanto se trata de singularidades evitables.}
y cada evento en el espacio puede ser causalmente conectado con esta singularidad (en particular por geod\'esicas que son curvas integrales
del campo de la 4-velocidad).

Nosotros solo hemos considerado reg\'imenes de colapso a partir de una hipersuperficie inicial de tiempo constante. As\'i, cada curva
temporaloidea futura (geod\'esica o no) comienza en una hipersuperficie de Cauchy de tiempo constante y termina en la singularidad de
colapso. Bajo estas condiciones, la singularidad es obviamente censurada.

%
%
\newpage

\section[ESTUDIO DE UN GAS MAGNETIZADO Y AUTO-GRAVITANTE DE NEUTRONES]{\large {ESTUDIO DE UN GAS MAGNETIZADO Y AUTO-GRAVITANTE DE NEUTRONES}}

\label{CAP_Neutrones}

\subsection{Introducci\'on}\label{intro}

En este cap\'itulo extendemos el trabajo previo hecho para un gas de electrones, al caso de neutrones. Seguimos una similar metodolog\'ia
basada en la reescritura de las ecuaciones de campo y de las ecuaciones de conservaci\'on, para el gas de neutrones en una geometr\'ia
Bianchi-I. Como sistema din\'amico, estudiamos su evoluci\'on en un espacio de fase de 4--dimensiones. Este sistema es analizado
cualitativamente y num\'ericamente.

Vale la pena resaltar las diferencias b\'asicas entre el gas magnetizado de electrones examinado en \cite{Shapiro1,AAS,Alain2} (ver
cap\'itulo anterior) y el gas de neutrones que nosotros consideraremos en este cap\'itulo.
 Los electrones interact\'uan con el campo magn\'etico a trav\'es de su carga el\'ectrica, contribuyendo al llamado diamagnetismo de Landau
caracterizado por un efecto de cuantizaci\'on asociado con los niveles de Landau. Sin embargo, los neutrones interact\'uan con el campo
magn\'etico mediante el Momento Magn\'etico An\'omalo (MMA), en el contexto del paramagnetismo de Pauli y de las ecuaciones de Pauli--Dirac.
Consecuentemente uno espera que la interacci\'on magn\'etica del neutr\'on sea d\'ebil, aunque un gas de neutrones degenerado en condiciones
de objeto compacto se espera que sea m\'as cr\'itico (debido a las altas densidades) que las de un gas degenerado en condiciones de baja
densidad. Por tanto, los efectos relativistas de la gravedad tendr\'an que tomarse en cuenta y podr\'ian dominar desde el punto de vista
din\'amico en el gas de neutrones. Tambi\'en es importante mencionar que un gas de neutrones auto--gravitante es un modelo simplificado de
una fuente para un objeto compacto, donde los protones y el potencial qu\'imico de equilibrio deben ser incluidos, para evaluar los efectos
locales ~\cite{Shapiro1,salgado1,salgado2}. Sin embargo, el gas de neutrones magnetizado \cite{ManrezaParet:2008vt} ya exhibe importantes
diferencias, tanto cualitativas como cuantitativas, en comparaci\'on con el gas de electrones previamente analizado.
\subsection{Ecuaci\'on de Estado de un gas magnetizado de neutrones}
\label{sec:1}

Propiedades importantes del gas de neutrones magnetizado y degenerado son reportadas en (ver \cite{Shapiro1,Aurora2,Guang}). Considerando el ensemble gran can\'onico, un subsistema puede ser pensado como un volumen local de un gas de neutrones bajo la influencia de un campo magn\'etico $\vec{H}$ asociado al resto del sistema (en un contexto astrof\'isico esta puede ser una buena aproximaci\'on a un volumen local dentro de un objeto compacto). Debido a este campo los subsistemas se pueden polarizar, permitiendo que surja una magnetizaci\'on que satisface la relaci\'on:\, $\vec{H}=\vec{\B}-4\pi \vec{\M}$. El campo $\vec{H}$ puede ser pensado como un campo ``externo'' al subsistema, mientras que $\vec{\B}$ ser\'ia el ``interno'' de cualquier part\'icula dentro del subsistema, el cual incluye (en adici\'on a $\vec{H}$) la contribuci\'on $4\pi\vec{\M}$ de part\'iculas del resto del subsistema.

La ecuaci\'on de estado para el gas de neutrones se obtiene de calcular el espectro de energ\'ia de las part\'iculas que forman el sistema. Nosotros podemos obtener este espectro de energ\'ia a partir de las ecuaciones de Dirac para part\'iculas neutras con momento magn\'etico an\'omalo,
\begin{equation}\label{Diraceq}
(\gamma^{\mu}\partial_{\mu}+m+iq\sigma_{\mu \nu}F^{\mu \nu})\Psi=0,
\end{equation}
donde $\sigma_{\mu\lambda}=\frac{1}{2}(\gamma_\mu\gamma_\lambda-\gamma _\lambda\gamma _\mu)$ es el tensor de esp\'in,\, $F^{\mu \nu}$ es el tensor del campo electromagn\'etico (nosotros hemos tomado $\hbar=\mathrm{c}=1$) y $\Psi$ es el campo de Dirac. Resolviendo la ecuaci\'on (\ref{Diraceq}) se obtiene el siguiente espectro de energ\'ia \cite{Guang},\cite{Bagrov},\cite{Wen:2005kf},
\begin{equation}
E_n(p,\B,\eta)=\sqrt{p_{\|}^2+(\sqrt{p_\bot^2+m_n^2}+\eta q\B)^2},
\label{eneutrones}
\end{equation}
donde $p_{\|}, p_\bot$ son respectivamente, las componentes del momentum en la direcci\'on paralela y perpendicular al campo magn\'etico $\B$, $m_n$ es la masa del neutr\'on, $q=-1.91\mu_N$ es el momento magn\'etico del neutr\'on ($\mu_N=e/2m_p$ es el magnet\'on nuclear), $\eta=\pm 1$ son los autovalores correspondientes a las dos orientaciones posibles (paralela y antiparalela) del Momento Magn\'etico An\'omalo (MMA) del neutr\'on respecto al campo magn\'etico.

Sustituyendo (\ref{eneutrones}) en la expresi\'on (\ref{omega_n}) obtenida en el cap\'itulo anterior podemos calcular todas las cantidades termodin\'amicas del sistema al igual que hicimos en ese mismo cap\'itulo para el gas de electrones magnetizado.

Tal como explicamos en dicho cap\'itulo el t\'ermino de vac\'io es divergente, y puede igualmente ser renormalizado, y para el gas de neutrones su contribuci\'on es relevante s\'olo para campos de intensidad $\B>10^{14}\,\mathrm{T}$ \cite{Aurora2}, por tanto nosotros no tomaremos en cuenta este t\'ermino\footnote{Realmente el t\'ermino de vac\'io pudiera ser relevante para cuando el sistema evoluciona hacia la singularidad y el campo magn\'etico crece a infinito pero, por simplicidad no lo tendremos en cuenta en este trabajo y suponemos que no influye de manera determinante.}.

La ecuaci\'on (\ref{omega_n}) puede ser integrada f\'acilmente\footnote{Recordar que en este caso de neutrones en (\ref{omega_n}) no se suma por los niveles de Landau sino por $-1,+1$, los dos estados posibles del MMA.} para
el caso degenerado $(T=0)$, y su forma expl\'icita es entonces,
\begin{equation}\label{omega}
\Omega_{sn}=-\lambda\sum_{\eta=1,-1}\biggl[\frac{\mu f_\eta^3}{12}+
\frac{(1+\eta \beta)(5\eta \beta-3)\mu f_\eta}{24}+\frac{(1+\eta
\beta)^3(3-\eta \beta)}{24}L_\eta-\frac{\eta \beta
\mu^3}{6}s_\eta\biggr],
\end{equation}
donde hemos introducido las siguientes expresiones,
\begin{equation}
f_\eta=\sqrt{\mu^2-(1+\eta\beta)^2}, \quad
s_\eta=\frac{\pi}{2}-\arcsin\biggr(\frac{1+\eta\beta}{\mu}\biggl),
\quad \mu=\frac{\mu_n}{m_n}, \label{efe}
\end{equation}
\begin{equation}
L_\eta=\ln\biggl(\frac{\mu+f_\eta}{1+\eta\beta}\biggr), \quad \beta=\frac{\B}{\B_c},
\end{equation}
con $\B_c=m_n/q \simeq 1.56\times10^{16}\,\mathrm{T}$ el campo cr\'itico para neutrones y
$\lambda=m_n^4/4\pi^2\hbar^3\mathrm{c}^3=4.11\times10^{35}\mathrm{\,J \,m^{-3}}$.

Calculando  la densidad de neutrones y la magnetizaci\'on obtenemos
que: $N=N_0\Gamma_N$, $\M=\M_0\Gamma_M$, donde $N_0=\lambda/m_n$,
$\M_0=N_0q$, y los coeficientes $\Gamma_N,\Gamma_M$ toman la forma,
\begin{eqnarray*}
\Gamma_N&=& \sum_{\eta=1,-1}^{}\biggl[\frac{f_\eta^3}{3}+\frac{\eta
\beta(1+\eta\beta)f_\eta}{2}-\frac{\eta\beta \mu^2}{2}s_\eta\biggr],\\
\Gamma_M&=&-\sum_{\eta=1,-1}^{}\eta\biggl[\frac{(1-2\eta \beta)\mu
f_\eta}{6}-\frac{(1+\eta\beta)^2(1-\eta\beta/2)}{3}L_\eta+\frac{\mu^3}{6}s_\eta
\biggr].
\end{eqnarray*}\label{gamma1}
Por consiguiente, dados (\ref{componentes de T_a})-(\ref{componentes de T_c}) y (\ref{omega}), nosotros podemos escribir la ecuaci\'on de estado para un gas de neutrones relativista y degenerado en presencia de un campo magn\'etico como,
%
\begin{eqnarray}
U&=&\mu_n N+\Omega=\lambda \Gamma_U(\beta,\mu),\label{EOS1}
\\
p&=&-\Omega=\lambda \Gamma_P(\beta,\mu),\label{EOS2}
\\
M&=&\B \M=\lambda\beta \Gamma_M(\beta,\mu),\label{EOS3}
\end{eqnarray}
%
donde,
\begin{eqnarray*}
\Gamma_P&=&\sum_{\eta=1,-1}^{}\biggl[\frac{\mu f_\eta^3}{12}+
\frac{(1+\eta \beta)(5\eta \beta-3)\mu f_\eta}{24}+
\\
&&\hspace{3cm}+\frac{(1+\eta\beta)^3(3-\eta\beta)}{24}L_\eta-\frac{\eta
\beta \mu^3}{6}s_\eta\biggr],
\\
\Gamma_U&=&\mu \Gamma_N-\Gamma_P.
\end{eqnarray*}

Note que estas ecuaciones de estado difieren de las obtenidas para el gas de electrones (\ref{EE_Electr_a})-(\ref{EE_Electr_b}) y
obviamente los resultados din\'amicos estar\'an marcados por estas diferencias. Las ecuaciones (\ref{EE_Electr_a})-(\ref{EE_Electr_b}) dependen de los niveles
de Landau por estar cuantificado el espectro. Las ecuaciones (\ref{EOS1})-(\ref{EOS3}) muestran su acoplamiento con el campo a trav\'es
de la presencia del MMA de los neutrones.

Puntualicemos que en (\ref{omega})-(\ref{EOS3}) nosotros estamos sumando sobre el momento magn\'etico paralelo ($\eta=-1$) o perpendicular ($\eta=1$) al campo magn\'etico (o sea estamos considerando el conocido Paramagnetismo de Pauli). La elecci\'on de $\eta=\pm 1$, \cite{Aurora2} es equivalente a considerar diferentes fases del
sistema. La aparici\'on de valores umbrales para el campo magn\'etico en cada caso pueden verse si analizamos las expresiones de las funciones $f_{\eta}$ y $s_{\eta}$ en (\ref{efe}). Si tomamos a $\beta\geq 0$, empezando con $f_{\eta}$,
\begin{equation}
\mu^2\geq (1+\eta \,\beta)^2=1+2\eta\beta+\eta^2\beta^2 .
\end{equation}
 Sin embargo, como $\eta=-1,1$ entonces siempre $\eta^2=1$ y nosotros podemos reescribir,
\begin{equation}
\mu^2\geq (\eta+\beta)^2,
\end{equation}
de donde,
\begin{equation}
(\beta+\eta-\mu)(\beta+\eta+\mu)\leq 0.
\end{equation}
Ahora tenemos dos posibilidades. Pero, es simple darse cuenta que la \'unica aceptable es;
\begin{equation}
-\mu-\eta\leq\,\beta\,\leq \mu-\eta.  \label{CasoB}
\end{equation}
 Esta es exactamente la restricci\'on que viene de la funci\'on,
$s_{\eta}$,
\begin{equation}
\mid \frac{1+\eta\beta}{\mu}\mid \leq 1.  \label{arcsin_restriccion}
\end{equation}
  Por tanto, ya sea de (\ref{CasoB}) o de (\ref{arcsin_restriccion}) los siguientes v\'inculos son obtenidos para el campo magn\'etico,
\begin{eqnarray}
\hbox{Si} \quad \eta=1 \,  \Rightarrow \qquad -1-\mu \leq \beta \leq \mu-1,  \\
\hbox{Si} \quad \eta=-1 \, \Rightarrow \qquad 1-\mu \leq \beta \leq 1+\mu.
\end{eqnarray}

Estas \'ultimas desigualdades restringen los valores del campo magn\'etico. As\'i, sistemas de neutrones con MMA alineado en la direcci\'on del campo magn\'etico solo tienen como valores permitidos del campo magn\'etico el rango de $1-\mu \leq \beta \leq 1+\mu$. Similarmente, para sistemas de neutrones con MMA orientado antiparalelamente respecto al campo magn\'etico, solo pueden tomar valores de campo en el intervalo $-1-\mu \leq \beta \leq \mu-1$. En particular para el caso de $\mu=1$ tenemos que,
\begin{eqnarray}
   \hbox{Si} \quad \eta=-1, \qquad \, 0 \leq \beta \leq 2,\\
   \hbox{Si} \quad \eta=1,  \qquad \, -2 \leq \beta \leq 0.
\end{eqnarray}
Esto significa que para campos menores (o iguales) que dos veces el campo magn\'etico cr\'itico, los neutrones estar\'an alineados todos con $\eta=-1$ (el MMA es paralelo al campo magn\'etico). O sea neutrones con $\eta=1$ ser\'an forzados a invertir su sentido.

\subsection{Ecuaciones de Einstein--Maxwell}
\label{sec:2}

Si deseamos estudiar la evoluci\'on din\'amica de un volumen local magnetizado y auto--gravitante de neutrones bajo las condiciones f\'isicas impuestas en el interior de un objeto compacto, entonces tenemos que contar con los efectos relativistas. Y esto implica que la din\'amica local debe ser estudiada bajo el marco de la TGR, por medio de las ecuaciones de campo de Einstein,
\begin{equation}
G_{a b}=R_{a b}-\frac{1}{2}R\, g_{a b}=\kappa\,\mathcal{T}_{a b}, \label{EE1}
\end{equation}
 junto con la ecuaci\'on de conservaci\'on del tensor energ\'ia--momentum y de las ecuaciones de Maxwell,
%
\begin{eqnarray}
  \label{Ebal}\mathcal{T}^{a b}\,_{;\,b}=0,\\
  \label{Maxwell_eq} F^{a b}\,_{;\,b}=0, \qquad F_{[\,a b\,;\,c\,]}=0,
\end{eqnarray}
%
donde $\kappa=8\pi \mathrm{G_N}$, con $\mathrm{G_N}$ constante de gravitaci\'on de Newton, el corchete cuadrado denota anti--simetrizaci\'on
en $a b;c$. El tensor energ\'ia--momentum $\mathcal{T}^{a}\,_{\,\,b}$ asociado al gas magnetizado de neutrones viene dado por
(\ref{componentes de T_a})-(\ref{componentes de T_c}), conectado al gran potencial termodin\'amico obtenido mediante la mec\'anica estad\'istica, y acoplado
apropiadamente a la ecuaci\'on de estado. Este tensor tambi\'en puede ser escrito en t\'erminos de la 4-velocidad del campo $u^{a}$ como
\footnote{Notar que el tensor (\ref{TE-M}), no contiene, ni el t\'ermino de vac\'io $B^2\,log (B/B_{c})$, ni se adiciona el campo cl\'asico
$B^2$ (de \'arbol). Ya que suponemos que la contribuci\'on del vac\'io es peque\~na y similarmente a como hicimos en el cap\'itulo
\ref{Cap_de_Electrones} hemos seguido trabajos en donde este t\'ermino cl\'asico no se toma en cuenta. Aunque debido a su importancia
factible en la evoluci\'on cercana al colapso, s\'i los asumiremos en el cap\'itulo \ref{Cap_Perturbado}.},
\begin{equation}
\mathcal{T}^{a}\,_{b}=(U+\widetilde{P})u^{a}u_{b}+\widetilde{P}\,\delta^{a}\,_{b}+\Pi^{a}\,_{b},
\qquad \widetilde{P}=p-\frac{2\B\M}{3}. \label{TE-M}
\end{equation}

Nosotros consideramos las ecuaciones de campo (\ref{EE1})--(\ref{Maxwell_eq}) con (\ref{TE-M}) como la fuente de materia para un
modelo Bianchi-I descrito en la representaci\'on de la m\'etrica de Kasner,
\begin{equation}
{ds^2}=Q_{1}(t)^2dx^{2}+Q_{2}(t)^2dy^2+Q_{3}(t)^2 d{z}^2-dt^2,
\label{Metrica-K}
\end{equation}
lo cual sugiere escoger la 4--velocidad en una representaci\'on com\'ovil $u^{a}=\delta^{a}\,_{t}$, as\'i el tensor de presiones anis\'otropas en (\ref{TE-M}) y en las coordenadas $[x,y,z,t]$ toma la forma,
\begin{equation}
\Pi^{a}\,_{b}=\hbox{diag}\,[\Pi,\Pi,-2\Pi,0], \qquad
\Pi=-\frac{\B\M}{3},\qquad \Pi^{a}\,_{a}=0. \label{TAn}
\end{equation}

Las ecuaciones de campo (\ref{EE1}) para (\ref{TE-M}),(\ref{Metrica-K}) y (\ref{TAn}) toman la forma,
%
\begin{eqnarray}
-G^{x}\,_{x}&=&\frac{\dot{Q_{2}}\dot{Q_{3}}}{Q_{2}Q_{3}}+\frac{\ddot{Q_{2}}}{Q_{2}}+\frac{\ddot{Q_{3}}}{Q_{3}}=-\kappa(p-\B\M),
\label{EE2_Gxx} \\
-G^{y}\,_{y}&=&\frac{\dot{Q_{1}}\dot{Q_{3}}}{Q_{1}Q_{3}}+\frac{\ddot{Q_{1}}}{Q_{1}}+\frac{\ddot{Q_{3}}}{Q_{3}}=-\kappa(p-\B\M),
\\
-G^{z}\,_{z}&=&\frac{\dot{Q_{1}}\dot{Q_{2}}}{Q_{1}Q_{2}}+\frac{\ddot{Q_{1}}}{Q_{1}}+\frac{\ddot{Q_{2}}}{Q_{2}}=-\kappa
p,
\\
-G^{t}\,_{t}&=&\frac{\dot{Q_{1}}\dot{Q_{2}}}{Q_{1}Q_{2}}+\frac{\dot{Q_{1}}\dot{Q_{3}}}{Q_{1}Q_{3}}+
\frac{\dot{Q_{2}}\dot{Q_{3}}}{Q_{2}Q_{3}}=\kappa U, \label{EE2_Gtt}
\end{eqnarray}
%
donde $\dot A= A_{;\alpha}\,u^{\alpha}=A_{,t}$. De la ecuaci\'on de conservaci\'on del tensor energ\'ia--momentum
(\ref{Ebal}) tenemos que,
\begin{equation}\label{eq_U[t]}
\dot{U}=\frac{\dot{Q_{3}}}{Q_{3}}(p+U)-(\frac{\dot{Q_{1}}}{Q_{1}}+\frac{\dot{Q_{2}}}{Q_{2}})(-\B\M+p+U),
\end{equation}
mientras que de las ecuaciones de Maxwell (\ref{Maxwell_eq}) obtenemos que,
\begin{equation}\label{Maxwell_eq_1}
\frac{\dot{Q_{1}}}{Q_{1}}+\frac{\dot{Q_{2}}}{Q_{2}}+\frac{1}{2}\frac{\dot{\B}}{\B}=0.
\end{equation}

Las ecuaciones de Einstein-Maxwell (\ref{EE2_Gxx})--(\ref{EE2_Gtt}),(\ref{eq_U[t]}) y (\ref{Maxwell_eq_1}) son ecuaciones diferenciales ordinarias no--lineales
y de segundo orden para las funciones m\'etrica $Q_{1}, Q_{2},Q_{3}$ y $U$. Para tratar este sistema num\'ericamente, es necesario
introducir nuevas variables que transformar\'an el sistema, en un sistema de ecuaciones de primer orden. Sin embargo, antes de tomar este
camino, veremos que ocurre cuando el campo magn\'etico es d\'ebil.
\subsection{L\'imite de campo magn\'etico d\'ebil}

Esta breve discusi\'on para el campo magn\'etico d\'ebil es importante pues ilustra la conexi\'on entre nuestro campo magn\'etico de origen cu\'antico y el campo magn\'etico Maxweliano en el contexto de un tratamiento magnetohidrodin\'amico. En la secci\'on \ref{sec:1} nosotros hemos mostrado ecuaciones fuertemente conectadas con un campo magn\'etico cu\'antico.

Es importante enfatizar que el t\'ermino ``campo magn\'etico cu\'antico'' contiene la interacci\'on semi--cl\'asica entre el campo magn\'etico y el momento magn\'etico an\'omalo. Esta aproximaci\'on implica una conexi\'on te\'orica entre la ecuaci\'on de estado introducida en la secci\'on anterior y el cuadro de trabajo en EDC\footnote{Electrodin\'amica Cu\'antica}. Como una condici\'on de consistencia, este marco de trabajo debe permitir en el caso l\'imite de una expansi\'on en series alrededor de $\beta=0$, llegar al l\'imite cl\'asico Maxweliano. Los t\'erminos l\'ideres en esta expansi\'on deben coincidir con los t\'erminos del tensor de energ\'ia-momentum para el campo magn\'etico Maxwelliano \cite{MTW}. En general, este tipo de expansi\'on en series puede efectuarse de la forma,
\begin{eqnarray}
p_{\perp}&=& \sum_{n=0}^{\infty}
(\frac{\partial^{n}{p_{\perp}}}{\partial{\beta^n}})\mid_{_{\beta=0}}\frac{\beta^{n}}{n!}\simeq
p_{1}-a_{1}\beta^2+\mathcal{O}(\beta^4),\label{WMF1}\\
p_{\|}&=& \sum_{n=0}^{\infty}
(\frac{\partial^{n}{p_{\|}}}{\partial{\beta^n}})\mid_{_{\beta=0}}\frac{\beta^{n}}{n!}\simeq
p_{1}+a_{3}\beta^2+\mathcal{O}(\beta^4),\label{WMF2}
\\
U&=& \sum_{n=0}^{\infty}
(\frac{\partial^{n}U}{\partial{\beta^n}})\mid_{_{\beta=0}}\frac{\beta^{n}}{n!}\simeq
U_{0}+a_{o}\beta^2+\mathcal{O}(\beta^4), \label{WMF3}
\end{eqnarray}
donde es f\'acil ver que: $p_{1}=p_{\|}(\beta=0)\equiv
p_{\perp}(\beta=0),\,a_{0}=({\partial^{2}U}/{\partial{\beta^2}})\mid_{_{\beta=0}}/2
,\,a_{1}=({\partial^{2}{p_{\perp}}}/{\partial{\beta^2}})\mid_{_{\beta=0}}/2
,\,a_{3}=({\partial^{2}{p_{\|}}}/{\partial{\beta^2}})\mid_{_{\beta=0}}/2,\,
U_{0}=U(\beta=0)$, y todas estas funciones dependen solamente del potencial qu\'imico adimensional $\mu$.

Como se muestra en los ejemplos encontrados en la literatura, los campos magn\'eticos cl\'asicos sobre un fondo geom\'etrico tipo Bianchi-I ~\cite{Tsagas:1999tu,Barrow:2006ch}, los t\'erminos l\'ideres son t\'erminos cuadr\'aticos. Esto sugiere que las series en (\ref{WMF1})-(\ref{WMF3}) deben ser truncadas a nivel cuadr\'atico, los t\'erminos de mayor orden como $\beta^4,\,\beta^6,...$, son en general contribuciones multipolares muy peque\~nas. Si asumimos que las fluctuaciones de las velocidades del plasma tienden a cero como promedio macrosc\'opico, y que el medio no sufre de movimientos de volumen internos, entonces estas contribuciones pueden ser despreciadas (aunque t\'ipicamente, velocidades altas podr\'ian surgir a partir de fluctuaciones t\'ermicas o desordenes cu\'anticos). Bajo estas suposiciones, el tensor de energ\'ia-momentum de un gas de neutrones m\'inimamente acoplado al campo magn\'etico, puede siempre ser escrito en la forma \cite{Tsagas:1999tu},
\begin{equation}
T_{a b}=(U_{0}+U_{mag})u_{a}u_{b}+(p_{0}+p_{mag})h_{a b}+\Pi^{mag}_{a b},\label{T_Roy}
\end{equation}
donde $p_{mag}$ y $U_{mag}$ son, respectivamente, la presi\'on magn\'etica y la densidad de energ\'ia del campo magn\'etico.

Es importante tener en cuenta que $p_{0}$ en (\ref{T_Roy}) es la contribuci\'on isotr\'opica a la presi\'on del sistema, la cual (en general) puede depender del potencial qu\'imico y del campo magn\'etico. Por otro lado, $p_{1}$ es la presi\'on para el caso de campo magn\'etico nulo $\beta=0$. En nuestro caso el tensor de materia tiene la forma,
\begin{equation}
T^{a}\,_{b}=\textrm{diag}[p_{1}-a_{1}\beta^2,p_{1}-a_{1}\beta^2,p_{1}+a_{3}\beta^2,-U_{0}-a_{0}\beta^2],\label{T_AUR_DEBIL}
\end{equation}
y comparando (\ref{T_AUR_DEBIL})-(\ref{T_Roy}) tenemos que,
\begin{eqnarray}
p_{mag}&=&\frac{H^{2}}{6}=a_{0}\beta^2/3,
\\
U_{mag}&=&\frac{H^{2}}{2}=a_{0}\beta^2,
\end{eqnarray}
donde $H^2=H^{i}H_{i}=2a_{0}\beta^2$ y $H_{i}$ son las componentes del campo magn\'etico, la cuales hemos asumido que apuntan en la direcci\'on $z$. Esta suposici\'on es consistente con el hecho de que un peque\~no volumen en el centro de un objeto compacto es aproximadamente homog\'eneo, y la rotaci\'on puede efectuarse en una direcci\'on privilegiada, que nosotros tomamos como la direcci\'on del eje $z$,
\begin{equation}
H^{i}=(0,0,\sqrt{2a_{0}}\frac{\beta}{Q_{3}},0), \qquad
\,H_{i}=(0,0,\sqrt{2a_{0}}\beta Q_{3},0) .
\end{equation}

El tensor $\Pi^{mag}_{a b}$ es entonces el tensor sim\'etrico de traza nula, proyectado con $h_{a b}$, y representa las presiones anis\'otropas debidas al campo magn\'etico. Este tensor puede ser escrito como,
\begin{equation}
\hspace{0.8cm}(\Pi^{mag})^{a}\,_{b}=\textrm{diag}[-\frac{1}{3}(a_{1}+a_{3})\beta^{2},-\frac{1}{3}(a_{1}+
a_{3})\beta^{2},\frac{2}{3}(a_{1}+a_{3})\beta^{2},0].\label{PI_mag}
\end{equation}

Es importante puntualizar que $p_{1}$ en (\ref{T_AUR_DEBIL}) es en general, diferente de $p_{0}$. As\'i,
\begin{equation}
p_{0}=p_{1}-(a_{0}+2a_{1}-a_{3})\frac{\beta^2}{3},
\end{equation}
donde $p_{1}=p_{(\beta=0)}=\widetilde{P}_{(\beta=0)}$ en (\ref{EOS2}) y (\ref{TE-M}) es la presi\'on sin campo magn\'etico. Solo cuando el campo magn\'etico se anula $\beta=0$, las presiones coinciden: $p_{0}=p_{1}$ y el tensor de energ\'ia--momentum toma la forma del tensor del fluido perfecto con presi\'on isotr\'opica. En este caso $p_{0}$ y $U_{0}$ corresponden a la presi\'on y la densidad de energ\'ia de un gas cl\'asico de neutrones.
\subsection{Din\'amica de cantidades covariantes}

Como estamos interesados en la evoluci\'on de un elemento de volumen para el gas de neutrones asociado con la fuente
(\ref{TE-M}), similarmente a como se hizo para el caso de electrones debemos reescribir las ecuaciones din\'amicas de Einstein-Maxwell en t\'erminos de par\'ametros covariantes asociados con la cinem\'atica local de un elemento de volumen como es descrito por un observador con 4--velocidad $u^{a}$. Para la m\'etrica de Kasner en una representaci\'on com\'ovil de geod\'esicas, los par\'ametros locales cinem\'aticos no nulos son la expansi\'on $\Theta$, y la deformaci\'on $\sigma^{a b}$, dada por,
\begin{eqnarray}
 \Theta &=& u^{a}\,_{;a}\,,   \label{expsc}\\
  \sigma_{a b} &=& u_{(a; b)}-\frac{\Theta}{3}\,h_{a b},  \label{shear}
\end{eqnarray}
donde $h_{a b}=u_{a}\,u_{b}+g_{a b}$ es el tensor de proyecci\'on y los par\'entesis denotan simetrizaci\'on respecto a los \'indices $a,b$.  El escalar de expansi\'on y las componentes del tensor de deformaci\'on para la m\'etrica de Kasner tiene la forma,
\begin{eqnarray}\label{ec_teta_K}
\Theta &=& \frac{\dot{Q_{1}}}{Q_{1}}+\frac{\dot{Q_{2}}}{Q_{2}}+\frac{\dot{Q_{3}}}{Q_{3}},\\
\sigma^{a}\,_{b} &=& \hbox{diag}\,[\sigma^x\,_x,\,\sigma^y\,_y,\,\sigma^z\,_z,\,0] =
\hbox{diag}\,[\Sigma_1,\,\Sigma_2,\,\Sigma_3,\,0],
\end{eqnarray}
donde, \footnote{Notar que a diferencia del cap\'itulo anterior aqu\'i, escribimos los resultados de una forma gen\'erica, o sea $A,B,C$
ahora son $Q_{1},Q_{2},Q_{3}$ esto nos resultar\'a muy conveniente.}
\begin{equation}
\Sigma_{\alpha}=\frac{2\dot{Q_{\alpha}}}{3Q_{\alpha}}-\frac{\dot{Q_{\beta}}}{3Q_{\beta}}-\frac{\dot{Q_{\gamma}}}
{3Q_{\gamma}},
\qquad \alpha \neq \beta \neq \gamma,\, (\alpha,\beta,\gamma=1,2,3). \label{sigma_comp}
\end{equation}

El tensor de deformaci\'on $\sigma^{a}\,_{b} $ es de traza nula: $\sigma^{a}\,_{a}=0$, entonces podemos eliminar una de las cantidades
$(\Sigma_1,\,\Sigma_2,\,\Sigma_3)$ y escribirlas en t\'erminos de las otras dos. En efecto, para el Bianchi-I con m\'etrica Kasner, una
de estas cantidades es suficiente para representar a $\sigma^{a}\,_{a}$, aunque para nosotros es conveniente mantener dos de estas
variables, eliminar $\Sigma_1$ y escribirla en t\'erminos de $(\Sigma_2,\,\Sigma_3)$. Por medio de (\ref{ec_teta_K}) y (\ref{sigma_comp}),
todas las derivadas de segundo orden de las funciones m\'etricas en (\ref{EE2_Gxx})-(\ref{EE2_Gtt}), (\ref{eq_U[t]}) y (\ref{Maxwell_eq_1})
pueden ser reescritas como derivadas de primer orden de $\Theta,\ \Sigma_2$ y $\Sigma_3$. Despu\'es de varias manipulaciones podemos
reescribir (\ref{EE2_Gxx})-(\ref{EE2_Gtt}), (\ref{eq_U[t]}) y (\ref{Maxwell_eq_1}) como un sistema de ecuaciones de evoluci\'on de primer
orden, de la siguiente forma,
%
%
\begin{eqnarray}
\dot{U}&=&-(U+p-\frac{2}{3}{\B}{\M})\Theta-\B\M \Sigma_3,\label{EE30}
\\
\label{EE31} \dot{\Sigma_2}&=&-\frac{\kappa \B\M}{3}-\Theta\Sigma_2,
\\
\label{EE32}\dot{\Sigma_3}&=&\frac{2}{3}\kappa \B\M- \Theta\Sigma_3,
\\
\label{EE33}\dot{\Theta}&=&\kappa (\B\M+\frac{3}{2}(U-p))-\Theta^2,
\\
\label{EE34}\dot{\beta}&=&\frac{2}{3}\beta (3\Sigma_3-2\Theta),
\end{eqnarray}
%
%
junto con estas ecuaciones, tenemos el siguiente v\'inculo,
\begin{equation}\label{constrain}
-\Sigma_2^2-\Sigma_2\Sigma_3+\frac{\Theta^2}{3}-\Sigma_3^2=\kappa U,
\end{equation}
que proviene de (\ref{EE2_Gtt}). Este sistema de ecuaciones de primer orden de las variables $U,\, \beta,\, \Theta,\,
\Sigma_2,\,\Sigma_3$ y el v\'inculo (\ref{constrain}) est\'an completamente determinados\footnote{Note que el sistema
(\ref{EE30})--(\ref{EE34}), junto con el v\'inculo (\ref{constrain}) es id\'entico al sistema (\ref{KMWAP8a})--(\ref{KMWAP8e}), cambiando
$\Sigma^{y},\Sigma^{z}$ por $\Sigma_{2},\Sigma_{3}$.}.
La soluci\'on de este sistema describe la evoluci\'on din\'amica de un volumen local en un gas de neutrones magnetizado y representa un
modelo aproximado a un subsistema gran can\'onico para esta fuente en el centro de un objeto compacto.

\subsubsection{Ecuaciones din\'amicas}

El sistema de ecuaciones de evoluci\'on (\ref{EE30})--(\ref{EE34}) y (\ref{constrain}) puede ser transformado en un sistema din\'amico, de la misma forma
a como se hizo en el cap\'itulo anterior (ver secci\'on {\ref{DCC_Cap2}}), introduciendo las siguientes variables adimensionales,
\begin{equation}\label{def_tau}
H=\frac{\Theta}{3}, \qquad
\frac{d}{d\tau}=\frac{1}{H_0}\frac{d}{dt},
\end{equation}
junto con las funciones adimensionales,
\begin{equation}\label{adim_var}
\HH=\frac{H}{H_0}, \ \ S_2=\frac{\Sigma_2}{H_0}, \ \
S_3=\frac{\Sigma_3}{H_0}, \ \ \beta=\frac{\B}{\B_c},
\end{equation}
donde $H_0$ es una constante inversa a la longitud de escala, la cual hemos elegido por conveniencia como $3H_0^2=\kappa\lambda \Rightarrow
|H_0|=1.66\times10^{-2}\,\mathrm{m^{-1}}$ \footnote{Note que $H_0$, no es la constante de Hubble, dada por $H_0^{\rm{cosm}}=0.59\times
10^{-26}\mathrm{m^{-1}}$.}. La constante $H_{0}$, provee una longitud de escala de $1/H_0\sim 6\,\times 10^{3}\mathrm{m}$ que es adecuada
con la longitud caracter\'istica del sistema que estamos considerando. Las funciones $S_2$ y $S_3$ son las componentes del tensor de
deformaciones normalizadas a esta escala, el tiempo adimensional $\tau$ puede ser negativo o positivo, dependiendo del signo de
$H_0=\pm\sqrt{\kappa\lambda/3}$.

Sustituyendo las variables (\ref{adim_var}) en el sistema (\ref{EE30})--(\ref{EE34}) y en el v\'inculo (\ref{constrain}) obtenemos un
sistema id\'entico al sistema (\ref{DV3d}) y (\ref{DV3a})-(\ref{DV3f})\footnote{Claramente debemos cambiar en (\ref{DV3d}) y
(\ref{DV3a})-(\ref{DV3f}), $S^{y},S^{z}$ por $S_{2},S_{3}$ respectivamente. Adem\'as, de recordar que $\dot{U}$ se reemplazo por
$\dot{\mu}$ debido a que de (\ref{EE30}) tenemos que:\, $U=U(\beta,\mu) \Rightarrow U_{,\tau}=\lambda(\Gamma_{U,\mu}\mu_{,\tau}+
\Gamma_{U,\beta}\beta_{,\tau})$, lo cual nos permite obtener $\mu_{,\tau}$ a partir de $U_{,\tau}$.}
, por lo que no vemos necesidad de reescribirlo.
\subsection{Soluciones num\'ericas y discusi\'on f\'isica}

Estamos interesados en estudiar el colapso de una configuraci\'on de un gas de neutrones magnetizados. Necesitamos considerar un elemento
de volumen local y resolver el v\'inculo (\ref{DV3d}). De las dos soluciones que aporta la ecuaci\'on cuadr\'atica del v\'inculo para
$\HH$, la ra\'iz negativa es la que esta relacionada con soluciones de colapso. De modo que para asegurarnos del colapso local debemos
imponer en las soluciones del sistema (\ref{DV3a})-(\ref{DV3f}), que la expansi\'on inicial $\Theta$, sea negativa lo cual implica que en la
hipersuperficie de tiempo inicial $\tau=0$ que $\HH(0)<0$. Por tanto, a partir de (\ref{ec_teta_K}) y (\ref{adim_var}) expresadas en
t\'erminos del volumen local $V=\sqrt{\det{g_{\alpha\beta}}}=Q_{1}Q_{2}Q_{3}$ como,
\begin{equation}\label{vol_local}
V=V(0)\exp\left(3\int_{\tau=0}^\tau\HH d\tau\right).
\end{equation}

Para investigar el colapso en dependencia de las direcciones en que se puede manifestar, podemos relacionar por medio de (\ref{ec_teta_K}),
(\ref{sigma_comp}), (\ref{def_tau}) y (\ref{adim_var}) a las componentes espaciales de la m\'etrica y a la combinaci\'on $\HH+S_{\beta}$,
ya que,
\begin{equation}
Q_{\beta}(\tau)=Q_{\beta}(0)\exp[\int(\HH+S_{\beta})d\tau], \qquad \beta=1,2,3,
\end{equation}
donde $Q_{\beta}(0)$, son constantes que pueden ser identificadas con los valores iniciales de $Q_{\beta}(\tau)$. Para resolver el sistema
(\ref{DV3a})-(\ref{DV3f}) usamos una gran cantidad de condiciones iniciales asociadas a las condiciones t\'ipicas que existen en ENs
\cite{Shapiro1,bocquet,Peng:2007uu,Reisenegger:2008et,Suh:2000ni,Lattimer1,salgado1,salgado2}. Por ejemplo:
$\mu=2\Rightarrow \rho\sim10^{18}\,\mathrm{kg/m^3} $, $\beta_0=10^{-2}-10^{-5}$, para campos magn\'eticos entre $10^{14}\,T$ y
$10^{11}\,T$. Adem\'as, debemos imponer en todas las pruebas num\'ericas la condici\'on de colapso para el volumen: $\HH(0)<0$,
junto con: $S_2(0)=0,\pm1$, $S_3(0)=0,\pm1$, la cual corresponde con casos de deformaci\'on inicial nula y deformaci\'on inicial no
nula en las direcciones $y$ o $z$.

La soluci\'on num\'erica para la funci\'on $\HH$, es ploteada en la Figura \ref{Hbeta}, para diferentes condiciones iniciales, muestra
que $\HH\rightarrow-\infty$, independientemente de las condiciones iniciales. El campo magn\'etico tiende ha aumentar, pero siempre se
mantiene por debajo del campo cr\'itico $\B=\B_c$. Este comportamiento es mostrado en la Figura \ref{Hbeta}, para todo un rango de
condiciones iniciales.
\FIGURE{\epsfig{file=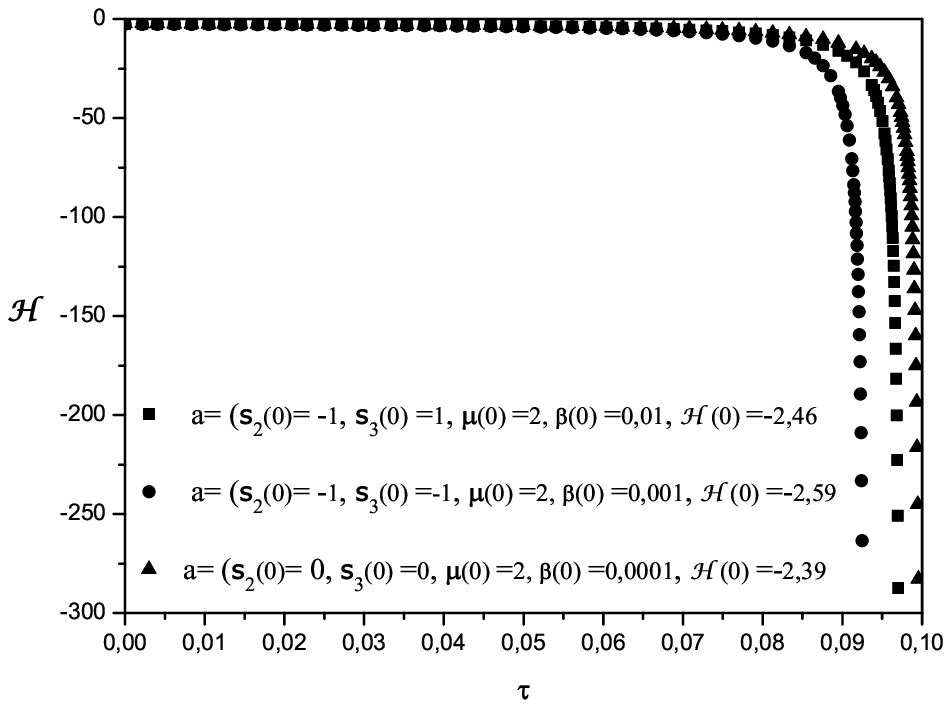,height=8cm,width=8cm,angle=0}\epsfig{file=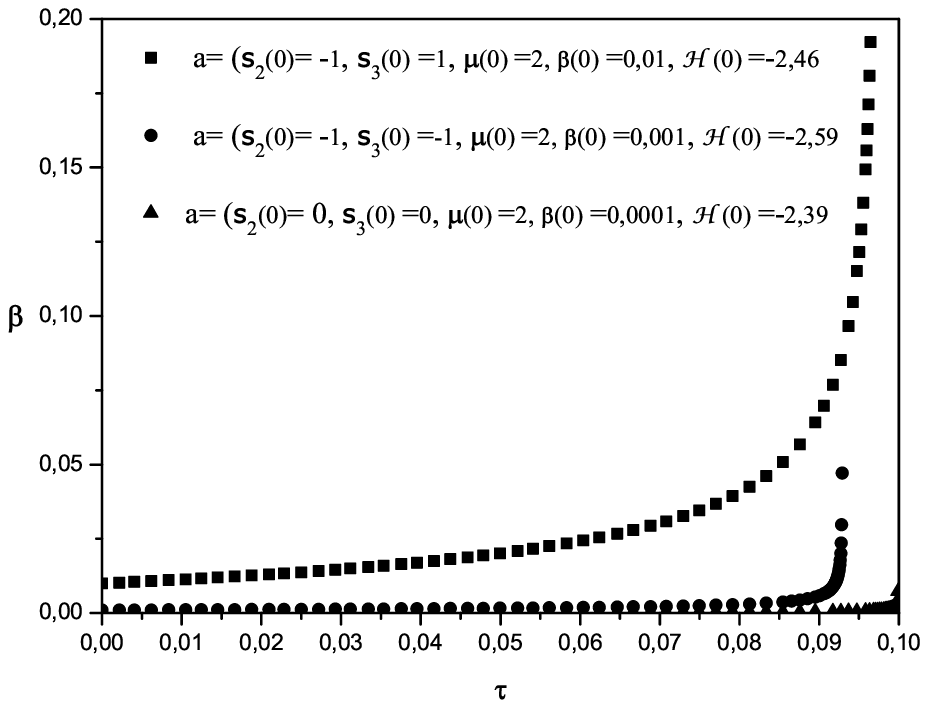,height=8cm,width=8cm,angle=0} \caption[Ejemplo
figura1]{\footnotesize{En el gr\'afico de la izquierda se muestra el comportamiento de $\HH$ vs $\tau$ para diferentes condiciones iniciales.
En el gr\'afico de la derecha se muestra la intensidad del campo magn\'etico adimensional $(\beta=\B/\B_c$), este tiene una tendencia a
crecer, pero siempre se mantiene por debajo del campo cr\'itico.}} \label{Hbeta}}

\FIGURE{\epsfig{file=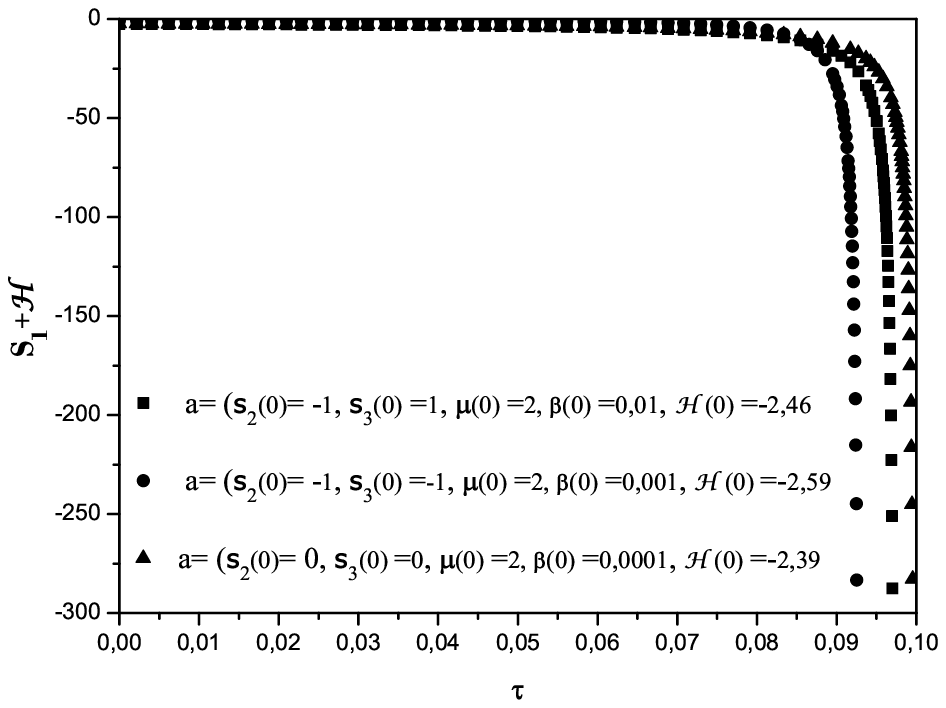,height=8cm,width=8cm,angle=0}\epsfig{file=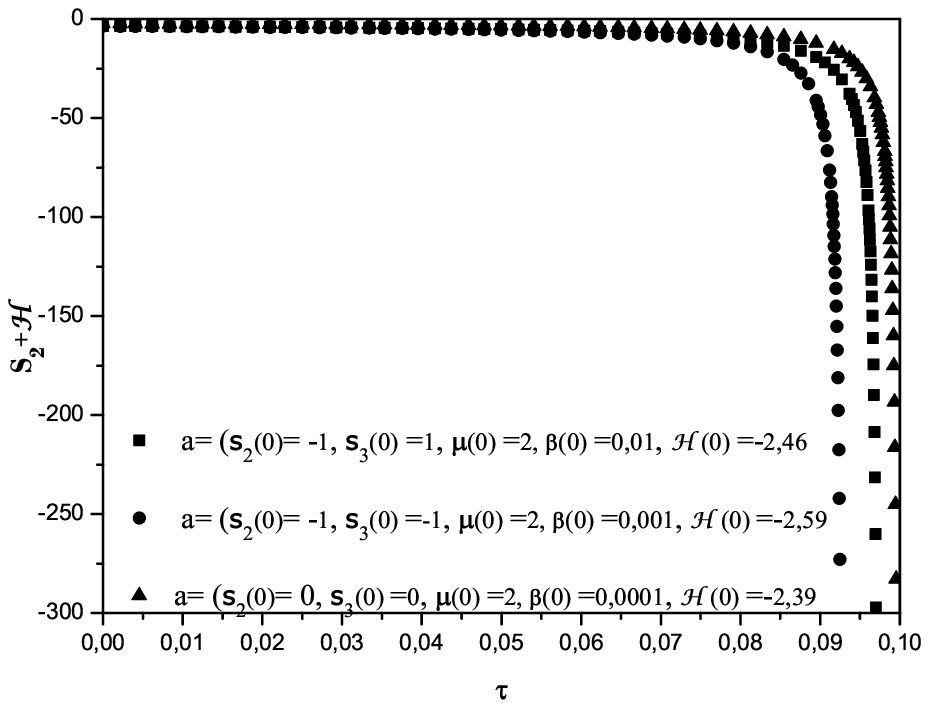,height=8cm,width=8cm,angle=0} \caption[Ejemplo
figura2]{\footnotesize{Es ploteado el comportamiento de $(S_1+\HH)$ y $(S_2+\HH)$ versus $\tau$. En estos gr\'aficos se puede apreciar
como estas cantidades tienden a $-\infty$, adem\'as de tener diferentes tiempos de colapso para diferentes condiciones iniciales.}}
\label{S1S2}}

Es evidente concluir de estas gr\'aficas que las cantidades $S_i+\HH\,\rightarrow\,-\infty$, as\'i como que los coeficientes m\'etricos
tiende a cero ($Q_{1}, Q_{2}, Q_{3}\,\rightarrow0$), lo cual muestra claramente que el elemento de volumen colapsa a una singularidad
isotr\'opica tipo ``punto''.
\subsubsection{Espacio de fase}

Del mismo modo a como se hizo en el cap\'itulo \ref{Cap_de_Electrones}, podemos usar el v\'inculo (\ref{DV3d}) para transformar el sistema
(\ref{DV3a})-(\ref{DV3f}) a un sistema de ecuaciones en funci\'on de las variables $(S_3,\beta, \mu, \HH)$. Si hacemos esto llegaremos al
sistema (\ref{PSaCS3a})-(\ref{PSaCS3d}) de la secci\'on \ref{EFySubC_Cap2}, donde solo tenemos que intercambiar $S^{z}$ por $S_{3}$.

Las trayectorias en la subsecci\'on 3--dimensional del espacio de fase, parametrizada por $(S_3, \beta, \mu)$, son mostradas en la Figura
\ref{espafase_Neutrones}, del Ap\'endice \ref{Ap_Esp_Fase}. La evoluci\'on del sistema es determinada por el signo de $H_0$. Para
$\tau<0 \Rightarrow H_0=-\sqrt{\kappa\lambda/3}$, y $\HH>0$ el sistema evoluciona tendiendo al atractor estable ``a'', mientras que si
$\tau>0 \Rightarrow H_0=\sqrt{\kappa\lambda/3}$, y $\HH<0$ las trayectorias evolucionan hacia una singularidad ``puntual''. Un estudio
similar fue hecho en las restantes subsecciones 3--dimensionales del espacio de fase, obteniendo resultados similares.

\FIGURE{\epsfig{file=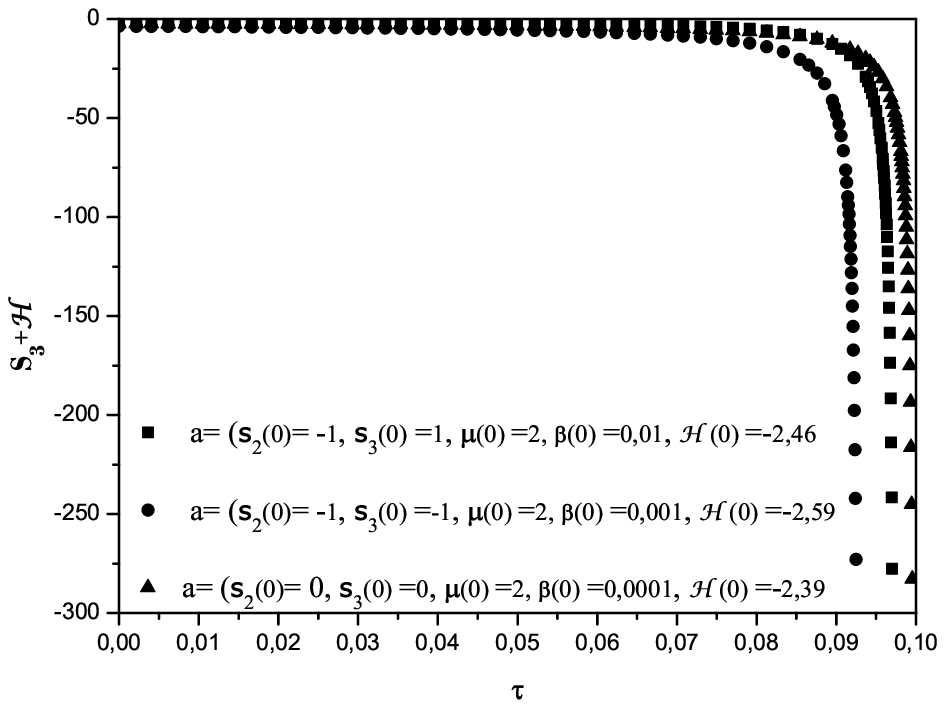,height=7cm,width=7cm,angle=0} \caption[Ejemplo figura3]{\footnotesize{El gr\'afico muestra el
comportamiento de $(S_3+\HH)$ versus $\tau$. Este comportamiento es similar a las trayectorias en la Fig\,\ref{S1S2}, por consiguiente
la cantidad $(S_3+\HH) \rightarrow -\infty$.}}
\label{S3espfase}}

\subsection{Conclusiones del cap\'itulo}

Hemos usado en el modelo Bianchi-I para estudiar la evoluci\'on de un elemento de volumen de un gas magnetizado de neutrones, caracterizado
f\'isicamente por una ecuaci\'on de estado completamente relativista.
Por otro lado, en nuestros estudios te\'oricos argumentamos la simplicidad de la geometr\'ia Bianchi como una aproximaci\'on a un subsistema
gran can\'onico formado por una fuente magnetizada de neutrones y sometida a las condiciones existentes en el centro de un objeto compacto.
Por este motivo, nuestro tratamiento debe ser concebido como una primera aproximaci\'on al problema, ya que cerca del centro del objeto
compacto se espera que las m\'etricas tengan un comportamiento aproximado, y esta descripci\'on ser\'ia \'util para entender la evoluci\'on
local de elementos de volumen bajo estas condiciones.

Sin embargo, un apropiado examen sobre las limitaciones espec\'ificas de la din\'amica de este modelo de juguete y su conexi\'on concreta
con los estudios astrof\'isicos actuales de los objetos compactos van m\'as all\'a del alcance de este cap\'itulo. En un futuro cercano,
se trabajar\'a con m\'etodos perturbativos, m\'etodos num\'ericos m\'as elaborados y fuentes menos idealizadas.

Las ecuaciones de campo de Einstein-Maxwell para un gas magnetizado de neutrones en una geometr\'ia Bianchi-I fueron transformadas en un
sistema de ecuaciones no--lineales de evoluci\'on, las cuales fueron resueltas num\'ericamente para una variedad de condiciones iniciales
de colapso. Estas soluciones fueron analizadas cualitativamente como un sistema din\'amico propio. As\'i los resultados que encontramos
fueron:
\begin{itemize}
  \item El estado final en la evoluci\'on de colapso de un elemento de volumen del sistema es una singularidad tipo ``punto''. Este estado
final ocurre para un ancho rango de condiciones iniciales asociadas con valores que t\'ipicamente existen en los objetos compactos.
  \item El campo magn\'etico aumenta r\'apidamente a medida que el sistema evoluciona hacia la singularidad. Aunque su intensidad siempre
se mantiene por debajo del campo magn\'etico cr\'itico. Este resultado es consistente con los valores obtenidos num\'ericamente para el
campo magn\'etico m\'aximo en modelos con configuraciones magnetizadas y en rotaci\'on estable (ver \cite{bocquet}).
  \item El estudio del espacio de fase asociado con las ecuaciones din\'amicas muestran que el sistema evoluciona, para $H>0$, a un punto de
equilibrio, (o sea, a una configuraci\'on estable). Es posible introducir una dependencia con la temperatura en la ecuaci\'on de estado.
Al introducir la temperatura, la evoluci\'on del gas de neutrones podr\'ia asociarse con fuentes de neutrones a altas temperaturas en el
contexto de universos magnetizados primordiales en modelos cosmol\'ogicos \cite{King:2006cy}.
\end{itemize}
Es importante tener en cuenta que, al contrario del estudio din\'amico para un gas de electrones en el cap\'itulo anterior (ver tambi\'en
\cite{Alain2}), las singularidades anis\'otropas tipo ``cigarro'' no ocurren para todas las condiciones iniciales impuestas. Sin embargo,
el efecto din\'amico del campo magn\'etico bajo condiciones cr\'iticas es un estado final de singularidad anis\'otropa alineada en la
direcci\'on del campo magn\'etico. La exclusiva aparici\'on de singularidades tipo ``punto'' para el gas de neutrones magnetizado sugiere
que el estado final de evoluci\'on de este gas es dominado m\'as intensamente por la fuerte gravedad en comparaci\'on con el gas de
electrones. Esto es consistente con el hecho que los electrones est\'an fuertemente acoplados al campo magn\'etico mediante su carga
el\'ectrica, sin embargo los neutrones tiene un acoplamiento m\'as d\'ebil asociado con su momento magn\'etico an\'omalo.

Los modelos con geometr\'ia Bianchi I en una representaci\'on com\'ovil tiene una severa limitaci\'on relacionada con los efectos din\'amicos
asociados con el campo magn\'etico. Esto es importante cuando se consideran a los neutrones como fuente para los cuales la carga el\'ectrica
es cero, no as\'i su momento magn\'etico an\'omalo.

Siendo estos espacio-tiempos espacialmente homog\'eneos con 4--velocidad ortonormal a las hipersuperficies de simetr\'ia maximal, la fuerza
de Lorentz es cero: $f^{b}=qu_{a}F^{a b}=0$. Tambi\'en, en efecto los modelos Bianchi-I son espacialmente planos y esto hace al modelo
inadecuado para estudiar (incluso como modelo juguete) el colapso local de dicho volumen elemental magnetizado dentro de objetos localizados.
O sea en tales objetos compactos la curvatura espacial es fuertemente positiva y no plana. Sin embargo, en nuestro caso tales
inconsistencias pueden ser sanadas (al menos parcialmente) si consideramos perturbaciones generales al modelo Bianchi-I, en los cuales la
curvatura escalar y la 4--aceleraci\'on no son nulas a orden perturbativo. En estos casos es posible examinar los efectos de acoplamiento
tipo magneto--curvatura asociados con fuerza de Lorentz y par\'ametro de desaceleraci\'on no nulos en la ecuaci\'on de Raychaudhuri (ver
\cite{Tsagas}). El uso de tales modelos Bianchi-I perturbados para la descripci\'on de un gas con fuente de materia en general ser\'a
presentado en el pr\'oximo cap\'itulo.

A la par de introducir posibles perturbaciones al modelo Bianchi-I, otra posibilidad que mejorar\'ia el modelo est\'a sobre la idea de
una mejor descripci\'on din\'amica de la fuente. Nuevas consideraciones para mejorar esto, podr\'ia ser suponer modelos Bianchi I, V, VII
\'o IX con 4--velocidad inclinada, la cual introduce m\'as grados de libertad din\'amicos, incluyendo hasta la posibilidad de rotaci\'on
(ver \cite{Coley:2008gh} y referencias incluidas). Estos modelos podr\'ian permitirnos un mejor estudio(menos restrictivo) de los efectos
din\'amicos. Estos efectos son reportados en \cite{Tsagas}, en cuyo trabajo se muestra que la tensi\'on magn\'etica y el colapso gravitatorio
pueden presentar un acoplamiento no nulo con la 4--aceleraci\'on y vorticidad distinta de cero (incluso no perturbativa) en presencia de
campo magn\'etico.

Finalmente, un gas magnetizado conteniendo solamente neutrones puede ser te\'oricamente interesante, pero es muy idealizado como fuente para
un objeto compacto. As\'i, se puede considerar como una extensi\'on a este trabajo, adicionar adem\'as de los grados de libertad din\'amicos
(mencionados arriba), mezclas de neutrones, electrones y protones, acompa\~nados con sus respectivas ecuaciones de balance y potenciales
qu\'imicos adecuados, en comparaci\'on con otras ecuaciones de estado \cite{bocquet,salgado1,salgado2}. Todas estas extensiones est\'an
bajo consideraci\'on para trabajos futuros.


%
\newpage
\section[CONDICIONES PARA FRENAR EL COLAPSO]{\large {CONDICIONES PARA FRENAR EL COLAPSO}}

\label{Cap_Perturbado}
%
%

Es posible mostrar,  a partir de la ecuaci\'on de Raychaudhuri (ecuaci\'on (\ref{raych}) del Cap\'\i tulo \ref{Capitulo_1_RG}), como el acoplamiento relativista entre magnetismo y geometr\'ia genera una tensi\'on magn\'etica que puede ``enrollar'' las l\'\i neas de fuerza del campo magn\'etico. Este efecto tambi\'en se identifica con una ``elasticidad'' de estas l\'ineas de fuerza del campo, las cuales tienden a mantenerse rectas. Concretamente, esta tensi\'on magn\'etica  genera fuerzas de magneto-curvatura cuya manifestaci\'on cinem\'atica  es la posibilidad de prevenir la convergencia de l\'ineas de universo (no necesariamente geod\'esicas) que inicialmente convergen.  Evidentemente, la consecuencia principal de este efecto es la posibilidad de frenar el colapso gravitacional de las fuentes asociadas a estas l\'ineas de universo, lo cual (en caso de suceder) ocurrir\'ia sin violar las condiciones de energ\'ia est\'andares.\footnote{Adem\'as de la tensi\'on magn\'etica, el tensor de deformaci\'on distorsiona la distribuci\'on de campos gradientes y, por lo tanto, tambi\'en podr\'\i a contribuir a resistir o favorecer al colapso.}

Un resultado interesante que trata sobre el colapso gravitacional de un fluido magnetizado sin simetr\'\i a esf\'erica fue obtenido por K. Thorne en el llamado ``Universo magnetizado de Melvin'' \cite{Krasinski:2006sb,Germani:2005ar,Melvin1964,Tsagas:2004kv,Thorne1965}. En dicho modelo el colapso del sistema se frena antes de alcanzar la singularidad, lo cual sucede debido al cambio de la densidad de energ\'\i a de la materia magn\'etica. Mas recientemente, C.G. Tsagas ha encontrado \cite{Tsagas} que existe un acoplamiento intr\'inseco entre el magnetismo y la geometr\'ia que surge de las propiedades de la tensi\'on de las l\'ineas de fuerza magn\'eticas (la ``elasticidad''), y cuya consecuencia es   el incremento a la resistencia al colapso. Esta tensi\'on magneto-geom\'etrica es la generalizaci\'on relativista del mismo efecto en la f\'\i sica newtoniana, apareciendo algunas veces en la literatura como ``frenado magn\'etico''.

En el presente cap\'itulo examinamos, haciendo uso de la ecuaci\'on de Raychaudhuri, el efecto del campo magn\'etico sobre el colapso gravitatorio. Para llevar a cabo este objetivo tomamos en cuenta el trabajo de Tsagas en \cite{Tsagas} y extendemos, al contexto astrof\'isico, el formalismo desarrollado por C.G.Tsagas y R.Martens \cite{Tsagas,Tsagas:1999tu,Tsagas:2004kv,Tsagas:1997vf,Tsagas:1998jm}. Retomando las limitaciones planteadas en los dos cap\'itulos anteriores, pasaremos a calcular las cantidades din\'amicas covariantes que aparecen en la ecuaci\'on de Raychaudhuri (ver Cap\'\i tulo \ref{Capitulo_1_RG}) para un modelo Bianchi-I perturbado. Trataremos la cuesti\'on del colapso de la fuente magnetizada en forma cualitativa,  sin resolver las ecuaciones de Einstein-Maxwell. El objetivo de dicho estudio cualitativo es obtener informaci\'on sobre el rol de las tensiones magn\'eticas y entender las implicaciones potenciales de estas en modelar la evoluci\'on de fuentes magnetizadas en situaciones menos idealizadas.

\subsection{Espacio-tiempo Bianchi-I perturbado}

Usando 1-formas diferenciales\footnote{Las 1-formas diferenciales constituyen la base de cualquier tensor $n$-veces covariante o sea $T=T_{a_{1}a_{2}...a_{n}}\emph{w}^{a_{1}}\emph{w}^{a_{2}}...\emph{w}^{a_{n}}$, para mayor informaci\'on ver \cite{MTW}}, el elemento de l\'inea de un espacio-tiempo Bianchi I perturbado puede ser escrito como,
\begin{equation}
ds^{2}=-(\emph{w}^{0})^2+(\emph{w}^{x^{1}})^2+(\emph{w}^{x^2})^2+(\emph{w}^{x^3})^2,
\end{equation}
donde,
\begin{eqnarray}
 \emph{w}^{0}= \,\,\sqrt{1+q_{0}}\,dx^{0}\,,\\
 \emph{w}^{x^1}=Q_{1}\sqrt{1+q_{1}}\,dx^{1}\,,\\
 \emph{w}^{x^2}=Q_{2}\sqrt{1+q_{2}}\,dx^{2}\,,\\
 \emph{w}^{x^3}=Q_{3}\sqrt{1+q_{3}}\,dx^{3}\,,
\end{eqnarray}\label{Set of one forms}
y las $q_{\mu}=q_{\mu}(t,\vec{x}),\,\, \mu=0,1,2,3$ son las funciones que representan las perturbaciones a la m\'etrica de Kasner, por lo que satisfacen $|q_{\mu}|\ll 1$, mientras que las  $Q_{j}=Q_{j}(t), j=1,2,3.$\footnote{En este cap\'itulo, a diferencia de los otros cap\'itulos, usamos \'\i ndices griegos para el espacio--tiempo e \'\i ndices latinos para la parte espacial. Adem\'as, usaremos indistintamente  para la derivada covariante el punto y coma ``;'' o el s\'imbolo ``$\nabla$''.} son las funciones no perturbadas, de modo que recuperamos la m\'etrica de Kasner si $q_{\mu}=0$. Elegimos la 4--velocidad  $u^{\alpha}$ ortogonal a las hipersuperficies espaciales, entonces,
\begin{equation}
u^{\alpha}=(1+q_{0})^{-1/2}\,\delta^{\alpha}_{0}, \qquad
u^{\alpha}u_{\alpha}=-1,
\end{equation}
por lo que la 4-aceleraci\'on $\dot{u}_{\alpha}=u^{\beta}u_{\alpha;\beta}$ toma la forma,
\begin{equation}
\dot{u}_{\alpha}=\frac{1}{2(1+q_{0})}\left[0,\,{{\partial_{_{x_1}}}{q_0}},\,{\partial_{_{x_2}}}{q_0},{\partial_{_{x_3}}}{q_0}\right].
\end{equation}

\subsection{C\'alculo de cantidades din\'amicas y ecuaci\'on de Raychaudhuri}

Tomando en cuenta la m\'etrica del modelo Bianchi I perturbado, las variables din\'amicas asociadas a la ecuaci\'on de Raychaudhuri (ver cap\'\i tulo \ref{Capitulo_1_RG}) pueden ser directamente calculadas. El escalar de expansi\'on y el tensor de deformaciones ser\'an,
\begin{eqnarray}
\Theta&=&\frac{1}{\sqrt{1+q_{0}}}\sum^{3}_{i=1}\left(\frac{\dot{Q_{i}}}{Q_{i}}+\frac{\dot{q_{i}}}{2(1+q_{i})}\right),\nonumber\\
&=&
\frac{1}{\sqrt{1+q_{0}}}\sum^{3}_{i=1}\frac{\partial}{\partial\,t}\left[\ln\left(Q_{i}\sqrt{1+q_{i}}\right)\right],
\end{eqnarray}
%
%
\begin{eqnarray}
\sigma^{\mu}_{\,\,\nu}&=& \hbox{diag}[\,\Sigma_{1},\,\Sigma_{2},\,\Sigma_{3},\,0],\\
\Sigma_{i}&=&\frac{1}{3\sqrt{1+q_{0}}}\left(\frac{\dot{2Q_{i}}}{Q_{i}}-\frac{\dot{Q_{j}}}{Q_{j}}-
\frac{\dot{Q_{k}}}{Q_{k}}+\frac{\dot{q_{i}}}{1+q_{i}}-\frac{\dot{q_{j}}}{2(1+q_{j})}-
\frac{\dot{q_{k}}}{2(1+q_{k})}\right),\nonumber\\
\qquad \qquad \qquad \qquad &i&\neq j \neq k \,\,(i,j,k=1,2,3).\nonumber
\end{eqnarray}
 El par\'ametro de deceleraci\'on $A\equiv u^{i}\,_{;i}$ y el escalar de deformaci\'on $\sigma^2$ toman la forma,
\begin{equation}
A=\frac{1}{2(1+q_{0})}\sum^{3}_{k=1}\frac{1}{Q^2_{k}(1+q_{k})}\left[{\partial^{2}_{k}q_{0}}-\frac{\partial_{k}q^2_{0}}
{2(1+q_{0})}+\frac{\partial_{k}q_{0}}{2}\left(-\frac{\partial_{k}q_{k}}{1+q_{k}}+\sum^{_{(l\neq
k)}}_{l=1}\frac{\partial_{k}q_{l}}{1+q_{l}}\right)\right], \label{BianPert_Param_Desacc}
\end{equation}
\begin{eqnarray}
\sigma^2 &=& \frac{1}{2}\sigma^{\mu \nu}\sigma_{\mu
\nu}=\frac{\lambda^2}{3}=\frac{1}{3(1+q_{0})}\sum^{3}_{k=1}\Biggl\{\left(
\frac{\dot{Q_{k}}}{Q_{k}}+\frac{\dot{q}_{k}}{2(1+q_{k})}\right)^{2}-\nonumber \\
&-&\frac{1}{2}\sum^{_{(l\neq
k)}}_{l=1}\left(\frac{\dot{Q_{k}}}{Q_{k}}+\frac{\dot{q}_{k}}{2(1+q_{k})}\right)\left(\frac{
\dot{Q_{l}}}{Q_{l}}+\frac{\dot{q}_{l}}{2(1+q_{l})}\right)\Biggr\}.
\end{eqnarray}
Por otra parte, nos ser\'a \'util definir el siguiente par\'ametro adimensional de impacto de la deformaci\'on, 
\begin{eqnarray}
\zeta &=&\frac{\lambda}{\Theta}=\Biggl\{\sum^{3}_{k=1}\Biggl\{\left(
\frac{\dot{Q_{k}}}{Q_{k}}+\frac{\dot{q}_{k}}{2(1+q_{k})}\right)^{2}-\nonumber
\\&-&\frac{1}{2}\sum^{_{(l\neq k)}}_{l=1}\left(\frac{\dot{Q_{k}}}{Q_{k}}+\frac{\dot{q}_{k}}{2(1+q_{k})}\right)\left(\frac{
\dot{Q_{l}}}{Q_{l}}+\frac{\dot{q}_{l}}{2(1+q_{l})}\right)\Biggr\}
  \Biggr\}^{\frac{1}{2}}\frac{1}{\sum^{_{3}}_{i=1}
\left(\frac{\dot{Q_{i}}}{Q_{i}}+\frac{\dot{q_{i}}}{2(1+q_{i})}\right)}.\nonumber
\end{eqnarray}

Dada la existencia de una fuente magnetizada (con campo el\'ectrico nulo), el tensor de energ\'\i a-momento viene dado por,\footnote{Note que aqu\'i, a diferencia de los tratamientos anteriores, s\'i hemos incluido la aproximaci\'on de \'arbol, o sea la contribuci\'on cl\'asica del campo magn\'etico $B^{2}$.}
\begin{equation}
T_{\mu \nu}=(e_{0}+\frac{B^2}{2})u_{\mu}u_{\nu}+(p_{0}+\frac{B^2}{6})h_{\mu \nu}+\Pi_{\mu \nu},  \label{TensorMater_Perturb}
\end{equation}
el cual corresponde al tipo de fuentes conocido como el fluido ideal ``conductor perfecto'' o de ``conductividad infinita''\cite{Trab_Numer_Flui_Cond_Infi}. Como $p_{0}$ y $\mu$ son las presiones y la densidad de energ\'ia del gas is\'otropo, y $H^2$ la densidad de energ\'ia del campo magn\'etico $[J\,m^{-3}]$, entonces los factores que miden las anisotrop\'ias son,
\begin{equation}
w=\frac{p_{0}}{\mu}\equiv p_{0}(t,\vec{\mathbf{x}})/e_{0},\qquad
h=\frac{H^2}{\mu}\equiv B^2(t,\vec{\mathbf{x}})/e_{0}.
\end{equation}

Sin embargo, la existencia de las perturbaciones $q_{k}$ implica un entorno anis\'otropico e inhomog\'eneo que generaliza la anisotrop\'\i a del modelo Bianchi I. Por lo tanto, la ecuaci\'on de conservaci\'on asociada al tensor energ\'ia-momento que hemos definido en (\ref{TensorMater_Perturb}) viene a ser  (ver \cite{Tsagas:1997vf}),
\begin{equation}
  (e_{0}+p_{0}+B^{2})\dot{u}^{\mu}+h^{\mu \nu}(p_{0}+\frac{1}{2}B^{2})_{;\nu}-h^{\mu}\,_{\nu}B^{\nu}\,_{;\gamma}B^{\gamma}-B^{\mu}B^{\nu}\,_{;\nu}=0,
\label{Ec_Balance_Pert_1}
\end{equation}
la cual tambi\'en se puede escribir como,
\begin{equation}
  (e_{0}+p_{0}+\frac{2}{3}B^{2})\dot{u}_{\mu}=-D_{\mu}p_{0}-a_{\mu}-\Pi_{\mu \nu}\dot{u}^{\nu},
\label{Ec_Balance_Pert_2}
\end{equation}
donde $D_{\mu}p_{0}=h_{\mu}^{\nu}\nabla_{\nu}p_{0}$ denota el gradiente de presi\'on medido por observadores en reposo com\'oviles con el fluido. Es importante enfatizar que todos los t\'erminos en (\ref{Ec_Balance_Pert_2}) se deben exclusivamente a la inhomogeneidad introducida por las perturbaciones.

En el modelo Bianchi I ``puro'' sin perturbaciones, al ser homog\'eneo y al no haber flujos t\'ermicos disipativos o efectos de rotaci\'on que puedan generar perturbaciones fuera de las condiciones de equilibrio, es evidente  que la anisotrop\'\i a de un fluido ideal debe ser causada solo por el campo magn\'etico. Sin embargo, la existencia de las perturbaciones implica que las desviaciones de la homogeneidad y la isotrop\'\i a no solo se deben al campo magn\'etico, por lo que podemos estudiar el acoplamiento de la geometr\'\i a con el campo magn\'etico en un contexto mucho m\'as general que incorpora tambi\'en a la inhomogeneidad (adem\'as de la anisotrop\'\i a).

Si llevamos a cabo un an\'alisis gravito-magn\'etico, vemos que $a_{\mu}$ es el vector que determina la aceleraci\'on asociada a la fuerza de Lorentz, la cual siempre es normal a la direcci\'on de las l\'ineas del campo $a_{\mu}B^{\mu}=0$ (adem\'as se cumple  $a_{\mu}u^{a}=0$), y surge cuando el modelo magn\'etico distorsiona las condiciones de equilibrio local. Este vector aceleraci\'on tiene la forma,
  \begin{equation}
   a_{\mu}=-\frac{1}{2}D_{\mu}B^{2}+B^{\nu}D_{\nu}B_{\mu}. \label{Vect_acele_Pert}
  \end{equation}
El primer t\'ermino del lado derecho de (\ref{Vect_acele_Pert}) viene dado por la ``presi\'on magn\'etica'' y el segundo viene de la ``tensi\'on del campo''. Si este \'ultimo no es balanceado aparecer\'a una fuerza neta sobre las part\'iculas del fluido. Para el tensor energ\'ia-momento (\ref{TensorMater_Perturb}), el cual es anis\'otropico y corresponde a un fluido conductor infinito, todos los t\'erminos en (\ref{Ec_Balance_Pert_2}) son no nulos.
 La ecuaci\'on cinem\'atica de propagaci\'on, es la ecuaci\'on de Raychaudhuri vista en el cap\'\i tulo \ref{Capitulo_1_RG} con $\Lambda=0$, la cual reescribimos como,
\begin{equation}
\dot{\Theta}+
{\frac{\Theta^2}{3}}+2\sigma^2+\frac{\mu}{2}(1+3w+h)-A=0,
\end{equation}
\begin{equation}
\dot{\Theta}= \frac{1}{\sqrt{1+q_{0}}}\sum^{3}_{k=1}\left[
\frac{\ddot{q_{k}}}{2(1+q_{k})}+\frac{\ddot{Q_{k}}}{Q_{k}}-\frac{\dot{Q^{2}_{k}}}{Q^{2}_{k}}-
\frac{1}{2}\frac{\dot{q}^{2}_{k}}{(1+q_{k})^2}-\frac{\dot{q}_{0}}{2(1+q_{0})}\left(\frac{\dot{q}_{k}}{2(1+q_{k})}+
\frac{\dot{Q}_{k}}{Q_{k}} \right) \right], \nonumber
\end{equation}
\begin{equation}
\frac{\Theta^{2}}{3}=
\frac{1}{3(1+q_{0})}\sum^{3}_{k=1}\left[\frac{\dot{Q^{2}_{k}}}{Q^{2}_{k}}+\frac{\dot{q}^{2}_{k}}{4(1+q_{k})^2}+\sum^{_{j\neq
k}}_{j=1}\left(\frac{\dot{Q}_{k}}{Q_{k}}\frac{\dot{Q}_{j}}{Q_{j}}+\frac{\dot{q}_{k}\dot{q}_{j}}{4(1+q_{k})(1+q_{j})}
\right)\right],\nonumber
\end{equation}
\begin{eqnarray}
 2\sigma^{2}&=& \frac{2}{3(1+q_{0})}\sum^{3}_{k=1}\Biggl\{\frac{\dot{Q}^{2}_{k}}{Q^{2}_{k}}+\frac{\dot{q}^{2}_{k}}{4(1+q_{k})^2}+
\frac{\dot{Q}_{k}\dot{q}_{k}}{Q_{k}(1+q_{k})}- \nonumber\\
&-&\frac{1}{2}\sum^{_{l\neq k}}_{l=1}\left(\frac{\dot{Q_{k}}\dot{Q_{l}}}{Q_{k}Q_{l}}+\frac{\dot{q}_{k}\dot{q}_{l}}{4(1+q_{k})(1+q_{l})}+
\frac{\dot{q}_{k}\dot{Q}_{l}}{(1+q_{k})Q_{l}}\right)\Biggr\}, \nonumber\\
&+&\frac{\mu}{2}(1+3w+h)=\frac{1}{2}(e_{0}+3p_{0}+B^2)= \nonumber \\
&\equiv&
\frac{1}{2(1+q_{0})}\sum^{3}_{k=1}\Biggl\{-\frac{\ddot{q}_{k}}{(1+q_{k})}-\frac{2\ddot{Q}_{k}}{Q_{k}}+\frac{1}{2}\frac{\dot{q}^{2}_{k}}{(1+q_{k})^{2}}+\frac{1}{2}\frac{\dot{q}_{0}}{(1+q_{0})}\frac{\dot{q}_{k}}{(1+q_{k})}\nonumber \\
&+&\frac{\dot{Q}_{k}}{Q_{k}}\left(\frac{\dot{q}_{0}}{(1+q_{0})}-\frac{2\dot{q}_{k}}{(1+q_{k})}\right)\Biggr\}+A.
\nonumber 
\end{eqnarray}
Note que tanto el escalar de expansi\'on como el tensor de deformaciones son cantidades positivas, y si se cumple la condici\'on de energ\'ia fuerte, entonces el escalar de Raychaudhuri (${\cal{R}}_{ab}u^au^b$) debe ser mayor que cero, o sea, se cumple que $\mu\,(1+3w+h)>0$. Entonces, el \'unico t\'ermino que se opone al colapso gravito--magn\'etico es el coeficiente de deceleraci\'on $A$ (si es que resulta ser positivo).

De las ecuaciones de Einstein obtenemos para este caso, que las componentes $T^{1}_{1}=T^{2}_{2}=T^{3}_{3}$ toman la forma,
\begin{equation}
\sum^{3}_{k,j=0}\frac{\partial_{x}q_{0}}{4Q^{2}_{k}(1+q_{0})}\Biggl\{
\frac{\partial_{k}q_{j}}{(1+q_{k})}-\frac{\partial_{k}q_{k}}{(1+q_{k})^2}\Biggr\}=\kappa\,(p_{0}+\frac{B^{2}}{6})\,.
\end{equation}
El t\'ermino $T^{4}_{4}$ de la ecuaci\'on de Einstein ser\'a,
\begin{eqnarray}
\frac{\kappa}{2}(2e_{0}&+& B^{2})=\sum^{_{k\neq
l}}_{k,l=1}\Biggl\{-\frac{1}{2}\frac{\partial^{2}_{k}q_{l}}{(1+q_{k})(1+q_{l})Q^{2}_{k}}+\frac{1}{8(1+q_{0})}
\frac{\dot{q}_{k}\dot{q}_{l}}{(1+q_{k})(1+q_{l})}\nonumber
\\
&+&\frac{1}{(1+q_{0})}\frac{\dot{Q}_{k}}{2Q_{k}}\frac{\dot{q}_{l}}{(1+q_{l})}+\frac{1}{8}\frac{\partial_{k}q^{2}_{l}}{(1+q_{k})(1+q_{l})^{2}Q^{2}_{k}}+ \\
&+&\frac{1}{4}\frac{\partial_{k}q_{k}\partial_{k}q_{l}}{(1+q_{k})^{2}(1+q_{l})Q^{2}_{k}}+\frac{1}{2(1+q_{0})}\frac{\dot{Q}_{k}\dot{Q}_{l}}{Q_{k}Q_{l}}+
\nonumber \\
&+&\sum^{_{l\neq k\neq
m}}_{m=1}\frac{\partial_{k}q_{m}}{4Q^{2}_{k}(1+q_{k})(1+q_{m})}\left(-\frac{\partial_{k}q_{l}}{(1+q_{l})}+\frac{\partial_{k}q_{k}}{(1+q_{k})}+
\frac{\partial_{k}q_{m}}{(1+q_{m})}\right)\Biggr\},  \nonumber
\end{eqnarray}
y la curvatura escalar 3-dimensional toma la forma no nula,
\begin{equation}
{}^{_{(3)}}{\cal{R}}=\sum^{3}_{_{k\ne
j=1}}\frac{1}{(1+q_{k})(1+q_{j})Q^{2}_{j}}\left\{-\partial^{2}_{j}
q_{k}+\frac{\partial_{j}q^{2}_{k}}{2(1+q_{k})}+\frac{\partial_{j}q_{k}\partial_{j}q_{j}}{2(1+q_{j})}-\sum^{3}_{_{m\neq
j\neq k=1}}\frac{\partial_{j}q_{k}\partial_{j}q_{m}}{2(1+q_{m})}\right\}.\label{BianPert_Curv3dim}
\end{equation}
Note que la curvatura 3--dimensional (al igual que $A$) solo depende de las perturbaciones, y sus derivadas espaciales. Si pasamos a calcular el escalar de Ricci 4--dimensional, obtendremos expresiones similares, pero tambi\'en depender\'an de las funciones $Q_{k}$ y de sus derivadas, tanto espaciales como temporales. Para el caso no perturbado, se obtiene la simple expresi\'on,
\begin{equation}
  {\cal{R}}=\,2\sum^{3}_{_{k=1}}\frac{\ddot{Q}_{k}}{Q_{k}}+\sum^{3}_{_{m\neq j\neq k=1}} \frac{\dot{Q}_{k}\dot{Q}_{j}}{Q_{k}Q_{j}}.
\end{equation}

\subsection{Conclusiones del cap\'itulo}
Considerando perturbaciones a la m\'etrica de Kasner (modelo Bianchi I), hemos calculado la serie completa de par\'ametros cinem\'aticos que contiene la ecuaci\'on de Raychaudhuri, para una 4--velocidad com\'ovil. 
Este resultado es importante desde el punto de vista te\'orico, pues nos permite examinar las condiciones de colapso para fuentes magnetizadas en el contexto de un espacio-tiempo inhomog\'eneo (aunque la inhomogeneidad es perturbativa).
Como se puede apreciar, al introducir las perturbaciones se complica en gran medida la forma expl\'icita de los resultados anal\'iticos, pero estos nos aportan informaci\'on cualitativa importante.

Se puede notar inmediatamente que las perturbaciones aparecen en todas las expresiones, pero una dependencia directa de las perturbaciones solo se manifiesta en el par\'ametro de deceleraci\'on $A$ y en la curvatura 3-dimensional $^{_{(3)}}{\cal{R}}$. O sea, estas son las variables  que nos traen informaci\'on de los efectos perturbadores del campo magn\'etico sobre las fuentes de materia-energ\'ia del sistema.

El solo hecho de tener una curvatura 3-dimensional no nula (aunque sea perturbativa) ya nos dice que este modelo perturbado (como supon\'iamos desde las conclusiones del cap\'itulo anterior) es m\'as realista que un Bianchi-I puro, el cual como ya sabemos es plano en sus hipersuperficies tridimensionales. Si exigimos que la cantidad (\ref{BianPert_Curv3dim}) sea positiva entonces logramos una mejor aproximaci\'on a una descripci\'on ``realista'' de un volumen elemental dentro de un objeto compacto. Por otra parte, si la cantidad (\ref{BianPert_Param_Desacc}) resulta ser positiva, entonces los efectos de las perturbaciones tienden a frenar el colapso, y en caso contrario a  ayudar a colapsar al sistema.

Si proponemos formas anal\'\i ticas expl\'icitas de las perturbaciones $q_k$, podr\'iamos usar las formas generales de las soluciones anal\'iticas que hemos dado para explorar los efectos cinem\'aticos y din\'amicos mencionados. Esto constituye un tema interesante a tomar en cuenta para investigaciones futuras, pero antes debemos encontrar (o al menos suponer) cual es la forma expl\'icita de las perturbaciones que mejor corresponda a una condici\'on f\'isica menos idealizada del colapso gravitacional de un sistema magnetizado.

Un punto en el cual no podemos ofrecer predicci\'on alguna es en el tipo de singularidad que pudiera resultar del colapso gravitacional: singularidades tipo ``cigarro'' o ``punto''. El escalar de Ricci 4-dimensional tambi\'en depende de las perturbaciones, de sus primeras derivadas espaciales, y de sus primeras y segundas derivadas temporales, y tambi\'en depende de derivadas primeras y segundas temporales de los coeficientes m\'etricos. Por tanto, para analizar su comportamiento en condiciones singulares (cuando este y los dem\'as escalares de curvatura divergen) tendr\'iamos que conocer la forma expl\'icita de las perturbaciones (o al menos proponer formas espec\'\i ficas). El an\'alisis que hemos realizado cumple con el objetivo de este cap\'itulo: encontrar resultados anal\'iticos y cualitativos, lo m\'as generales posibles, que contribuyan al entendimiento de la interacci\'on entre la geometr\'ia y el campo magn\'etico, asi como faciliten la elaboraci\'on de estudios num\'ericos para casos particulares que podr\'\i an ser llevados a cabo en el futuro.

La limitaci\'on principal del modelo Bianchi I perturbado es que las funciones $q_{k}$ deben ser mucho menores que la unidad, de modo que desde el inicio ya estamos imponiendo una fuerte limitaci\'on a la geometr\'\i a y a la din\'amica de las fuentes (como sucede en todo tipo de an\'alisis perturbativo). Si las perturbaciones influyen sobre la intensidad del campo magn\'etico, entonces dicho campo nunca llegar\'ia a vencer a la intensidad del campo gravitatorio. Por esta raz\'on, para lograr una mejor descripci\'on y entendimiento de la din\'amica del colapso gravito-magn\'etico nos vemos en la necesidad de considerar para trabajos futuros, modelos Bianchi con m\'as grados de libertad que el modelo I.  
%
%
%
%
%
%

\part{PEQUE\~NOS AGUJEROS NEGROS}
\newpage
\section[ENTROP\'IA DE AGUJEROS NEGROS]{\large ENTROP\'IA DE AGUJEROS NEGROS}

\subsection{Entrop\'ia de agujeros negros: pelos cl\'asicos}

  Existe una simp\'atica expresi\'on sobre los ANs, que sol\'ia repetir John Wheeler, ``los ANs no tienen pelos''. Con esto quer\'ia decir que, excepto por unas pocas caracter\'isticas que los distinguen, todos los ANs resultan parecidos no exhiben ning\'un ``peinado'' personal, algo caracter\'istico que nos permita diferenciar un AN de otro. Lo \'unico que los caracteriza son su masa, su carga el\'ectrica, y su momento angular. Nadie podr\'ia distinguir dos ANs con estos mismos valores caracter\'isticos, y, precisamente, esta similitud de los rasgos definitorios ha hecho creer, a lo largo de los a\~nos, la ``extra\~na suposici\'on'' de que los ANs podr\'ian ser gigantescas part\'iculas elementales.

La Teor\'ia de Cuerdas nos ha permitido avanzar en esta direcci\'on pues variedades como la de Calabi-Yau\footnote{Es una variedad con 6-dimensiones enrolladas (compactificadas), y presenta dos esferas empotradas una de 2-dimensiones y otra de 3-dimensiones que din\'amicamente pueden interactuar.} presentan interacciones internas entre sus subvariedades, que apantalladas por Branas\footnote{Es un objeto extendido, sobre el cual viven las cuerdas.} envolventes, permiten obtener soluciones de AN sin masa, esto es lo m\'as similar a un fot\'on o sea pura radiaci\'on electromagn\'etica en nuestro mundo.

Por otro lado y menos especulativo, los ANs deber\'ian estar rodeados por un anillo de luz que revela sus propiedades fundamentales. Muy pocos astr\'onomos ponen en duda la existencia de los ANs, aunque la evidencia de su existencia es indirecta, a partir del comportamiento de otros objetos cercanos a los ANs, tales como estrellas brillantes. \\
Estos objetos literalmente podr\'ian ser engullidos por el AN y en caso de dar se\~nales medibles de tal comportamiento, entonces esto constituir\'ia una prueba de su existencia. Obviamente, para la comunidad de Astr\'onomos y para los F\'isicos, la detecci\'on directa contin\'ua siendo un desaf\'io.

 Por ejemplo, en los dos \'ultimos a\~nos, se ha utilizado la interferometr\'ia para crear im\'agenes del AN central de nuestra galaxia. Estas im\'agenes junto con las que se esperan obtener en los pr\'oximos a\~nos deber\'ian ser capaces de resolver detalles sobre la escala del horizonte de sucesos del AN.

Retornemos entonces al conocido ``teorema del no pelo'' que naturalmente impone unos l\'imites estrictos a la influencia que el AN pueda ejercer en su entorno. Im\'agenes de un AN o m\'as bien de la materia que lo circunda, podr\'ian refutar el teorema, si se prueba que dicha materia se distorsiona de alguna manera. Tim Johannsen y Psaltis Dimitrios \cite{JohannsenDimitrios}, de la Universidad de Arizona, en Tucson, dicen que hay una manera f\'acil de probar el teorema. En la actualidad, calculan que los ANs deben estar rodeados por un anillo de luz.

Esta luz proviene de los fotones que han quedado atrapados en una \'orbita circular alrededor del AN, en las afueras del horizonte de sucesos, que luego se dispersa por el gas y el polvo que cae en el AN. Este anillo debe ser mucho m\'as brillante que el gas y el polvo que lo rodea, debe tener un di\'ametro que es unas diez veces el tama\~no del AN, lo que significa que deber\'ia ser visible en las im\'agenes que pronto estar\'an disponibles y que proporcionar\'an una medida directa de la masa del AN. La forma de este anillo, depende de las propiedades del AN, y no de la estructura de la materia que cae en el AN. Eso significa que la forma del anillo es la medida de las propiedades del AN y cualquier asimetr\'ia en el anillo ser\'a una violaci\'on directa del teorema del no pelo.
Los astr\'onomos no tienen que mirar muy lejos para encontrar tal anillo. ``El AN del centro de la V\'ia L\'actea'', es el candidato ideal para probar el teorema del no pelo debido a su brillo, a su gran tama\~no, momento angular, y relativa cercan\'ia.

La soluci\'on estacionaria de Schwarzschild (1916) la discutimos en el cap\'itulo \ref{Capitulo_1_RG} y la ecuaci\'on (\ref{Sol_Schwarzschild}), muestra el estado final que debe tener todo colapso gravitacional con simetr\'ia central. Hist\'oricamente, despu\'es surgieron otras dos familias de soluciones, la soluci\'on con carga el\'ectrica ``$e$'' de Reissner (1916)--Nordstrom (1918) y posteriormente, la soluci\'on de Kerr (1963) con momento angular ``$J=a\,M$''. Una generalizaci\'on de la familia de Kerr, cargada fue obtenida por Newman en 1965. Esta \'ultima soluci\'on de Kerr cargada est\'a formada por tres--par\'ametros, y la m\'etrica del espacio-tiempo junto con el vector potencial electromagn\'etico son dados matem\'aticamente por la ecuaci\'on siguiente,     
%
\begin{eqnarray}
  d^{2}s &=& -\Bigl(\frac{\bigtriangleup -a^2\,sin^{2}\theta}{\Sigma}\Bigr)\,dt^2-\frac{2a\,sin^{2}\theta (r^2+a^2-\bigtriangleup)}
{\Sigma}dt d\phi+\nonumber\\
&+&\Bigl[\frac{(r^2+a^2)^2-\bigtriangleup a^2 sin^{2}\theta }{\Sigma}\Bigr]sin^{2}\theta\,d\phi^2+\frac{\Sigma}{\bigtriangleup} dr^{2}+
\Sigma\, d\theta^{2},  \label{Metrica_KN}\\
A_{a}&=&-\frac{er}{\Sigma}[(dt)_{a}-a\, sin^{2}\theta (d\phi)_{a}],
\end{eqnarray}
 donde, 
   \begin{eqnarray}
      \Sigma &=& r^2+a^2\,cos^2 \theta \, ,\, \\
      \bigtriangleup &=& r^2+a^2+e^2-2Mr,
   \end{eqnarray}
 y $e,a,$ y $M$ son los tres par\'ametros de la familia. Cuando $e=0$, tenemos a $A_{a}=0$, la m\'etrica del espacio-tiempo se reduce a la
familia de soluciones de Kerr. Cuando $a=0$ se recupera la soluci\'on de Reissner-Nordstrom y cuando $e=a=0$, la ecuaci\'on
(\ref{Metrica_KN}) se reduce a la soluci\'on de Schwarzschild.
  As\'i, todas las soluciones de ANs, estacionarias conocidas est\'an contenidas dentro de la familia 3-param\'etrica, seg\'un la TGR.

En todas las soluciones que tienen una interpretaci\'on f\'isica directa, aparecen los 3-par\'ametros $e,a$ y $M$. Para cualquier 2-esfera,
$S$, en la regi\'on asint\'otica, tenemos que,
%
 \begin{eqnarray}
    \frac{1}{2}\int_{S}\epsilon_{abcd}F^{cd}\,dx^{a}\,\wedge dx^{b} &=& 4\pi\, e,\label{Ec_Carga_Masa_Momento_1}\\
    -\frac{1}{8\pi}\int_{S}\epsilon_{abcd}\nabla^{c}\xi^{d}\,dx^{a}\,\wedge dx^{b} &=& M,\label{Ec_Carga_Masa_Momento_2}\\
    \frac{1}{16\pi}\int_{S}\epsilon_{abcd}\nabla^{c}\psi^{d}\,dx^{a}\,\wedge dx^{b} &=& M\,a,\label{Ec_Carga_Masa_Momento_3}
 \end{eqnarray}
donde, $\xi^{a}=(\partial/\partial t)^{a},$ y $\psi^{a}=(\partial/\partial\phi)^{a}$ son los vectores de Killing
(ver \cite{RMW_GR})\footnote{Para m\'as informaci\'on sobre la notaci\'on usada en
(\ref{Ec_Carga_Masa_Momento_1})--(\ref{Ec_Carga_Masa_Momento_3}), ver p\'ag 151, \cite{MTW} o Ap\'endice B de \cite{RMW_GR}.}.   
\subsection{Mecanismo de Sen}
Es conocido que la Teor\'ia de Cuerdas en el l\'imite de bajas energ\'ias surgen acoplamientos entre la gravedad y otros campos. Estas teor\'ias tambi\'en contienen soluciones tipo AN. As\'i la teor\'ia de cuerdas brinda un cuadro de trabajo para estudiar las propiedades de los ANs, sean cl\'asicos o cu\'anticos \cite{Sen:2007qy}.

Cl\'asicamente los ANs son soluciones de las ecuaciones de Einstein con propiedades especiales. Todos los ANs tienen una superficie hipot\'etica que los rodea, llamada horizonte de eventos de donde ning\'un objeto puede escapar despu\'es de atravesar su frontera. En la teor\'ia cu\'antica los ANs se comportan como cuerpos negros con temperatura finita conocida como ``temperatura de Hawking''. As\'i mismo emiten radiaci\'on Hawking en correspondencia con las leyes de radiaci\'on del cuerpo negro, e interact\'uan con la materia como un sistema termodin\'amico. Si nos basamos en leyes aproximadas donde se ignoran t\'erminos de derivaci\'on superior a segundo orden, la entrop\'ia de los ANs vendr\'a dada por la expresi\'on de Bekenstein--Hawking,
\begin{equation}
   S_{BH}=A/(4G_{N}),
\end{equation}
donde $A$ es el \'area del horizonte de eventos y $G_{N}$ es la constante de Newton. Una importante pregunta es si:
``\textquestiondown Podremos entender esta entrop\'ia desde el punto de vista estad\'istico, o sea como el logaritmo del n\'umero de
microestados cu\'anticos asociados al AN ?''.

Aunque no se tiene una respuesta definitiva a esta pregunta, para una clase especial de AN en teor\'ia de cuerdas conocido como ``AN
extremal'', esta pregunta ha sido contestada y en forma afirmativa. Estos ANs tienen temperatura cero por lo que no emiten radiaci\'on
Hawking y son usualmente estables. Algunas veces (no siempre) ellos son invariantes ante cierto n\'umero de transformaciones de
supersimetr\'ia y en ese caso se llaman ``ANs BPS''. Debido a la estabilidad y a las propiedades supersim\'etricas, se tiene cierto
control sobre la din\'amica de la configuraci\'on microsc\'opica (esencialmente envuelven D-branas, cuerdas fundamentales y otros objetos
solit\'onicos) de estos ANs. Esto permite calcular la degeneraci\'on de tales estados con un acoplamiento d\'ebil donde los efectos de
``backreaction''\footnote{En f\'isica te\'orica la ``backreaction'' es frecuentemente utilizada para calcular el comportamiento de una
part\'icula de prueba u objeto, en un campo externo. Por ejemplo, si la part\'icula de prueba tiene masa o carga infinitesimal entonces
se ignora la ``backreaction'', sin embargo dicha part\'icula posee carga y masa no nula lo que altera a su vez el asumir campo externo
fijo. Por esto, en ciertos casos se debe tomar en cuenta su efecto, entonces los efectos de ``backreaction'' no se podr\'an despreciar.}
del sistema pueden ser ignorados. La supersimetr\'ia nos permite extender el resultado al caso de acoplamiento fuerte donde el
``backreaction'' gravitacional pasa a ser importante y el sistema puede ser descrito como un AN. En teor\'ia de cuerdas nos encontramos
con una gran variedad de ANs extremos tipo BPS, que en el l\'imite de AN grande \cite{Strominger:1996sh}, la entrop\'ia toma la forma,
\begin{equation}
  S_{BH}(Q)=S_{est}(Q),\label{Igualdad_Entr_BH_est}
\end{equation}
donde $S_{BH}(Q)$ es la entrop\'ia de Bekenstein-Hawking de un AN extremal debido a las cargas ``$Q$'', y $S_{est}(Q)$ es definida como,
\begin{equation}
 S_{est}(Q)=ln\,d(Q),
\end{equation}
donde $d(Q)$ es la degeneraci\'on de los estados BPS en la teor\'ia debido al conjunto de cargas ``Q''. Esto nos brinda un buen entendimiento de la entrop\'ia de Bekenstein-Hawking desde el punto de vista microc\'opico.

Las comparaciones entre $S_{BH}$ y $S_{est}$ fueron hechas primeramente en el l\'imite de cargas grandes. En este l\'imite el horizonte
es grande de modo que la curvatura y otros campos de fuerza en \'el son peque\~nos y podemos calcular la entrop\'ia usando
(\ref{Igualdad_Entr_BH_est}) sin preocuparnos por correcciones con t\'erminos de orden superior de derivaci\'on\footnote{Se les conoce
por ``t\'erminos de orden superior de derivaci\'on'', porque introducen derivadas de la m\'etrica superiores al segundo orden, en las
ecuaciones de campo.} a la acci\'on efectiva de cuerdas. Adem\'as el c\'alculo de $S_{est}(Q)$ se simplifica porque la din\'amica del
sistema microsc\'opico es frecuentemente descrita por una Teor\'ia de Campo Conforme(TCC), $1+1$ dimensional con coordenada espacial
compactificada en un c\'irculo. Un AN extremo fuertemente cargado, se corresponde con estados en esta TCC con los autovalores $L_{0}$
(o $\bar{L}_{0}$) y cero $\bar{L}_{0}$ (o $L_{0}$). La degeneraci\'on de tales estados puede ser calculada usando la f\'ormula de Cardy en
t\'erminos de la carga central izquierda y derecha $(c_{L},c_{R})$, sin conocer los detalles de la teor\'ia TCC, o sea, 
%
  \begin{eqnarray}
 S_{est}(Q) \simeq \,2\pi \sqrt{\frac{c_{L}L_{0}}{6}}  \qquad \hbox{para} \qquad \bar{L}_{0}=0,\\
            \simeq \,2\pi \sqrt{\frac{c_{R}\bar{L}_{0}}{6}}  \qquad \hbox{para} \qquad {L}_{0}=0.
  \end{eqnarray}
%

Entonces constituye una sorpresa que dos c\'alculos completamente diferentes (uno para $S_{BH}(Q)$ y otro para $S_{est}(Q)$), den resultados
iguales. De este modo cuando nos movamos lejos del l\'imite de cargas grandes, la curvatura y los restantes campos de fuerza en el horizonte
no ser\'an despreciables. As\'i tendremos que tomar en cuenta el efecto de los t\'erminos de derivaci\'on superior en la acci\'on efectiva
de la teor\'ia. T\'erminos de derivaci\'on superior son por ejemplo potencias cuadr\'aticas y de mayor orden del tensor de Riemann. Nosotros
esperamos que para los ANs grandes (no infinitos) los efectos de estos t\'erminos de derivaci\'on superior sobre el horizonte sean 
peque\~nos pero no nulos, estos efectos introducir\'an peque\~nas modificaciones a la geometr\'ia del horizonte y a la entrop\'ia de los ANs. Por otro lado, la f\'ormula de Cardy para la entrop\'ia estad\'istica, si se calcula fuera del l\'imite de AN grande, tambi\'en recibir\'a modificaciones inversas a potencias de la carga del AN, acentu\'andose sorpresivamente cada vez m\'as la similitud entre $S_{BH}$ y $S_{est}$, incluso para cuando tenemos en cuenta t\'erminos de derivaci\'on superior.

La entrop\'ia $S_{BH}$ de Bekenstein-Hawking fue calculada originalmente en TGR. Sin embargo, pretendemos aqu\'i incorporarle al resultado
t\'erminos de derivaci\'on superior. Varias dualidades\footnote{Las dualidades son transformaciones matem\'aticas que relacionan las
5--teor\'ias de cuerdas existentes.} en teor\'ias de cuerdas pueden mapear contribuciones cl\'asicas a la teor\'ia efectiva cu\'antica,
por tanto, no tiene sentido restringir nuestro an\'alisis a la teor\'ia cl\'asica. Una elecci\'on es elegir la acci\'on efectiva que genera
los diagramas irreducibles de una part\'icula (1PI), esta respeta todas las simetr\'ias de dualidad. Sin embargo, la teor\'ia de cuerdas
tiene part\'iculas no masivas, y si vamos a \'ordenes mayores en t\'erminos de derivaci\'on superior, la acci\'on 1PI contendr\'a
contribuciones no--locales. Como hemos visto aqu\'i, para cualquier teor\'ia de gravedad con t\'erminos de derivaci\'on superior y con una
densidad Lagrangiana local existe un buen algoritmo para calcular la entrop\'ia de ANs, hasta el presente no existen t\'ecnicas para tratar
con teor\'ias con acci\'on no-local. Esto causa un potencial problema para definir entrop\'ia de un AN en teor\'ias de cuerdas, m\'as all\'a
de los primeros \'ordenes. Se podr\'ia esquivar este problema usando una acci\'on efectiva local de Wilson, pero esta no respecta todas las
simetr\'ias de dualidad de la teor\'ia.

Tambi\'en se presentan ambiguedades en el lado estad\'istico, donde las correcciones a la entrop\'ia dependen del inverso de la carga del AN,
y se sabe que tales correcciones dependen del ensemble que elegimos para definir entrop\'ia. Por ejemplo podemos usar ensembles
micro-can\'onicos o gran-can\'onicos invariantes de dualidad, o podemos usar ensembles mixtos no invariantes de dualidad \cite{Ooguri:2004zv}. 

Esperamos que los ejemplos expl\'icitos para el c\'alculo de las correcciones a la entrop\'ia de ANs, puedan resolver las ambiguedades
anteriores y hacer una formulaci\'on m\'as precisa sobre la relaci\'on entre las dos entrop\'ias.    Nosotros nos dedicaremos al c\'alculo
de la entrop\'ia $S_{BH}$ de ANs teniendo en cuenta t\'erminos de derivaci\'on superior, bas\'andonos en el formalismo entr\'opico
\cite{Examples_EF_1} que es una adaptaci\'on de un formalismo m\'as general \cite{Wald:1993nt} al caso especial de ANs extremales. La
segunda l\'inea de trabajo o sea calcular la entrop\'ia estad\'istica se puede encontrar en \cite{David:2006yn,David:2006ru,David:2006ud},
y \cite{Sen:2007qy} donde incluso se comparan las dos v\'ias para el caso de un AN en una teor\'ia de cuerda supersim\'etrica en
4-dimensiones.

\subsection{Definici\'on de AN extremal}

Analizemos un AN de Reissner-Nordstrom (RN), que describe un AN esf\'ericamente sim\'etrico con carga el\'ectrica en la teor\'ia usual de Einstein-Maxwell,
\begin{equation}
  S=\int d^{4}x\, \sqrt{-det\, g}\,\mathcal{L}, \qquad \mathcal{L}=\frac{1}{16\pi G_{N}}R-\frac{1}{4}F_{ab}F^{ab}.
\label{Accion_de_RN}
\end{equation}
La soluci\'on de RN en esta teor\'ia viene dada por,
%
\begin{eqnarray}
 ds^{2}&=&-(1-a/\rho)(1-b/\rho)d\tau^{2}+\frac{d\rho^{2}}{(1-a/\rho)(1-b/\rho)}+\rho^{2}(d\theta^{2}+sin^{2}\theta d\phi^{2}),\\
 F_{\rho \tau}&=&\frac{q}{4\pi\rho^{2}}, \qquad F_{\theta \rho}=\frac{p}{4\pi}sin \theta,
\end{eqnarray}
%
donde $\rho,\theta,\phi$ y $\tau$ son las coordenadas del espacio-tiempo, $a$ y $b$ dos constantes dadas por,
\begin{equation}
  a+b=2\,G_{N}\,M, \qquad ab=\frac{G_{N}}{4\pi}(q^{2}+p^{2}),
\end{equation}
donde $q$, $p$ y $M$ que denotan la carga el\'ectrica, magn\'etica y la masa del AN. Cuando $a>b$ entonces el horizonte interno estar\'ia en $r=b$ y el externo en $r=a$. El l\'imite extremo ser\'a cuando $r=a=b$, entonces,
\begin{equation}
  M^{2}=\frac{1}{4\pi G_{N}}(q^2+p^2), \qquad a=b=\sqrt{\frac{G_{N}}{4\pi}(q^2+p^2)}.
\end{equation}

Ahora si definimos, 
\begin{equation}
t=\lambda \tau/a^{2}, \qquad r=\lambda^{-1}(\rho-a),
\end{equation}
donde $\lambda$ es una constante arbitraria, y reescribimos la soluci\'on extrema en estas nuevas coordenadas,
obtenemos que, 
%
\begin{eqnarray}
 ds^{2}&=&-\frac{r^2\,a^4}{(a+\lambda\,r)^{2}}dt^{2}+\frac{(a+\lambda\,r)^{2}}{r^2}dr^2+(a+\lambda\,r)^{2}(d\theta^2+sin^{2}\theta\,d\phi^{2}),\\
F_{rt}&=&\frac{q\,a^2}{4\pi(a+\lambda r)^2}, \qquad F_{\theta \phi}=\frac{p}{4\pi}sin \theta,
\end{eqnarray}
%
y en el l\'imite muy ``cercano al horizonte'' $\lambda \rightarrow 0$, la soluci\'on tomar\'a la forma,
%
  \begin{eqnarray}
    ds^2 &=& a^2 \Bigl(-r^2\,dt^2 +\frac{dr^2}{r^2}\Bigr)+a^2\,(d\theta^2+sin^2\theta\,d\phi^2), \label{RN_AdS}\\
     F_{rt} &=& \frac{q}{4\pi}, \qquad F_{\theta \phi}=\frac{p}{4\pi}sin\theta.
  \end{eqnarray}
%
La entrop\'ia que se obtiene dividiendo el \'area del horizonte por $4G_{N}$ es,
\begin{equation}
 S_{BH}=\frac{1}{4}(q^2+p^2),
\end{equation}
y tiene las propiedades siguientes:
\begin{itemize}
  \item En el l\'imite $\lambda \rightarrow 0$ manteniendo a $r$ fija, la coordenada original $\rho$ se aproxima a $a$.
As\'i (\ref{RN_AdS}) describe el campo de configuraci\'on de un AN cercano al horizonte.
  \item El espacio de (\ref{RN_AdS}) se desdobla en dos espacios. Uno etiquetado por $(\theta,\phi)$ descrito por una esfera $S^2$
de 2-dimensiones, y otro etiquetado por $(r,t)$ descrito por un espacio-tiempo Anti de Sitter (AdS) de 2-dimensiones. Este \'ultimo
resulta de la soluci\'on de las ecuaciones de Einstein en 2-dimensiones con constante cosmol\'ogica negativa.
  \item El reflejo de la simetr\'ia esf\'erica del AN original hace que en (\ref{RN_AdS}) se presente una isometr\'ia $SO(3)$ que
act\'ua sobre la esfera $S^2$. En (\ref{RN_AdS}) tambi\'en se presenta una isometr\'ia $SO(2,1)$ actuando sobre el espacio $AdS_{2}$,
esta es generada por, 
\begin{equation}
  L_{1}=\partial_{t}, \,\, L_{0}=t\partial_{t}-r\partial_{r},\,\,L_{-1}=\frac{1}{2}(\frac{1}{r^2}+t^2)\partial_{t}-tr\partial{r}. 
\end{equation}
  As\'i, tanto la m\'etrica como los campos de calibraci\'on son invariantes bajo una transformaci\'on del tipo    $SO(2,1)\times SO(3)$.
\end{itemize}

Entonces si postulamos que la geometr\'ia del AN no cambiar\'a si a\~nadimos t\'erminos de derivaci\'on de orden superior, y en el caso de D-dimensiones se puede generalizar a que los ANs tengan isometr\'ias del tipo $SO(2,1)\times SO(D-1)$, entonces todo esto constituye nuestra definici\'on de ``AN extremo''\footnote{ Para 4 y 5-dimensiones, estos postulados han sido probados \cite{James2}.}.

\subsection{M\'etodo de Sen}

Siguiendo \cite{Kraus:2005vz,Astefanesei:2006dd,Examples_EF_1}, supongamos una teor\'ia acoplada a campos de calibraci\'on $A^{(i)}_{\mu}$ y a campos escalares ${\phi_{s}}$. La densidad Lagrangiana ser\'a expresada en funci\'on de la m\'etrica, de estos campos, y de las derivadas covariantes de estos campos. Adem\'as, suponiendo ANs extremos, esf\'ericamente sim\'etricos con invariancia tipo $SO(2,1)\times SO(D-1)$ cerca del horizonte, la configuraci\'on de campos m\'as general y consistente con esta isometr\'ia es,
%
  \begin{eqnarray}
    ds^2 &=& g_{ab}dx^{a}dx^{b}= v_{1}\Bigl(-r^2\,dt^2+\frac{dr^2}{r^2}\Bigr)+v_{2}\,(d\theta^2+sin^2\theta\,d\phi^2), \label{BH_AdS_extremo_general}\\
     \phi_{s}&=& u_{s},\,\,F^{(i)}_{rt}=e_{i},\,\, F^{(i)}_{\theta \phi}=\frac{p_{i}}{4\pi}sin\theta,
  \end{eqnarray}
%
donde $v_{1},v_{2},u_{s},e_{i}$ y $p_{i}$ son constantes. Para esta m\'etrica las componentes del tensor de Riemann son,  
 \begin{eqnarray}
  R_{abcd}&=& -v^{-1}_{1}(g_{ac}g_{bd}-g_{ad}g_{bc}), \qquad a,b,c,d=r,t,\\
  R_{mnpq}&=& v^{-1}_{2}(g_{mp}g_{nq}-g_{mq}g_{np}), \qquad m,n,p,q=\theta,\phi.
 \end{eqnarray}
%

Las derivadas covariantes de los campos escalares $\phi_{s}$, los campos de calibraci\'on $F^{(i)}_{ab} $ y del tensor de Riemann $R_{abcd}$ se anulan cerca del horizonte de geometr\'ia. Debido a consideraciones generales de simetr\'ia, todas las contribuciones a las ecuaciones de movimiento de cualquier t\'ermino en $\mathcal{L}$ que contenga derivadas covariantes de los campos en general se anulan para esta m\'etrica, entonces podemos concentrarnos en aquellos t\'erminos que no contengan derivadas covariantes de los campos.

Definamos la densidad de lagrange $f(\vec{u},\vec{v},\vec{e},\vec{p})$ evaluada cerca de la geometr\'ia del horizonte
(\ref{BH_AdS_extremo_general}) e integrada por las variables angulares, 
\begin{equation}
 f(\vec{u},\vec{v},\vec{e},\vec{p})=\int d\theta d\phi \sqrt{-det\, g} \, \mathcal{L}.\label{Ec_f_integrales}
\end{equation}
Las ecuaciones de campo cerca del horizonte corresponden a extremar $f$ respecto a las variables $\vec{u}$ y $\vec{v}$,
\begin{equation}
  \frac{\partial f}{\partial u_{s}}=0,\qquad \frac{\partial f}{\partial v_{i}}=0. \label{Ec_Mov_Formalis_AS} 
\end{equation}

Las componentes no triviales de las ecuaciones del campo de calibraci\'on y de las identidades de Bianchi para la soluci\'on completa del AN toman la forma,
\begin{equation}
 \partial_{r}\Bigl(\frac{\delta \mathcal{S}}{\delta F^{(i)}_{rt}}\Bigr)=0, \qquad \partial_{r}F^{(i)}_{\theta \phi}=0,
\end{equation}
donde $\mathcal{S}=\int d^{4}x\,\sqrt{-det \,g}\,\mathcal{L}$ es la accci\'on. Estas ecuaciones son satisfechas cerca del horizonte de (\ref{BH_AdS_extremo_general}) y podemos extraer m\'as informaci\'on de ellas, pues evaluando estas integrales,
\begin{equation}
 \int d\theta d\phi \frac{\delta \mathcal{S}}{\delta F^{(i)}_{rt}}=a_{i}, \qquad \int d\theta d\phi F^{(i)}_{\theta \phi}=b_{i},
\end{equation}
cerca del horizonte (\ref{BH_AdS_extremo_general}), nos da que,
\begin{equation}
 a_{i}=\frac{\partial f}{\partial e_{i}}, \qquad b_{i}=p_{i}.
\end{equation}
Ellas son justamente la carga el\'ectrica $a_{i}$, y la carga magn\'etica $b_{i}$. Denotemos por,
\begin{equation}
 q_{i}=\frac{\partial f}{\partial{e_{i}}},  \label{Ec_Carga_Elec_AN}
\end{equation}
como la carga el\'ectrica del AN. Ahora el conjunto (\ref{Ec_Mov_Formalis_AS}) y (\ref{Ec_Carga_Elec_AN}), forma un sistema de ecuaciones con igual n\'umero de inc\'ognitas $\vec{u},\vec{v}$ y $\vec{e}$. En la mayor\'ia de los casos este sistema se puede resolver y quedar determinado en funci\'on de $\vec{q}$ y $\vec{p}$. Esto es consistente con el mecanismo atractor en supersimetr\'ia el cual enuncia que cerca de la configuraci\'on de un AN, las magnitudes dependen solamente de la carga el\'ectrica y magn\'etica del AN y no del valor de sus campos escalares.

Definamos ahora,
\begin{equation}
 \mathcal{E}(\vec{u},\vec{v},\vec{e},\vec{q},\vec{p})=2\,\pi(e_{i}q_{i}-f(\vec{u},\vec{v},\vec{e},\vec{p})).
\label{DEF_de_E}
\end{equation}
Las ecuaciones (\ref{Ec_Mov_Formalis_AS}) y (\ref{Ec_Carga_Elec_AN}), determinando $\vec{u},\vec{v}$ y $\vec{e}$ pueden ser escritas como,
\begin{equation}
 \frac{\partial \mathcal{E}}{\partial u_{s}}=0,\,\,\frac{\partial \mathcal{E}}{\partial v_{1}}=0,\,\,\frac{\partial \mathcal{E}}{\partial v_{2}}=0,\,\,\frac{\partial \mathcal{E}}{\partial e_{i}}=0.
\end{equation}

O sea, todos los par\'ametros cerca del horizonte pueden ser calculados extremando la funci\'on $\mathcal{E}$.
Por otro lado, la f\'ormula general planteada en \cite{Wald:1993nt} y escrita para ANs esf\'ericamente sim\'etricos, tiene la forma,
\begin{equation}
  S_{BH}=-8\pi \int_{H} d\theta d\phi \frac{\delta \mathcal{S}}{\delta R_{rtrt}}\sqrt{-g_{rr}g_{tt}},
  \label{Wald_Entr_AN_extremos}
\end{equation}
donde $H$ denota el horizonte del AN. Despu\'es de una serie de pasos a partir de (\ref{Wald_Entr_AN_extremos}) se llega a que (ver \cite{Sen:2007qy}, p\'ag 14),
\begin{equation}
  S_{BH}=2\pi\,\Bigl(e_{i}\frac{\partial f}{\partial e_{i}}-f \Bigr). \label{Wal_Entr_AN_Extre_Terminos_f}
\end{equation}

Comparando (\ref{Wal_Entr_AN_Extre_Terminos_f}) con (\ref{DEF_de_E}) se obtiene que,
\begin{equation}
 S_{BH}=\mathcal{E}(\vec{u},\vec{v},\vec{e},\vec{q},\vec{p}).
\end{equation}

O sea $S_{BH}(\vec{q},\vec{p})/2\pi$ puede ser interpretado como la transformaci\'on de Legendre de la funci\'on $f(\vec{u},\vec{v},\vec{e},\vec{p})$ con respecto a la variable $e_{i}$, despu\'es de eliminar $\vec{u}$ y $\vec{v}$ mediante el sistema (\ref{Ec_Mov_Formalis_AS}).

As\'i el formalismo de la funci\'on entr\'opica, reduce el problema del c\'alculo de la entrop\'ia de un AN extremal a un problema de solucionar un sistema de ecuaciones. Veamos un caso muy simple, el AN de RN,
 %
  \begin{eqnarray}
       ds^2 &=& v_{1}\Bigl(-r^2\,dt^2+\frac{dr^{2}}{r^2} \Bigr)+v_{2}(d\theta^2+sin^2\theta\,d\phi^2),\\
     F_{rt} &=& e, \qquad F_{\theta \phi}=p\, sin\theta/4\pi. 
  \end{eqnarray}
%
Usando (\ref{Accion_de_RN}), (\ref{Ec_f_integrales}) tenemos que,
\begin{equation}
     f(v_{1},v_{2},e,p)=4\pi v_{1}v_{2}\Bigl[\frac{1}{16\pi G_{N}}\Bigl(-\frac{2}{v_{1}}+\frac{2}{v_{2}}\Bigr)+\frac{1}{2}v^{-2}_{1}e^2-\frac{1}{2}v^{-2}_{2}(\frac{p}{2\pi})^2\Bigr]\,, 
\end{equation}
y que,
\begin{equation}
 \mathcal{E}(v_{1},v_{2},e,q,p)=2\pi(q\,e-f)=2\,\pi \Bigl[q\,e-\frac{1}{4G_{N}}(2v_{1}-2v_{2})-2\pi\,v_{2}v^{-1}_{1}e^2+2\pi v_{1}v^{-1}_{2}(\frac{p}{4\pi})^2 \Bigr].
\end{equation}
Aqu\'i $\mathcal{E}$ es extrema si,
\begin{equation}
   v_{1}=v_{2}=G_{N}\frac{q^2+p^2}{4\pi}, \qquad e=\frac{q}{4\pi}.
\end{equation}
Sustituyendo en la expresi\'on para $\mathcal{E}$, tenemos que,
\begin{equation}
  S_{BH}=\mathcal{E}=\frac{1}{4}(q^2+p^2),
\end{equation}
que coincide con al entrop\'ia cl\'asica de Bekenstein-Hawking para un AN de RN.
%

\newpage
\section[ENTROP\'IA DE AGUJEROS NEGROS EXTREMOS]{\large ENTROP\'IA DE AGUJEROS NEGROS EXTREMOS}

\subsection{Nuevos resultados en la aplicaci\'on del formalismo de Sen}

Motivados ya sea por lograr la total unificaci\'on de las fuerzas fundamentales de la Naturaleza o por tratar de explicar fen\'omenos que
hoy en d\'ia permanecen sin resolver en la f\'isica te\'orica, la TGR ya ha sido generalizada de muchas formas. Unas de estas v\'ias es
apoyarse en el contenido geom\'etrico que posee toda teor\'ia relativista y de esta forma lograr dicha generalizaci\'on. Esto es
espec\'ificamente lo que hacemos en este cap\'itulo aplicado al caso particular de ANs extremos.

En la f\'isica es muy importante conocer los invariantes que posee la teor\'ia si se piensa proponerla como una teor\'ia nueva, as\'i se
hace por ejemplo en electrodin\'amica, EDC o CDC\footnote{EDC (Electrodin\'amica Cu\'antica) y CDC (Cromodin\'amica Cu\'antica).} donde
el invariante fundamental es,
\begin{equation}
\frac{1}{\alpha}\,F^{a b}\,F_{a b},
\end{equation}
donde $\alpha$ es la masa del gap.\footnote{Determinar la masa del gap es un problema abierto, y est\'a dentro de los 23 problemas
de Hilbert (Axiomatizar toda la f\'isica), tambi\'en se encuentra entre los problemas del milenio. Esta masa se corresponde con la primera
excitaci\'on o estado b\'asico fundamental con respecto al vac\'io.} Si sobre ellos aplicamos el m\'etodo extremal se obtienen las ecuaciones
de campo de la teor\'ia. Ya sea las ecuaciones de Maxwell para la EDC o las de Yang Mills en CDC. Similarmente en la TGR el invariante
principal es el escalar de Ricci $R$ y las ecuaciones de Einstein se obtienen extremando su acci\'on. Aunque este invariante es
indispensable para la formulaci\'on de toda la teor\'ia de la TGR, no constituye el \'unico invariante posible, existen muchos m\'as que se
pueden construir. Por ejemplo, usando el tensor de Ricci $R_{a b}$, y el tensor de Riemann $R_{a b c d}$, se obtienen los invariantes,
\begin{equation}
   R^2, \quad R_{a b}R^{a b}, \quad R_{a b c d}R^{a b c d}.
\end{equation}

Al incorporarlos a ellos se define el sector geom\'etrico (o gravitatorio) de la teor\'ia. Claramente esto mismo se puede hacer con la
acci\'on de la EDC y en el marco de las teor\'ias de campo se define como trabajar en el sector de calibraci\'on. Adem\'as cuando decimos
que trabajamos con invariantes de orden superior (o de orden superior de derivaci\'on) en el sector geom\'etrico, significa que dicho
invariante es de mayor orden que $R$, la interpretaci\'on es similar en el sector de calibrado para invariantes de \'ordenes 
mayores\footnote{As\'i mismo como $R^2=R_{a b}R^{a b}$, el t\'ermino $F^{2}=F_{a b}F^{a b}$, se define similarmente.} que $F^{2}$.
Hoy en d\'ia la incorporaci\'on de invariantes de orden superior es un tema abierto.

En este cap\'itulo calculamos la entrop\'ia de ANs extremos y estacionarios teniendo en cuenta t\'erminos de derivaci\'on superior. La
inclusi\'on de estos t\'erminos en la teor\'ia de alta gravedad puede hacerse usando por ejemplo los t\'erminos de Gauss-Bonnet(GB)
\cite{Morales:2006gm,Astefanesei2,Astefanesei3}. Estos t\'erminos pueden aparecer en varios escenarios te\'oricos como; Teor\'ias de Cuerdas,
Teor\'ias de Branas \cite{Maldacena:1996ky,Duff:1999rk,Peet:2000hn} y Gravedad Cu\'antica Semicl\'asica. En particular, t\'erminos de
derivaci\'on superior aparecen en la teor\'ia de cuerdas cuando se eval\'ua el l\'imite efectivo de baja energ\'ia.

Por otro lado, se conoce el problema de no-renormalizaci\'on que sufren las teor\'ias de gravedad. En especial la teor\'ia de GB s\'i puede ser
renormalizada pero a su vez aparecen otros problemas de igual importancia, emergen fantasmas (part\'iculas de $spin=2$ masivas) o
gravitones masivos \cite{Farhoudi,B.Zwiebach}.

En este cap\'itulo presentaremos una serie de soluciones aproximadas para la entrop\'ia de ANs extremales cerca del horizonte de geometr\'ia
para cuatro y cinco dimensiones. As\'i para alcanzar esta meta, usaremos el formalismo de Sen. La teor\'ia de partida ser\'a detallada en
el ep\'igrafe \ref{TEORIA_GENERALIZADA_CAP5}, pero en efecto ella est\'a compuesta por una teor\'ia de Einstein-Maxwell con constante
cosmol\'ogica, adem\'as del conjunto completo de los invariantes de Riemann. Es importante se\~nalar que en aquellos casos en que las
ecuaciones de movimiento (en la cercan\'ia del horizonte), no se puedan resolver anal\'iticamente, entonces efectuaremos una aproximaci\'on
por series de potencias bas\'andonos en un par\'ametro inverso a la carga el\'ectrica del AN.

Por otro lado, si queremos incluir el conjunto completo de los invariantes de Riemann como t\'erminos de derivaci\'on superior de la
teor\'ia, debemos considerar la versi\'on m\'as actual de este conjunto. O sea, el conjunto definido por Carminati y McLenagham (CM)
\cite{JCarminati}, m\'as el invariante $m_{6}$ introducido por Zakhary y McIntosh \cite{Zakhary_McIntosh}.

Los invariantes CM son escalares construidos a partir del tensor de Riemann, $R_{abcd}$, el tensor de Weyl $C_{abcd}$ 
(y su dual) y el tensor de Ricci de traza nula, definido como $S_{ab}=R_{ab}-(1/d) R\,g_{ab}$ (en $d$-dimensiones). Comenzando por seis
escalares reales $R,r_{1},r_{2},r_{3},m_{3},m_{4}$ y cinco escalares complejos $w_{1},w_{2},m_{1},m_{2},m_{5}$, ellos hacen un total de
diecis\'eis invariantes escalares (Note que los s\'imbolos $r,w$ y $m$ se asocian respectivamente a invariantes de Riemann, Weyl y
su mezcla). El conjunto de invariantes de CM contiene el n\'umero requerido de invariantes para los casos de Einstein-Maxwell y el fluido
perfecto. Adem\'as, con la inclusi\'on del invariante $m_{6}$, se ha probado que el conjunto de CM se convierte en un {\it conjunto
completo}. De esta forma este conjunto cubre los 90 casos posibles (6 tipo Petrov $\times$ 15 tipo Segre, ver \cite{Zakhary_McIntosh}).
En otras palabras, un conjunto completo de invariantes debe contener (al menos para el espacio-tiempo $d=4$) los bien conocidos
invariantes f\'isicos, as\'i como los invariantes geom\'etricos.

En el Cuadro \ref{Definicion_de_Invariantes}, (ver Ap\'endice \ref{Apend_Invariantes}) mostramos las definiciones y resultados obtenidos para el conjunto de invariantes no nulos de Riemann sobre una geometr\'ia $AdS_{2}\times S^{d-2}$ para $d=4$ y $d=5$. Es importante se\~nalar que para el caso de $d=5$ no es posible calcular los invariantes complejos de Riemann, ellos son desconocidos. Para esta dimensi\'on podr\'ia requerirse de m\'as invariantes y la pregunta ser\'ia, cuantos invariantes se necesitan para que estos conjuntos con $d>4$ sean {\it completos}. Este todav\'ia es un problema abierto.

Igualmente en el Cuadro \ref{Tabla_de_Invariantes}, (ver Ap\'endice \ref{Apend_Invariantes}), se muestran los invariantes organizados por su grado.
Siguiendo las definiciones dadas en \cite{Zakhary_McIntosh}, de un conjunto dado de invariantes, el $j-$\'esimo invariante escrito de la forma $I^{p}_{j}$, es llamado invariante de $orden$ $p$. Si existe otro invariante tal que pueda ser escrito de la forma $I^{p}_{j}I^{q}_{k}I^{r}_{l}$, entonces se dice que este es un invariante de $orden\,\,p-q-r$ y el resultado de la suma $p+q+r$ ser\'a su $grado$. Por ejemplo, en el Cuadro \ref{Definicion_de_Invariantes}, el invariante $\mathfrak{Re}(m_{1})$, denota la parte real del invariante $m_{1}$, tiene $orden\,\,1-2$ y es de $1+2=3$, tercer-$ grado$. Adem\'as, el invariante $I^{p}_{j}$ ser\'a $independiente$ si \'el no puede ser escrito en t\'erminos de otro invariante, ya sea de igual o menor grado. Tambi\'en, dos invariantes se dice que son $equivalentes$ si ellos pueden ser escritos en t\'erminos de otros, o como el producto de otros de menor grado. Las relaciones de $equivalencias$ o (syzygies) son dadas en el Cuadro \ref{Tabla_de_Invariantes}. Es f\'acil de ver que en 4 y 5 dimensiones todos los invariantes pueden ser escritos en t\'erminos de los invariantes $independientes$ $R,\,r_{1}$ y $r_{2}$.
Es importante se\~nalar que todos los invariantes no nulos fueron tomados en cuenta en nuestros c\'alculos, los restantes nulos no son incluidos en los Cuadros \ref{Definicion_de_Invariantes}-\ref{Tabla_de_Invariantes}.

Naturalmente, una pregunta sale a flote, y es: \textquestiondown por qu\'e no est\'an presentes los invariantes de orden superior de la
teor\'ia de calibraci\'on?, dado que en efecto pensamos incluir la teor\'ia de Einstein-Maxwell donde los campos de calibraci\'on est\'an
presentes. La respuesta es, que lo hacemos por simplicidad, o sea solo trabajaremos en el sector de gravedad pura, para los t\'erminos de
orden superior de la teor\'ia, quedando fuera invariantes de orden superior como: 
$(F_{ab}F^{ab})^2,\,F^{a}\,_{b}\,F^{b}\,_{c}F^{c}\,_{d}F^{d}\,_{a},\,R\,F^2$,..., los covariantes $F_{a b}\Box F^{a b}$,..., y los
invariantes de forma. Es decir solamente el sector gravitacional ser\'a tratado para el conjunto de CM.

Todos los c\'alculos fueron hechos con el paquete para tensores $GRTensor$ sobre el programa de c\'omputo algebraico Maple.
\subsection{Generalizaci\'on de la Teor\'ia de Einstein usando los invariantes de Riemann}
\label{TEORIA_GENERALIZADA_CAP5}

Consideraremos una teor\'ia de gravedad de orden superior en el sector geom\'etrico, introduciendo el conjunto completo de los invariantes de Riemann como t\'erminos de derivaci\'on superior en la acci\'on. Adem\'as incluimos constante cosmol\'ogica, y campos electromagn\'eticos. As\'i nuestra acci\'on toma la forma, 
\begin{equation}      {\mathcal{S}}=\frac{1}{16\pi
      G_{d}}\int{dx^{d}\sqrt{-g}\left(R+\Lambda-\frac{F^{2}}{4}+{\mathcal{L}}^{d}_{inv}\right)},  \label{AccionGeneralizada}\end{equation}
donde,
   \begin{eqnarray}
      {\mathcal{L}}^{d=4}_{inv}&=&a_{2}R^2+b_{2}R_{2}+a_{3}R^{3}+b_{3}R\,R_{2}+a_{4}R^4+b_{4}{R^{2}_{2}}+c_{4}{R^2}\,R_{2}+a_{5}R^{5}+
      b_{5}{R^{3}}\,R_{2}\,, \nonumber \\
      {\mathcal{L}}^{d=5}_{inv}&=&a_{2}R^2+b_{2}R_{2}+a_{3}R^{3}+b_{3}{R}\,R_{2}+c_{3}R_{3}+a_{4}R^4+b_{4}{R^{2}_{2}}+c_{4}{R^2}\,{R_{2}}+
      e_{4}{R}\,{R_{3}}\,,\nonumber
   \end{eqnarray}
siendo $G_{d}$ la constante de Newton $d$-dimensional, $R$ el escalar de Ricci, $\Lambda$ la constante cosmol\'ogica, $F_{a b}$ el tensor electromagn\'etico, y $F^{2}=F_{a b}F^{a b}$. $R_{2}$ y $R_{3}$ son los dos primeros invariantes de Riemann reales definidos en el Cuadro \ref{Definicion_de_Invariantes}, para la m\'etrica (\ref{Metrica_Extremal_en_4Dand5D}). Los par\'ametros $a_{i},b_{i},c_{i},e_{i}...$ son las constantes de acoplamiento para cada t\'ermino de derivaci\'on superior de $i$-\'esimo grado. ${\mathcal{L}}^{d}_{inv}$ denota los t\'erminos de derivaci\'on superior que se tendr\'an en cuenta. En ambos casos, en cuatro y cinco dimensiones, los invariantes forman un conjunto completo, y se tiene en cuenta todos los invariantes de alto orden dentro del conjunto de invariantes de Carminati-McLenagham.

El espacio tiempo m\'as general \cite{Sen:2007qy} que se puede escribir para una configuraci\'on de AN est\'atico y extremal con topolog\'ia $AdS_{2}\times S^{d-2}$ cerca del horizonte de geometr\'ia es, 
 \begin{eqnarray}
      ds^{2}&=&v_{1}(-r^2dt^{2}+\frac{dr^{2}}{r^2})+v_{2}d{\Omega_{d-2}}, \label{Metrica_Extremal_en_4Dand5D}\\
        e^{a_{I}\Psi_{I}}|_{H}&=&u_{I}, \, \, \, {F}^{d}_{0r}=e, \, \, \,
          {F}^{d=4}_{\theta\phi}=p \sin\theta,\,\,{F}^{d=5}_{\theta\phi}=0, \\
        d\Omega^{2}_{d-2}&=& d\theta^{2}_{1}+\sum^{d-2}_{i=2}\prod^{i-1}_{j=1}\sin^{2}\theta_{j} d\theta^{2}_{i}, \,\, 0\leq
      \theta_{i} \leq \pi, \, 0\leq \theta_{d-2} \leq 2\pi, \, (1\leq i\leq d-3), \nonumber
    \end{eqnarray}
donde $e$ y $p$ son funciones relacionadas con la carga el\'ectrica y magn\'etica, mientras que $v_{1}$ y $v_{2}$ son funciones conectadas con la garganta del AN. En general las constantes $u_{i}$ se relacionan con los valores de los campos escalares $\Psi_{i}$  sobre el horizonte, pero en nuestro caso ellos no est\'an presentes por tanto $u_{i}=0$. En los ep\'igrafes siguientes usaremos el formalismo de Sen~\cite{Sen:2007qy}, en el cual la funci\'on entr\'opica es definida como,
\begin{equation}
    \mathcal{E}(\vec{u},\vec{v},\vec{e},\vec{q},\vec{p})=2\pi(e_{i}q_{i}-f(\vec{u},\vec{v},\vec{e},\vec{p})),
\end{equation}
y $f(\vec{u},\vec{v},\vec{e},\vec{p})$ es la densidad Lagrangiana $\sqrt{{-det}\,{g}}{\mathcal{L}}$, evaluada cerca del horizonte de geometr\'ia. Todos estos par\'ametros pueden ser determinados extremando la funci\'on entr\'opica.
\begin{equation}
 \frac{\partial{\mathcal{E}}}{\partial{u_{i}}}=\frac{\partial{\mathcal{E}}}{\partial{v_{j}}}=
   \frac{\partial{\mathcal{E}}}{\partial{e}}=0,  \qquad {i=1,..;N}, {j=1,..,2}. \label{R4_4}
\end{equation}
Las ecuaciones (\ref{R4_4}) representan el conjunto de ecuaciones de movimiento del sistema escritas cerca del horizonte de geometr\'ia del fondo extremal (\ref{Metrica_Extremal_en_4Dand5D}). 
As\'i, la entrop\'ia del AN en el l\'imite extremal se obtiene resolviendo el sistema de ecuaciones (\ref{R4_4}) y sustituyendo estos par\'ametros en la funci\'on entr\'opica. 
Esto demuestra que $S_{BH}/2\pi$ puede ser considerada como la transformada de Legendre de la funci\'on $f(\overrightarrow{u},\overrightarrow{v},\overrightarrow{e},\overrightarrow{p})$, con respecto a las variables $e_{i}$. 
%
%
\subsubsection{Soluciones anal\'iticas para invariantes de $2^{do}$ orden en $d=4$}
Los c\'alculos para contribuciones de segundo orden a la entrop\'ia de ANs extremales corresponden a considerar solamente los invariantes de segundo orden. Sobre la base del principio de correspondencia, la soluci\'on de Gauss-Bonnet(GB) debe estar contenida en estos resultados \cite{Morales:2006gm}, as\'i como la soluci\'on de Reissner-Nordstrom (RN). Por tanto, en este caso tomamos $a_{i}=b_{i}=c_{i}=0$ para $i\geq 3$ en
(\ref{AccionGeneralizada}), entonces la funci\'on 
$f(\vec{v},\vec{e},\vec{p})$ ser\'a,
  \begin{equation}
     f(v_{1},v_{2},e,p)={\int_{S^{2}}}{\sqrt{-g}\left(R+\Lambda-\frac{F^{2}}{4}+a_{2}R^2+b_{2}R_{2} \right)d\theta
     d\phi},   \label{eRN}
  \end{equation}
mientras la funci\'on entr\'opica tomar\'a la forma,
\begin{eqnarray}
  \mathcal{E}(v_{1},v_{2},e,q,p)&=& \Biggl\{ 2\,q{e}G_{4}+
    2\left( 2-\,{\frac {{v_{1}}}{{v_{2}}}}-\,{\frac {v_{2}}{v_{1}}} \right){a_{2}}+ \left( -{\frac {v_{1}}{8v_{2}}}-{\frac{{v_{2}}}{{8v_{1}}}}
      -\frac{1}{4}\right) {b_{2}}+\nonumber \\
      &+& \left( {\frac {{{p}}^{2}}{2v_{2}}}-2-{v_{2}}\,\Lambda \right)\frac{v_{1}}{2}-{\frac {{v_{2}}\,{{e} }^{2}}
    {4v_{1}}}+{v_{2}} \Biggr\}\frac{\pi}{G_{4}}\,,
\end{eqnarray} 
y el sistema de ecuaciones de movimiento bas\'andonos en (\ref{R4_4}) ser\'ia,
 \begin{eqnarray}
    2\,q-\,\frac{v_{2}\,e}{2v_{1}G_{4}}=0,\\
        \,\frac{v_{1}\Lambda}{2}-1+2\,{\frac {a_{2}}{v_{1}}}-2\,{\frac {{
         v_{1}}\,{a_{2}}}{{{v_{2}}}^{2}}}+{\frac {{b_{2}}}{8v_{1}}}-{\frac {{v_{1}}\,{b_{2}}}{8{v_{2}}^{2}}}+{\frac
            {{{e}}^{2}}{4v_{1}}}+{\frac {{v_{1}}\,{p}^{2}}{4{v_{2}} ^{2}}}=0,\\
            -{v_{2}}\,\frac{\Lambda}{2}-1+2\,{\frac {{v_{2}}\,{a_{2}}}{{{v_{1}}}^{2
           }}}-2\,{\frac {{a_{2}}}{{v_{2}}}}+{\frac {{v_{2}}\,{b_{2}}}{{ {8v_{1}}}^{2}}}-{\frac {{b_{2}}}{{8v_{2}}}}+{\frac
       {{v_{2}}\,{{e}}^{2}}{{{4v_{1}}}^{2}}}+{\frac
       {{p}^{2}}{4v_{2}}}=0.
  \end{eqnarray}
Note que cuando resolvemos este sistema de ecuaciones, todas estas soluciones pueden ser escritas en t\'erminos de la funci\'on $v_{2}$ como sigue,
\begin{eqnarray}
 v_{1}&=&{\frac {{v_{2}}}{{v_{2}}\,\Lambda+1}}, \qquad
   q ={\frac{f}{{8G_{4}}}}, \qquad
   {e}={\frac{f}{2({v_{2}}\,\Lambda+1)}},\hspace{0.5cm} \label{Sol_2_Orden_d4} \\
 f&=&
 \sqrt{8\,(\Lambda{v_{2}}+2)v_{2}-2(16\,{a_{2}}+b_{2})\,({v_{2}}\Lambda+2)v_{2}\Lambda-4\,{{p}}^{2}}\,, \nonumber
\end{eqnarray}
y la entrop\'ia para un AN extremal y est\'atico (teniendo en cuenta invariantes de segundo orden) ser\'a, 
\begin{equation}
   {S}_{BH}=\left( 1-{\frac { \left( 16\,{a_{2}}+{b_{2}} \right)
     \Lambda}{{4}}} \right)\,\frac{\pi v_{2}}{G_{4}}-{\frac {\pi
   \,{b_{2}}}{{2G_{4}}}}.
   \label{SBH_Second_Order_d4}
\end{equation}

La soluci\'on de GB que fue encontrada por Morales y Samtleben en \cite{Morales:2006gm} puede ser obtenida sustituyendo $b_{2}=-8\alpha$, y $a_{2}=\alpha/2$ en (\ref{SBH_Second_Order_d4}), as\'i,
\begin{equation}
   S_{GB}=(v_{2}+4\alpha)\frac{\pi}{G_{4}}, \label{GB_4D}
 \end{equation} 
donde el par\'ametro $\alpha$ es la constante de acoplamiento de GB. De igual manera, si tomamos  $a_{2}=b_{2}=0$ en (\ref{SBH_Second_Order_d4}), entonces la entrop\'ia se convierte en la conocida entrop\'ia del AN de RN en el l\'imite extremal y con carga magn\'etica $p$,
\begin{equation}
   S_{BH}|_{a_{2}=b_{2}=0}=\frac{\pi v_{2}}{G_{4}}\equiv S^{d=4}_{RN}.
   \label{RN_entropy}
\end{equation}

Las variables $v_{1},e$ y $q$ ser\'an entonces,
\begin{equation}
   v_{1}={\frac {{v_{2}}}{{v_{2}}\,\Lambda+1}},\qquad
    q={\frac{\sqrt{2\,(\Lambda{v_{2}}+2)v_{2}-\,{{p}}^{2}}}{{4G_{4}}}},\qquad
   {e}={\frac{\sqrt{2\,(\Lambda{v_{2}}+2)v_{2}-\,{{p}}^{2}}}{{v_{2}}\,\Lambda+1}},\label{Sol_RN_Lambda}
\end{equation} donde directamente de (\ref{SBH_Second_Order_d4}) se puede ver que la constante cosmol\'ogica por s\'i misma no modifica la ley del \'area de Bekestein-Hawking. Por tanto ella necesita estar acompa\~nada de t\'erminos de derivaci\'on superior, de al menos segundo orden, para poder modificarla y obtener un resultado nuevo. O sea la constante cosmol\'ogica $\Lambda$ solo cambia la geometr\'ia de la garganta como se ve en (\ref{Sol_2_Orden_d4}) y (\ref{Sol_RN_Lambda}). Sin embargo, las constantes $a_{2}$ y $b_{2}$ en (\ref{SBH_Second_Order_d4}) representan las desviaciones de dicha ley. En la siguiente secci\'on se obtendr\'an soluciones aproximadas para gravedad de orden superior, teniendo en cuenta las contribuciones restantes $R^{3},R^{4}$ y $R^5$. 

\subsubsection{Soluciones aproximadas para el conjunto completo en $d=4$}\label{Section-4}

 Debido a la no linealidad de las ecuaciones de Einstein-Maxwell es muy dif\'icil encontrar soluciones exactas, m\'as a\'un si incluimos t\'erminos de alta derivaci\'on. En muchos casos se emplean m\'etodos aproximados o soluciones num\'ericas.
 Por tanto, si consideramos el conjunto completo de invariantes de Riemann entonces no podremos encontrar soluciones expl\'icitas del sistema de ecuaciones de movimiento (\ref{R4_4}). Nosotros resolveremos este problema introduciendo un par\'ametro $w$ para hacer una expansi\'on alrededor de \'el. En efecto en ({\ref{AccionGeneralizada}}), $w$ puede siempre ser extra\'ido de las constantes de acoplamiento $a_{i},b_{i},c_{i},e_{i}$ por re-escalamiento, teniendo en cuenta que debe ser extra\'ido con el orden apropiado. La soluci\'on de RN (caso $\mathcal{L}_{inv}=0$) es bien conocida anal\'iticamente, y las primeras constantes de acoplamiento para invariantes de orden superior son $a_{2}$ y $b_{2}$. Consecuentemente para este caso: $a_{2},b_{2} \Rightarrow w\,a_{2},w\,b_{2},...$ y en general, tendremos que,
\begin{equation}
a_{l},b_{m},c_{n},e_{p}\Longrightarrow
w^{l-1}{a}_{l},w^{m-1}{b_{m}},w^{n-1}{c_{n}},w^{p-1}{e_{p}}, \qquad
l,m,n,p=2,3,4,5,...
\end{equation}
El par\'ametro de expansi\'on $w$ puede ser considerado proporcional al inverso de la carga del AN, la cual es cero a primer orden. Esta es la forma en que $w$ debe aparecer en la funci\'on entr\'opica $\mathcal{E}$. Adem\'as, como estamos interesados en soluciones aproximadas, entonces debemos construir series de expansi\'on de las funciones $e,q$ y $v_{1}$. En general, cualquiera de estas funciones puede ser expandida alrededor de $w$, 
\begin{equation}
h(w)=\sum_{k=0}^{\infty}\frac{w^k}{k!} \left(\frac{\partial^{k}
h}{\partial w^k}\right)_{w=0}.\end{equation}
Esta expansi\'on nos permite escribir,
\begin{eqnarray}
  &e& \simeq e_{0}+w\Delta e_{1}+w^2\Delta e_{2}+w^3\Delta
  e_{3}+...,\nonumber  \\
    &q& \simeq q_{0}+w\Delta q_{1}+w^2 \Delta q_{2}+w^3 \Delta
    q_{3}+..., \nonumber \\
    &v_{1}& \simeq v_{1\,0}+w\Delta v_{1\,1}+w^2\Delta v_{1\,2}+w^3 \Delta v_{1\,3}+...,   \label{Approach} \\
  &\mathcal{E}& \simeq \mathcal{E}_{0}+w \Delta \mathcal{E}_{1}+w^2 \Delta \mathcal{E}_{2}+w^3 \Delta
  \mathcal{E}_{3}+..., \nonumber
\end{eqnarray}
donde en (\ref{Approach}) hemos tomado $\Delta h_{k}=(1/k!)({\partial^{k} h/\partial w^{k}})_{_ {\tiny w=0}}$.
As\'i, $w$ fija el nivel de aproximaci\'on, en el caso que $w=0$ tenemos soluciones de orden cero que ser\'an de tipo RN
(\ref{RN_entropy})-(\ref{Sol_RN_Lambda}). Finalmente incluyendo el conjunto completo de invariantes (o sea en (\ref{AccionGeneralizada})
tomando todos los t\'erminos), o contribuciones de quinto orden para la entrop\'ia de un AN extremo, y despu\'es de resolver el conjunto
de ecuaciones iteradas que aparecen al extremar la funci\'on entr\'opica, en este trabajo se obtienen la siguientes expresiones aproximadas
para la entrop\'ia $S_{BH}$,
\begin{equation}
  S_{BH}=
  S^{d=4}_{RN}+\left(S^{(2)}_{BH}+S^{(3)}_{BH}w+S^{(4)}_{BH}w^2+S^{(5)}_{BH}w^3
  \right){\it w}+{\it O(w^5)}, \hspace{1.5cm}
\label{SBH_All_Order_d4}
\end{equation}
donde,
\begin{equation}
    S^{(2)}_{BH} = -\left({\left( 16\,{a_{2}}+{b_{2}}
    \right)\frac{\Lambda}{4}}{v_{2}}+{\frac {{b_{2}}}{{2}}}\right)\frac{\pi}{G_{4}}, \label{SHB_All_Order_d4(2nd_Degree)} 
\end{equation}
\begin{equation}
     S^{(3)}_{BH}= \left(12\Lambda^{2}{v_{2}}\,a_{3}+(\frac{3}{4}v_{2}\Lambda^{2}+2\Lambda+
     \frac{1}{v_{2}}){b_{3}} \right)\frac{\pi}{G_{4}},\label{Approach_3nd_4D}\\
\end{equation}
 
\begin{eqnarray}
       S^{(4)}_{BH}&=& -\Biggl\{{32\,{v_{2}}\,{\Lambda}^{3}}\,a_{4}+{\frac { \left(
       {v_{2}}\,\Lambda+2 \right) ^{3}\,{b_{4}}}{8{v_{2}}^{2}}}+ {\frac {2\Lambda\, \left( {v_{2}}\,\Lambda+2
       \right)  \left( {v_{2}}\,\Lambda+1 \right) \,{c_{4}}}{{v_{2}}}}+(16a_{2}+b_{2})\times \nonumber \\
       &\times& \left(\frac{\Lambda\,(v_{2}\Lambda+2)^2\,b_{3}}{16v_{2}}+v_{2}a_{3}\Lambda^3 \right)  \Biggl\} \frac{\pi}{G_{4}},\label{Approach_4nd_4D}
\end{eqnarray}
\begin{eqnarray}
       {S}^{(5)}_{BH}&=& \Biggl\{80 v_{2}\Lambda^{2}{a_{5}}+(5v_{2}\Lambda^{2}+16\Lambda+
  \frac{12}{v_{2}}){b_{5}}+\,96v_{2}\Lambda^2a_{3}^2+4(\frac{6}{v_{2}}+3v_{2}\Lambda^{2}+8\Lambda)a_{3}b_{3}+\\ \nonumber
  &+&\frac{(3v_{2}\Lambda+4)(v_{2}\Lambda+2)^2}{8\Lambda v_{2}^2}b_{3}^2+(16a_{2}+b_{2})\left(4\Lambda^2a_{4}v_{2}+\frac{(v_{2}\Lambda+2)^2c_{4}}{4v_{2}}+
  \frac{(v_{2}\Lambda+2)^4b_{4}}{64\Lambda^2v_{2}^3}\right)\Biggr\}\frac{\pi\,\Lambda^2}{G_{4}}.
\label{Approach_5nd_4D} 
\end{eqnarray}

En caso que estudiemos una teor\'ia efectiva donde solo invariantes de tercer orden son necesarios, entonces debemos tomar en (\ref{AccionGeneralizada}), (\ref{SBH_All_Order_d4}) las constantes de acoplamiento con los valores $a_{j}=b_{j}=c_{j}=e_{j}=0$ para $j\geq 4$, y la soluci\'on para la entrop\'ia del AN extremo, ser\'a simplemente:
$S_{BH}=S_{RN}+(S^{(2)}_{BH}+S^{(3)}_{BH}w){\it w+O({w^3})}$. Esto mismo se puede hacer para cuarto orden.
Esta aproximaci\'on es m\'as que suficiente, pues se puede ver f\'acilmente que la soluci\'on anal\'itica de segundo orden (\ref{SBH_Second_Order_d4}) es exactamente reproducida en (\ref{SHB_All_Order_d4(2nd_Degree)}).
 Note que cada contribuci\'on entr\'opica de invariantes de $i-$\'esimo grado es etiquetada como $S^{(i)}$,
entonces el s\'uper-\'indice $(i)$ solamente indica el grado del invariante que produce esto y no el orden de aproximaci\'on.

  En el conjunto de soluciones aproximadas los t\'erminos no lineales con respecto a la constante de acoplamiento aparecen en $S^{(4)}_{BH}$ y $S^{(5)}_{BH}$ con contribuciones como $a_{2}a_{3},a_{2}b_{3},b_{2}a_{3},b_{2}b_{3}$ y $ a^{2}_{3},a_{3}b_{3},b^{2}_{3},a_{2}a_{4},a_{2}c_{4},
a_{2}b_{4},b_{2}a_{4},b_{2}c_{4},b_{2}b_{4}$ respectivamente.

\subsubsection{Soluciones anal\'iticas para invariantes de $2^{do}$ orden en $d=5$} 
En este caso, la funci\'on $f(\vec{v},\vec{e},\vec{p})$ toma la misma forma que (\ref{eRN}), pero la integraci\'on esta vez se realiza sobre $S^{3}$, entonces la funci\'on entr\'opica es dada por,
\begin{eqnarray}
  \mathcal{E}(v_{1},v_{2},e,q,p)&=&\,2\,\pi \,qe-{\frac {{\pi }^{2}
    \left(3\,v_{1}-v_{2} \right) ^{2}a_{2}}{ G_{5}\,v_{1}\,\sqrt
     {v_{2}}}}-{\frac3{40}}\,{\frac {{\pi } ^{2} \left(
      2\,v_{1}+v_{2}\right) ^{2}b_{2}}{G_{5}\,v_{1}\,\sqrt
    {v_{2}}}}-,\,\nonumber \\
  &-&\,{\frac {\sqrt {v_{2}} \left(2v_{1}v_{2}(\Lambda
  v_{1}-2)+{e}^{2}v_{2}+12\,{ {v_{1}}}^{2} \right) {\pi
 }^{2}}{{8G_{5}}\,{v_{1}}}}, \label{Entropy_Function5D}
\end{eqnarray}
y el sistema de ecuaciones de movimiento (\ref{R4_4}) cerca del horizonte ser\'an,
%
  \begin{eqnarray}
      0&=&{\frac {\pi \left( 3{v_{1}}+v_{2} \right) \left( 3{v_{1}}-{v_{2}} \right)  {a_{2}}}{\sqrt
       {v_{2}}{{v_{1}}}^{2}}}+{\frac {3\pi\left( 2\,{v_{1}}-{v_{2}} \right)
         \left( 2\, {v_{1}}+{v_{2}} \right){b_{2}}}{40\sqrt{v_{2}}{{v_{1}}}^{2}}}-\\
       &-&{\frac {\pi \,\sqrt{{v_{2}}}\left(-12\,{{v_{1}}}^{2}-2\,\Lambda\,{{v_{1}}}^{2}{v_{2}}+{e}^{2}{v_{2}}\right)}
            {8{v_{1}}^2}},\nonumber \\
              0&=&{\frac{3\pi \left( {v_{1}}+{v_{2}} \right)  \left(3{v_{1}}
              -{v_{2}} \right) {a_{2}}}{\sqrt {{v_{2}}}{{v_{1}}}^{2}}}+{\frac
            {3\pi\left( 2\,{v_{1}}+{v_{2}} \right) \left( 2\, {v_{1}}-3\,{v_{2}}
           \right) {b_{2}}}{40\sqrt {v_{2}}{v_{1}}^{2}}}-\\
&-&{\frac {3\pi \,\sqrt{{v_{2}}} \left(2{v}^{2}_{1}(2+\Lambda v_{2})+v_{2}(e^2-4v_{1})
       \right) }{{{8v_{1}}}^{2}}},\nonumber \\
      0&=&q-\,{\frac {\pi \,{{v_{2}}}^{\frac32}e}{{v_{1}}\,{8G_{5}}}}.
  \end{eqnarray}
El sistema puede ser resuelto expl\'icitamente, y entonces las funciones $e,v_{1}$ y $q$ pueden ser escritas en t\'erminos de $v_{2}$ como,
%
 \begin{eqnarray}
  {v_{1}}&=&{\frac{{v_{2}}\left(5\,{v_{2}}+20{a_{2}}-{b_{2}}
      \right) }{60{a_{2}}+2\,{b_{2}}+5v_{2}(4+v_{2}\Lambda)}}, \label{v1_2ndOrder_5D} \\
      q&=&{\frac {\pi\,\sqrt{v_{2}}\,\widetilde{f}}{8\left( 5{v_{2}}+20{a_{2}}-{b_{2}} \right){G_{5}}}},\,\,\,
  e\, = \,\frac {\,\widetilde{f}}{20{v_{2}}+60{a_{2}}+2{b_{2}}+5{v}^{2}_{2}\Lambda},
\end{eqnarray}
donde,
$$
      \widetilde{f} = \sqrt {5}\,{\left( {{v_{2}}}^{2}\Lambda+20{a_{2}}+6{v_{2}}
    \right)}^{\frac{1}{2}}{\left(10{v}^{2}_{2}-60a_{2}b_{2}-40a_{2}{v}^{2}_{2}\Lambda-2{b}^{2}_{2}-3b_{2}{v}^{2}_{2}\Lambda-10b_{2}v_{2}
  \right)}^{\frac{1}{2}}. \nonumber
$$
Por consiguiente, la entrop\'ia de un AN est\'atico y extremo en $d=5$, teniendo en cuenta t\'erminos de derivaci\'on superior de segundo grado, tiene la forma,
\begin{equation}
  {S}_{BH}={\frac {{\pi }^{2}{{v_{2}}}^{\frac{5}{2}} \left( 40{a_{2}}+3{b_{2} } \right)
   \Lambda}{ 4\left( -5{v_{2}}-20\,{a_{2}}+{b_{2}}\right) {G_{5}}}}+{\frac{5{\pi}^{2}
   {{v^{\frac{1}{2}}_{2}}}\left(-{{v_{2}}}^{2}+10{a_{2}}{b_{2}}+2{b_{2}}{v_{2}}\right)}{2\left(-5
   {v_{2}}-20{a_{2}}+{b_{2}} \right)
   {G_{5}}}}.\label{SBH_Second_Order_d5}
\end{equation}
De forma an\'aloga, la soluci\'on extremal de Gauss-Bonnet mostrada en \cite{Morales:2006gm} puede ser obtenida con las sustituciones, ${b_{2}}=-8\,{\it \alpha}$, y $a_{2}=\frac{3}{5}{\it \alpha}$, en la soluci\'on principal (\ref{SBH_Second_Order_d5}), as\'i,
\begin{equation}
   S_{GB}=(v_{2}+12\alpha)\frac{\pi^2{v_{2}}^{\frac{1}{2}}}{2G_{5}}.\label{GB_5D}
\end{equation}
Esto no pasa para una contribuci\'on gen\'erica tipo GB. En general, el establecimiento de una relaci\'on entre las soluciones de GB y las soluciones de segundo grado requieren de un v\'inculo (ver secci\'on \ref{GBsection}). Por otro lado, si $a_{2}=b_{2}=0$ en (\ref{SBH_Second_Order_d5}), la conocida soluci\'on de RN extremal tambi\'en es obtenida,
\begin{equation}
{S}_{BH}|_{a_{2}=b_{2}=0}=\frac{\pi^2{v}^{\frac{3}{2}}_{2}}{2G_{5}}\equiv
S^{d=5}_{RN}.
\end{equation}
As\'i como acurri\'o en (\ref{SBH_Second_Order_d4}), la constante cosmol\'ogica no cambia la entrop\'ia del AN por si misma: Ella podr\'ia
necesitar de mayores t\'erminos de derivaci\'on superior. Una prueba de esto sigue de tomar $\Lambda=0$ en (\ref{SBH_Second_Order_d5}),
donde se ve claramente que las contribuciones de $a_{2}$ y $b_{2}$ no desaparecen. Estas contribuciones tambi\'en producen una marcada
desviaci\'on de la ley del \'area. Note que, contrariamente al caso de $d=4$, no solo la constante cosmol\'ogica modifica la geometr\'ia de
la garganta del AN (ver (\ref{v1_2ndOrder_5D})), sino que las constantes de acoplamiento asociadas a los invariantes de segundo orden,
tambi\'en tienen su efecto sobre la topolog\'ia de la garganta.

\subsubsection{Soluciones aproximadas para el conjunto completo en $d=5$}

 Como hicimos en la secci\'on \ref{Section-4}, podemos construir aproximadamente soluciones para cinco dimensiones. As\'i, considerando un AN extremo en el cual el conjunto completo de invariantes de Riemann (o sea en (\ref{AccionGeneralizada}) se deben tomar las constantes de acoplamiento $a_{j},b_{j},c_{j},e_{j}\neq 0$) a diferentes niveles de aproximaci\'on tiene la siguiente forma,
\begin{equation}
S_{BH}={S}^{d=5}_{RN}+(S^{(2)}_{BH}+S^{(3)}_{BH}w+S^{(4)}_{BH}w^2){\it
w}+{\it O(w^4)}, \label{SBH_All_Order_d4}
\end{equation}
donde, 
\begin{equation}
S^{(2)}_{BH}= -\left( 2\,{{\left(1+\Lambda\,{v_{2}} \right)
{a_{2}}}}+{\frac{3}{20}}\,{{\left( 6+\Lambda\,{v_{2}} \right)
{b_{2}}}} \right)\frac{\pi^2{v^{\frac{1}{2}}_{2}}}{G_{5}}\,\,,  \label{Approach_2nd_5D} 
\end{equation}
\begin{eqnarray}
{S}^{(3)}_{BH}&=& \Biggl\{6\,{ { \left( 1+\Lambda\,{v_{2}} \right)
^{2}{a_{3}} }}+{\frac {3}{20}}\,{{\left( 6+\Lambda\,{v_{2}} \right)
\left( 3\,\Lambda\,{v_{2}}+8 \right) {b_{3}}}}-{\frac
{9}{400}}\,{{\left( 6+\Lambda\,{v_{2}} \right) ^{2}{c_{3}}}}+ \label{Approach_3nd_5D}\\
\nonumber &+& 8(1+\Lambda
v_{2})a_{2}^2+(-9+v_{2}\Lambda)\frac{a_{2}b_{2}}{5}-\frac{3}{100}(v_{2}\Lambda+6)b_{2}^2
\Biggr\}\frac{\pi^2}{{v^{\frac{1}{2}}_{2}}G_{5}}\,\,,  
\end{eqnarray}
\begin{eqnarray}
{S}^{(4)}_{BH}&=& -\Biggl\{ {{{16}\left( 1+\Lambda\,{v_{2}}
\right)^{3}{a_{4}}}}+{{\frac9{100}\left( 6+\Lambda\,{v_{2}}
\right)^{3}{b_{4}}}}+{{\frac35\left( 6+\Lambda\,{v_{2}} \right)
\left( 2\,\Lambda\,{v_{2}}+7 \right)\left( 1+\Lambda\,{v_{2}}
\right){c_{4}}}}- \nonumber \\
&-&{{\frac3{200}\left( 4\,\Lambda\,{v_{2}}+9 \right)\left( 6+
\Lambda {v_{2}} \right)
^{2}{e_{4}}}}-32(1+v_{2}\Lambda)a_{2}^3+\frac{4}{5}(11+v_{2}\Lambda)b_{2}a_{2}^2+\label{Approach_4nd_5D}\\
\nonumber
&+&(9+4v_{2}\Lambda)\frac{a_{2}b_{2}^2}{25}-8(10+v_{2}\Lambda)(1+v_{2}\Lambda)^2a_{2}a_{3}
-(270+248v_{2}\Lambda+3v_{2}^3\Lambda^3+56v_{2}^2\Lambda^2)\frac{a_{2}b_{3}}{5}+\\
\nonumber
&+&\frac{3}{100}(v_{2}\Lambda+6)(v_{2}^2\Lambda^2+16v_{2}\Lambda+30)a_{2}c_{3}-\frac{3}{500}(v_{2}\Lambda+6)b_{2}^3-
\frac{3}{5}(1+v_{2}\Lambda)(v_{2}\Lambda+5)\times \\ \nonumber
&\times&(v_{2}\Lambda-4)a_{3}b_{2}-\frac{3}{200}(v_{2}\Lambda+6)(3v_{2}^2\Lambda^2+8v_{2}\Lambda-20)b_{2}b_{3}
+\frac{9 \Lambda v_{2}}{4000}(v_{2}\Lambda+6)^2 b_{2}c_{3}
\Biggr\}{\frac{\pi^2\,}{G_{5}{v_{2}}^{\frac32}}}\,.  
\end{eqnarray}

 Es correcto decir que si la ecuaci\'on (\ref{SBH_Second_Order_d5}) se expande en series de segundo orden de $v^{-1}_{2}$, entonces logramos reproducir la ecuaci\'on (\ref{Approach_2nd_5D}). Si nosotros tomamos $a_{2}=\frac{3}{5}\alpha$ y $b_{2}=-8\alpha$ en (\ref{Approach_2nd_5D}) se obtiene la soluci\'on de GB (\ref{GB_5D}). En este caso, las contribuciones no-lineales de las constantes de acoplamiento aparecen en $S^{(3)}_{BH}$ y $S^{(4)}_{BH}$ como combinaciones de las constantes de acoplamiento de peque\~no orden como $a^{2}_{2}, a_{2}b_{2},b^{2}_{2},$ y $a^{3}_{2},$...etc. Note que la ausencia de t\'erminos $a^{2}_{2},a_{2}b_{2}$ y $b^{2}_{2}$ en (\ref{Approach_3nd_4D}) en contraste con (\ref{Approach_3nd_5D}) es simplemente un resultado del c\'alculo. Lo mismo pasa en (\ref{Approach_4nd_4D}) y (\ref{Approach_4nd_5D}) con los t\'erminos $a^{3}_{2},b_{2}a^{2}_{2},a_{2}b^{2}_{2},b^{3}_{2}$.

\subsubsection{Invariantes de Riemann y gravedad gen\'erica de Gauss-Bonnet}\label{GBsection}

En esta secci\'on examinaremos la teor\'ia de Einstein-Maxwell con una modificaci\'on a la cual llamaremos Gauss Bonnet gen\'erica (o sea
con tres constantes de acoplamiento diferentes). Esto es algo que debe hacerse con cuidado pues se sabe que en general una teor\'ia as\'i
no es renormalizable, pero en nuestro caso solo lo hacemos a modo de generalizar nuestro resultado. Adem\'as de todas formas se sabe que
las teor\'ias de Gauss-Bonnet, aunque superan el problema de la renormalizaci\'on, tienen problemas con la existencias de part\'iculas
fantasmas. As\'i la acci\'on para una teor\'ia de GB gen\'erica en $d$-dimensiones puede ser escrita como,
\begin{equation}
{\mathcal{S}}=\frac{1}{16\pi
      G_{d}}\int{dx^{d}\sqrt{-g}\left(R+\Lambda-\frac{F^{2}}{4}+\chi_{1}
      R_{\alpha \beta \gamma \delta}\,R^{\alpha \beta \gamma \delta}-4\,\chi_{2}R_{\mu \nu}\,R^{\mu \nu}+\chi_{3}R^{2}\right)}\,,
      \label{GB_with_3_coupling}
\end{equation}
de este modo, cuando se aplica el mecanismo de la funci\'on entr\'opica, la entrop\'ia de un AN extremo para $d=4$ toma la forma, 

\begin{eqnarray}
S^{(d=4)}_{GB}&=& -{\frac {2\pi \, \left(
\,{{v_{2}}}^{2}{\Lambda}^{2}+2\,{v_{2}}\, \Lambda+2 \right)
{\chi_{1}}}{{G_{4}}\, \left( {v_{2}}\,\Lambda+1
 \right) }}+{\frac {4\pi \, \left(2+\,{{v_{2}}}^{2}{\Lambda}^{2}+2
\,{v_{2}}\,\Lambda \right) {\chi_{2}}}{{G_{4}}\, \left( {v_{2}}\,
\Lambda+1 \right) }},  \label{Entropia_GB_3CC_4D}\\
&-& \,{\frac {2\pi \,{{v_{2}}}^{2}{\Lambda}^{2}{\chi_{3}}}{{G_{4}}\,
\left( {v_{2}}\,\Lambda+1 \right) }}+\frac{\pi\,v_{2}}{G_{4}}\,.
\nonumber
\end{eqnarray}
Las soluciones para las funciones $v_{1},\,e,$ y $q$ son,
\begin{eqnarray}
  v_{1}&=& \frac{v_{2}}{v_{2}\,\Lambda+1},\qquad e=\frac{{f_{0}}}{v_{2}\,\Lambda+1},
   \qquad q=-\frac{{f_{0}}}{4G_{4}},  \\
    f_{0}&=& \Biggl\{-8\chi_{1}v^{2}_{2}\Lambda^{2}-16\chi_{1}v_{2}\Lambda+16\,\chi_{2}v^{2}_{2}\Lambda^2+32\chi_{2} v_{2}\Lambda-8\chi_{3}v^{2}_{2}\Lambda^{2}  \nonumber \\
&-& 16\,\chi_{3}v_{2}\Lambda+2{v_{2}}^{2}\Lambda+4v_{2}-p^{2}\Biggr\}^{1/2},  \nonumber
\end{eqnarray}
entonces si $\chi_{1}=\chi_{2}=\chi_{3}=\chi$, en (\ref{Entropia_GB_3CC_4D}), se obtienen las soluciones (\ref{GB_4D}) con ($\chi=\alpha$). Sin embargo, si queremos encontrar un resultado general para GB (\ref{Entropia_GB_3CC_4D}) a partir de las soluciones anal\'iticas (\ref{SBH_Second_Order_d4}), (o viceversa) entonces los siguientes v\'inculos deben exigirse sobre ambos resultados,
\begin{equation}
 \chi_{1}-2\chi_{2}+\chi_{3}=0, \qquad
   \chi_{1}=2\chi_{2}+\frac{3b_{2}}{16}+a_{2}, \qquad
 \chi_{3}=a_{2}-\frac{b_{2}}{16}\,.
\end{equation}
 Igualmente para $d=5$ la entrop\'ia resultante viene dada por,
\begin{eqnarray}
 S^{(d=5)}_{GB}=
   \frac{\pi^2}{(v_{2}+4\chi_{3})G_{5}}(-8\sqrt{v_{2}}\chi^{2}_{1} &+&
    \chi_{1}(\,48\chi_{2}\sqrt{v_{2}}-
    2\sqrt{v_{2}}(4v_{2}+16\chi_{3}+\Lambda\,v^{2}_{2}))-\, \nonumber \\
   -64\sqrt{v_{2}}\chi^{2}_{2}&+&\,4\,\sqrt{v_{2}}(\,20\chi_{3}+\Lambda v^{2}_{2}+4v_{2})\chi_{2}-\frac{v^{\frac{5}{2}}_{2}}{2}(4\chi_{3}\Lambda-1)
 )\,,   \label{Entropia_GB_3CC_5D}
\end{eqnarray} 
para las funciones $v_{1},\,e$ y $q$, dadas por,
%
\begin{eqnarray}
v_{1}&=&\frac{v_{2}(v_{2}+4\chi_{3})}{4v_{2}+\Lambda
v^{2}_{2}-16\chi_{2}+12\chi_{3}+4\chi_{1}}\,,
\\
q\, &=& \,q(v_{2},\Lambda,\chi_{1},\chi_{2},\chi_{3}),\,\qquad
e\,=\,e(v_{2},\Lambda,\chi_{1},\chi_{2},\chi_{3})\,.
\end{eqnarray}
%
Tambi\'en, la relaci\'on entre las entrop\'ias (\ref{Entropia_GB_3CC_5D})
y (\ref{SBH_Second_Order_d5}) vienen dadas para cuando se cumplan los siguientes v\'inculos,
\begin{equation}
\chi_{1}-2\chi_{2}+\chi_{3}=0, \qquad
a_{2}=\frac{3}{5}(2\chi_{2}-\chi_{1}), \qquad
b_{2}=8(\chi_{1}-2\chi_{2})\,.
\end{equation}
Si entonces tomamos $\,\chi_{1}=\chi_{2}=\chi_{3}\equiv\chi$, en (\ref{Entropia_GB_3CC_5D}), se obtiene la ecuaci\'on (\ref{GB_5D}).

Estos resultados son interesantes porque ellos se corresponden con casos donde los invariantes de Riemann no pueden reproducir los
resultados de la teor\'ia gen\'erica de GB. Esto nos sugiere que a nuestra teor\'ia siempre le faltar\'an invariantes por adicionar. Aunque,
esta situaci\'on es irrelevante si se tiene en cuenta que es bien conocido que la teor\'ia de GB sobre una gravedad AdS es en general
inconsistente debido al principio variacional \cite{Aros:1999id} y al problema de la regularizaci\'on que presentan. Una discusi\'on
extensiva de este problema puede verse en \cite{Olea:2005gb},\cite{Kofinas:2006hr},\cite{Kofinas:2008ub}. 

\subsection{Conclusiones y perspectivas del cap\'itulo}

Nosotros hemos calculado la entrop\'ia de ANs extremos en el caso de $d=4$ y $d=5$ dimensiones, teniendo en cuenta t\'erminos de
derivaci\'on superior construidos a partir del conjunto completo de invariantes de Riemann, en una forma, que hasta donde sabemos, no se
ha realizado con anterioridad.

Aunque el caso de GB fue estudiado con anterioridad en \cite{Morales:2006gm}, en nuestro trabajo, hemos encontrado invariantes de segundo
orden que han generalizado los resultados del caso GB encontrando como caso particular el caso de RN. Tambi\'en se han obtenido los t\'erminos
principales para las aproximaciones en el caso de t\'erminos de derivaci\'on superior. Por consiguiente, el conjunto de soluciones obtenido
muestra un ejemplo concreto de que el formalismo de Sen funciona correctamente y es un proceso menos complicado que el uso de la
f\'ormula de Wald \cite{Matyjasek:2008yq}-\cite{Matyjasek:2006nu}. Para poder aplicar la f\'ormula de Wald se debe conocer la soluci\'on
exacta de las ecuaciones de campo del problema y esto solo se tiene para un n\'umero muy reducido de casos. 
\newpage
\section{CONCLUSIONES Y PERSPECTIVAS}

\section*{\Large{Conclusiones}}
En esta tesis nos planteamos dos objetivos inspirados en el estudio de la F\'isica de Objetos Compactos en presencia de campos magn\'eticos intensos. El primer objetivo, el estudio del colapso de un gas denso autogravitante de materia
fermi\'onica: electrones y neutrones sobre la base de un modelo construido para describir, al menos cualitativamente, el interior de una EB o una EN. Bajo las condiciones asumidas en el modelo, arribamos a las siguientes conclusiones:\\
\begin{enumerate}

\item El gas de electrones magnetizado dependiendo de las condiciones iniciales del problema colapsa a una singularidad tipo ``punto'' o ``cigarro''.

\item Para el gas de electrones magnetizado siempre existe un valor de campo magn\'etico para el cual el colapso tipo ``cigarro'' se extiende en la direcci\'on del campo magn\'etico.

\item Un gas de neutrones siempre colapsa a una singularidad tipo ``punto''.

\item El que el gas de electrones pueda tener una singularidad tipo ``cigarro'', y no as\'i para el gas de neutrones, se debe a que el campo magn\'etico se acopla de manera m\'as fuerte (mediante su carga el\'ectrica) con los electrones, a diferencia de los neutrones, ya que ellos no poseen carga el\'ectrica y se acoplan al campo magn\'etico mediante su MMA.

\item Hemos introducido perturbaciones enriqueciendo la din\'amica del espacio-tiempo Bianchi demostrando que este espacio-tiempo puede aproximarse m\'as a situaciones reales.

\end{enumerate}

Del segundo objetivo podemos destacar a manera de conclusiones lo siguiente:

\begin{enumerate}
 \item El m\'etodo de Sen nos permiti\'o encontrar nuevos resultados para la entrop\'ia de ANs extremos en 4 y 5--dimensiones con
topolog\'ia AdS y teniendo en cuenta t\'erminos de derivaci\'on superior. Se mostr\'o as\'i que estos resultados generalizan la
f\'ormula de \'Area encontrada por Bekenstein-Hawking ($S=A/4G$) para la TGR.

 \item El m\'etodo de Sen nos demostr\'o que es una herramienta realmente potente y funciona bien para teor\'ias con t\'erminos de
derivaci\'on superior. C\'alculos que anteriormente eran imposibles de efectuar, ahora son posibles de realizarlos.
\end{enumerate}

\newpage
\section*{\Large{Perspectivas}}

Como perspectivas podemos destacar varias relacionadas con los dos objetivos.

\begin{enumerate}
\item El tema de estudiar la evoluci\'on din\'amica de un gas degenerado y magnetizado autogravitante de fermiones, podr\'ia estudiarse en
condiciones m\'as realistas. Se podr\'ia enriquecer la din\'amica del espacio-tiempo, ya sea introduciendo perturbaciones como se intenta
hacer en el cap\'itulo \ref{Cap_Perturbado} o cambiando a otros Bianchis de ordenes superiores en donde se presentan condiciones de
4--velocidad inclinada con vorticidad no nula, as\'i como curvatura positiva que es la que naturalmente posee un objeto compacto.

El hecho de incorporar fuentes de materia de distinto tipo, o sea materia mezclada, electrones, protones, neutrones, tambi\'en incrementar\'ia
la validez del modelo. Incorporando las condiciones de neutralidad de carga y equilibrio beta que est\'an presentes en el interior de una
EB y de una EN, adem\'as entender si en esas condiciones prevalecen la singularidad de tipo punto o si las de tipo cigarro aparecen.

Examinar la posibilidad de incorporar la aproximaci\'on de \'arbol en el potencial termodin\'amico y las de un lazo para las ecuaciones
de Maxwell. 

Tambi\'en podr\'ia generalizarse este estudio al contexto cosmol\'ogico a\~nadiendo los efectos de la temperatura. Ello podr\'ia describir
la din\'amica de la nucleos\'intesis en presencia de campos magn\'eticos.

\item Para el segundo tema, se podr\'ia incorporar hasta orden cinco los invariantes relacionados con la teor\'ia de calibraci\'on o sea
el mismo orden m\'aximo que poseen los invariantes de Riemann y encontrar all\'i invariantes independientes, despu\'es con ellos calcular
la entrop\'ia de ANs extremos con topolog\'ia AdS de forma aproximada. Tambi\'en, se podr\'ian generalizar estos resultados a
6,7,8,...$n$-dimensiones.

\end{enumerate}

\newpage

\begin{appendices}
\noappendicestocpagenum
 \addappheadtotoc

\section[Convenios y Notaciones]{\Large{Convenios y Notaciones}}
\label{Convenios_Notaciones}

Constantes f\'isicas m\'as usadas,
\begin{eqnarray}
\hbox{Velocidad de la luz}\,\, c&=& 2.998 \times 10^{8}\,m\,s^{-1}, \nonumber \\
\hbox{Carga el\'ectrica del electr\'on}\,\, e&=& 1.60 \times 10^{-19}\,C, \nonumber \\
\hbox{Constante de Dirac}\,\, \hbar &=& 1.054 \times 10^{-34}\,J\,s, \nonumber    \\
\hbox{Masa en reposo del electr\'on}\,\, m_{e}&=& 9.109 \times 10^{-31}\,kg,  \nonumber   \\
\hbox{Constante de gravitaci\'on}\,\, G &=& 6.6742 \times 10^{-11}\,N\,m^{2}\,kg^{-2}, \nonumber     \\
\hbox{Acoplamiento gravitatorio}\,\,\kappa &=& \frac{8\pi G}{c^4}= 2.075 \times 10^{-43}\,m\,J^{-1},  \nonumber   \\
\hbox{Masa del Sol}\,\, M_{\odot} &=& 1.99 \times 10^{30}\,kg, \nonumber \\
\hbox{Radio del Sol}\,\, R_{\odot}&=& 6.96 \times 10^{5}\,km, \nonumber \\
\hbox{Temperatura del Sol}\,\, T_{\odot}&=& 15 \times 10^{6}\,K, \nonumber \\
\hbox{Campo Magn\'etico}\,\, \B_{\odot}&=& 1\,G, \nonumber \\ 
\hbox{Densidad del Sol}\,\, \rho_{\odot}&=& 1410\, kg\, m^{-3}.\nonumber
\end{eqnarray}
Matrices de Pauli,
\begin{equation} {\sigma}^{1} = \left(
\small\begin{array}{ll}
\\0 \ \ \ 1\\
1 \ \ \ 0\\
\end{array}
\right), \ \ \
 {\sigma}^{2} = \left(
\small\begin{array}{ll}
\\0 \ \ -i\\
i \ \ \ \ \ \ 0\\
\end{array}
\right), \ \ \
 {\sigma}^{3} = \left(
\small\begin{array}{ll}
\\1 \ \ \ \ \ \ 0\\
0 \ \ \ -1\\
\end{array}
\right).
 \end{equation}
Matrices gamma de Dirac,
\begin{equation} {\gamma}^{0} = \left(
\small\begin{array}{ll}
\\1 \ \ \ \ \ 0\\
0 \ \ \ -1\\
\end{array}
\right), \ \ \ \ \
 {\gamma}^{j} = \left(
\small\begin{array}{ll}
\\0 \ \ \ \ \ \sigma^{j}\\
\sigma^{j} \ \ \ 0\\
\end{array}
\right).
 \end{equation}

Si se usa el convenio $\hbar =c=1,$ en este sistema,
\begin{equation}
[longitud]=[tiempo]=[masa]^{-1}=[energ\acute{\i}a]^{-1},
\end{equation} 
se deduce que si se escribe una $m$ sola, entonces se entender\'a como energ\'{\i}a en reposo $mc^2$, as\'{\i} como su inverso se entender\'a, como la longitud de onda Compton $(\hbar/mc)$,

\begin{equation}
m_{e}=9.109 \times 10^{-31}kg \equiv 0.511\,MeV \equiv (3.862 \times 10^{-11}cm)^{-1}.
\end{equation} 
En el contexto de la teor\'{\i}a cu\'{a}ntica de campos, se introduce una derivada covariante de la forma,
\begin{equation}
D_{\mu}=\partial_{\mu}+ie A_{\mu}\,,
\end{equation}
donde $A_{\mu}$ es el 4-vector potencial del campo electromagn\'{e}tico y se toma el convenio para la m\'{e}trica del modo $[+ - - -]$. As\'i, $x^{\mu}=(x^{0},\textbf{x})$, y $x_{\mu}=g_{\mu \nu}x^{\nu}$ las que para el caso de una m\'{e}trica plana de Lorentz, toman la forma $x_{\mu}=(x^{0},-\textbf{x})$. As\'i mismo, para el operador derivada se define como es usual,

\begin{equation}
\partial_{\mu}=\frac{\partial}{\partial
x^{\mu}}=(\partial_{0},{\nabla}).
\end{equation}

Sin embargo, si estamos en la TGR, generalmente tomamos el convenio [- + + +] y la relaci\'on entre las constantes $G/c=1$. Usamos \'indices latinos para el espacio-tiempo ($a,b,c,d,...i,j,k,...=1,2,3,4$) e \'indices griegos para el espacio 3-dimensional ($\alpha,\beta,\gamma,...=1,2,3$), con la excepci\'on del Cap\'itulo-4 donde invertimos esta selecci\'on.

Tambi\'en, empleamos la notaci\'{o}n de coma ``,'' para la derivada normal,  
\begin{equation}
T^{ab}\,_{,\mu}=\frac{\partial T^{ab}}{\partial {x^{\mu}}}
\end{equation} y punto y coma $(;)$ o $\nabla$(antes del ente que se desea derivar) para la derivada covariante. Ejemplo para un tensor $T^{ab}$ dos veces contravariante,

\begin{equation}
T^{ik}\,_{;j}=\frac{\partial T^{ik}}{\partial x^{j}}+\Gamma^{i}_{jn}T^{nk}+\Gamma^{k}_{jm}T^{im}\equiv \nabla_{j} T^{ik},
\end{equation}

donde $\Gamma^{i}_{kj}$ son los \'{\i}ndices de Christoffel.
 
Adem\'as se recuerda el conjunto de definiciones:

\begin{itemize}
 \item $T^{a_{1}a_{2},...,a_{p}}\,_{b_{1},b_{2},..b_{q}}\qquad$\,  Componentes de un tensor ``$p$'' veces contravariante y ``$q$'' veces covariante.
 \item $u^{a}\qquad$  Componentes de un vector contravariante en el espacio-tiempo.
 \item $\vec{u}\qquad$ Componente de un vector del espacio 3-dimensional.
 \item $w^{i}\qquad$ 1-forma diferencial.
 \item $e_{j}\qquad$ base covariante.
\end{itemize}

 En general un tensor cualquiera se puede escribir rigurosamente de la forma:

\begin{equation}
  T=T^{a_{1}a_{2},...,a_{p}}\,_{b_{1},b_{2},..b_{q}} w^{b_{1}}\otimes w^{b_{2}}\otimes...w^{b_{q}}\otimes e_{a_{1}}\otimes e_{a_{2}}...\otimes e_{a_{p}},
\end{equation}
donde las bases deben cumplir con las siguientes relaciones:
%
  \begin{eqnarray}
     <e_{i},e_{j}>&=&\,g_{ij},\\
     <w^{i},w^{j}>&=&\, g^{ij},\\
     <w^{j},e_{i}>&=&\, \delta^{j}\,_{i},
  \end{eqnarray}
 donde $< , >$ denota el producto escalar.

%
%

\newpage

\section[Diagramas del Agujero Negro de Schwarzschild]{\Large{Diagramas del Agujero Negro de Schwarzschild}}

\label{Ap_AN}

%

\FIGURE{\epsfig{file=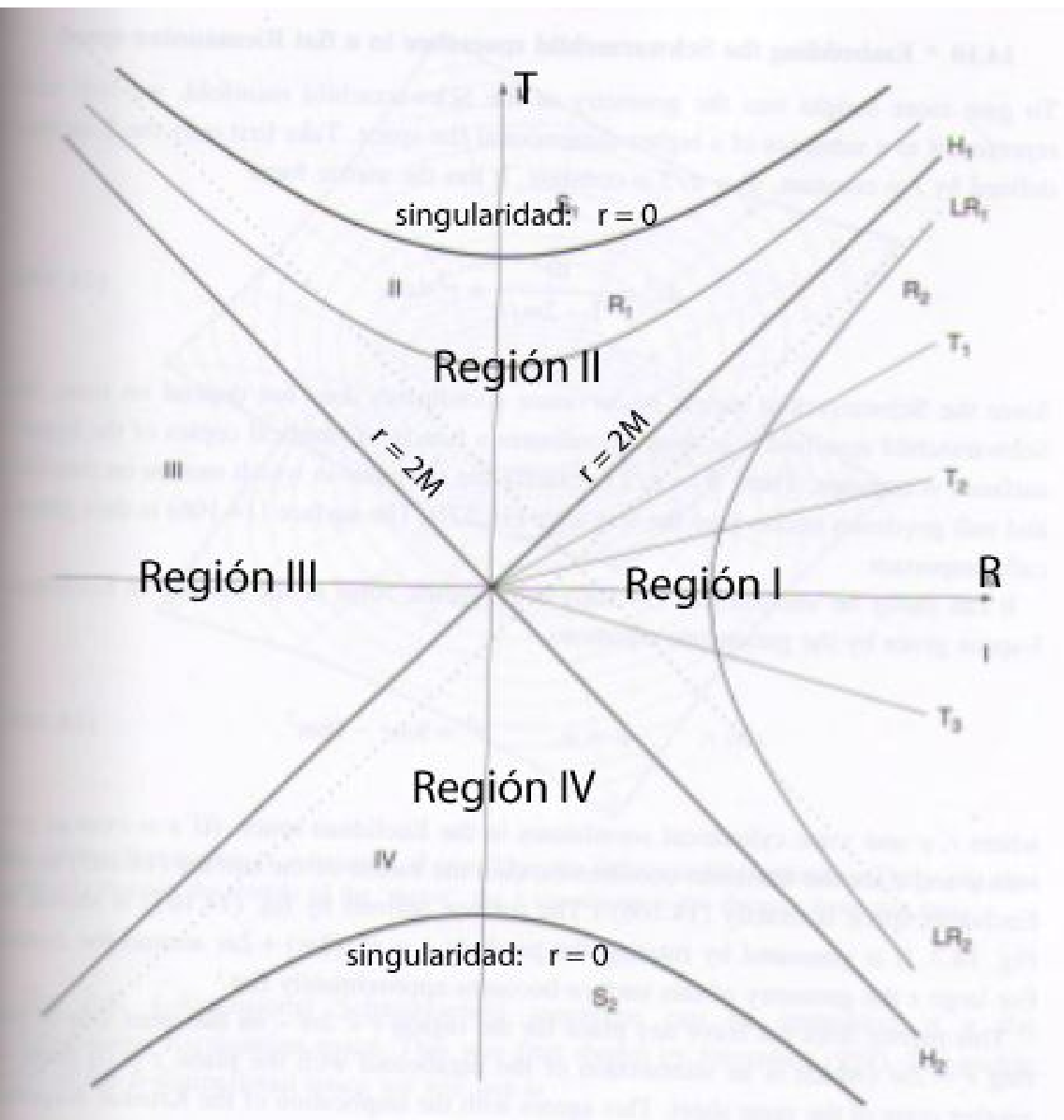,height=10cm,width=10cm,angle=0}[h]
\caption{Se muestra el diagrama del AN de Schwarzschild en las coordenadas de Kruskal-Szekeres.}
\label{KruskalDiagram_1}}

\newpage
 
\FIGURE{\epsfig{file=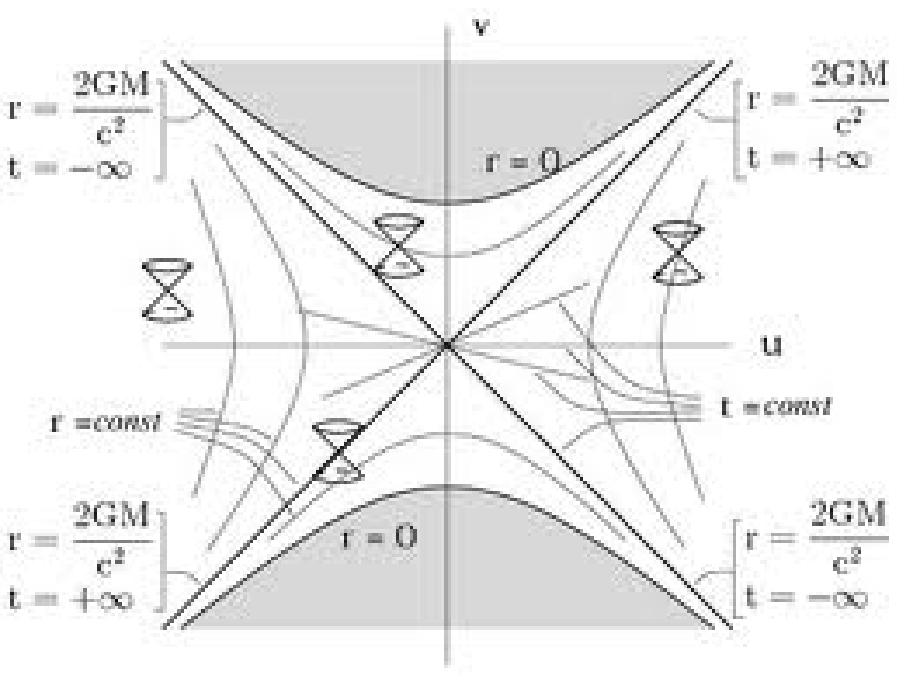,height=9cm,width=9cm,angle=0}
\caption{\footnotesize{Se muestra el diagrama del AN de Schwarzschild en las coordenadas de Kruskal-Szekeres.}}
\label{KruskalDiagram_2}
}

%
%

\newpage
\section[Fuentes de Materia]{\Large{Fuentes de Materia}}
\label{AP_Fuentes_Materia}

\subsection{Fluido Termodin\'amico}
La ecuaci\'on de estado m\'as general para un fluido termodin\'amico mono--componente puede ser escrita como,
\begin{equation}
  \rho=\rho(n,T), \qquad p=p(n,T), \label{Ecu_Estado_General}
\end{equation}
donde $n$ y $T$ son la densidad del n\'umero de part\'iculas y la temperatura. En la geometr\'ia RW este fluido cumple con la ecuaci\'on de conservaci\'on,
\begin{equation}
  \nabla_a(nu^a)=0,\qquad\Rightarrow\qquad \frac{\dot n}{n}+\frac{3\dot R}{R}=0,\qquad\Rightarrow\qquad n\propto R^{-3},  \label{Ec_Cons_Materia}
\end{equation}
 y la ecuaci\'on de equilibrio de Gibbs,  
\begin{equation}
T\hbox{d}S=\hbox{d}\left(\frac{\rho}{n}\right)+p\,\hbox{d}\left(\frac{1}{n}\right)\qquad \Rightarrow \qquad T\dot{S}=\frac{\hbox{d}}{\hbox{d}t}\left(\frac{\rho}{n}\right)+p\,\frac{\hbox{d}}{\hbox{d}t}\left(\frac{1}{n}\right), \label{Ec_Cons_Gibbs}
\end{equation}
donde $S$ es la entrop\'ia por part\'icula. Comparando (\ref{Ec_Cons_Materia}) y (\ref{Ec_Cons_Gibbs}), con (\ref{EEEB_1}) obtenemos $\dot S=0$, de modo que el fluido es isoentr\'opico (la entrop\'ia $S$ se conserva a lo largo de $u^a$), lo cual es de esperar para un fluido en equilibrio t\'ermico (bajo un proceso reversible y cuasi--est\'atico). Usando (\ref{EEEB_1}) y (\ref{Ec_Cons_Materia}), la condici\'on de integrabilidad (\ref{Ec_Cons_Gibbs}) permite obtener la ley de evoluci\'on,  
\begin{equation}
  \frac{\dot T}{T}+3\left[\frac{\partial p/\partial T}{\partial \rho/\partial
T}\right]_n\,\frac{\dot R}{R}=0, \label{Ley_Evoluc_Temp}
\end{equation}
donde el \'indice $_n$ indica evaluaci\'on a $n$ constante. La integraci\'on de (\ref{Ley_Evoluc_Temp}) para cualquier ecuaci\'on de estado como (\ref{Ecu_Estado_General}) permite una relaci\'on del tipo $T=T(R,n)$, pero $n\propto R^{-3}$, la relaci\'on $T=T(R,n)$ transforma a (\ref{Ecu_Estado_General}) en $\rho=\rho(R)$, permitiendo la integraci\'on de la ecuaci\'on de Friedmann (\ref{EEEB_2}).
 \subsection{Gas ideal relativista no-degenerado.}
 Un importante ejemplo de un sistema termodin\'amico es el gas ideal relativista de una componente, asociado con la distribuci\'on de Juttner, la cual es la generalizaci\'on relativista de la distribuci\'on de Maxwell-Boltzmann en la Teor\'ia Cin\'etica Relativista. Una descripci\'on hidrodin\'amica para este tipo de gas viene dada por la ecuaci\'on de estado,
\begin{equation}
 \rho=mc^2n\Gamma(\beta)-nk_{_B}T,\qquad p=nk_{_B}T,  \label{EEstado_gasRelat}
\end{equation}
donde, 
\[
\Gamma(\beta) \equiv{\frac{K_3(\beta)}{{K_2(\beta)}}},\qquad \beta\equiv{ \frac{mc^2}
{{kT}}},
\]
aqu\'i $m$ es la masa de la part\'icula, $k_{_B}$ es la constante de Boltzmann y $K_{2},K_{3}$ son funciones de Bessel modificadas de segundo tipo, de segundo y tercer orden. La ecuaci\'on de estado (\ref{EEstado_gasRelat}) es un poco complicada, aunque se simplifica para dos situaciones extremas de temperatura y energ\'ia: el r\'egimen ultra-relativista (UR) y el no-relativista (NR), caracterizados por $\beta\ll 1$ y $\beta\gg 1$ respectivamente. Es \'util apreciar estos rangos de temperaturas para estos casos l\'imites. Considerando masas de orden nuclear ($\approx 10^{-27}$ kg), con $k_{_B}\approx 10^{-23}\,\hbox{J/K}$, nosotros tenemos que $\beta\approx 1$ para $T\approx 10^{12}$ K. De modo que podemos modelar materia c\'osmica bas\'andonos en un gas ideal no relativista. El caso ultra-relativista podr\'ia ser v\'alido para eras c\'osmicas muy calientes. Considerando el comportamiento de $\Gamma(\beta)$ para estos casos l\'imites, nosotros tenemos,
%
 \begin{eqnarray}
   \beta\ll 1,\qquad \Gamma\approx{\frac{4}{\beta}}+{\frac{\beta}{2}}+ \hbox{O}
 (\beta^3),\qquad \rho\approx 3nk_{_B}T \qquad \mbox{UR} , \\
    \beta\gg1,\qquad \Gamma\approx 1+{\frac{5}{{2\beta}}}+\hbox{O} (\beta^{-2}),\qquad 
\rho\approx mc^2n+{\frac{3}{2}}nk_{_B}T \qquad \mbox{NR}.
 \end{eqnarray}
%
 %
Identificando la ecuaci\'on de estado para el caso del gas ideal ultra-relativista (UR),
\begin{equation}
 \rho=3nk_{_B}T,\qquad p=nk_{_B}T=\frac{1}{3}\rho,  \label{EE_GasUltraRelat}
\end{equation}
y para el gas cl\'asico no-relativista (NR),
\begin{equation}
 \rho=mc^2n+\frac{3}{2}nk_{_B}T,\qquad p=nk_{_B}T,
\end{equation}
de las ecuaciones de conservaci\'on (\ref{Ec_Cons_Materia}), y (\ref{Ec_Cons_Gibbs}) se obtiene inmediatamente para el gas NR,
%
  \begin{eqnarray}
  n&=&n_0\left(\frac{R_0}{R}\right)^3, \qquad T=T_0\left(\frac{R_0}{R}\right)^2, \\
  p(R)&=& n_0k_{_B}T_0\left(\frac{R_0}{R}\right)^5, \\
  \rho(R)&=& mc^2n_0\left(\frac{R_0}{R}\right)^3\left[1+\frac{3}{2}\frac{k_{_B}T}{mc^2}\right], \label{EE_DE_GasRelat_c} 
  \end{eqnarray}\label{EE_DE_GasRelat}
%
donde $R_0,\,n_0,\,T_0$ son $R,\,n,\,T$ evaluados en el valor fijo $t=t_0$. La ecuaci\'on (\ref{EE_DE_GasRelat_c}) es la relaci\'on necesaria $\rho=\rho(R)$ que transforma a (\ref{EEEB_2}) en una cuadratura integrable.\\
La ecuaci\'on de estado del gas ideal relativista (\ref{EEstado_gasRelat}) es aplicable a gases diluidos en los que las distancias medias entre las part\'\i culas (entre colisiones) son mucho mayores que la longitud de onda t\'ermica. Por lo tanto, los efectos cu\'anticos del tipo de estad\'\i stica, pueden ser despreciados (gases no--degenerados). En los cap\'\i tulos \ref{Cap_de_Electrones} y \ref{CAP_Neutrones} se tratar\'a el caso de gases de fermiones, en los cuales estas suposiciones no son v\'alidas. Aunque las ecuaciones de estado resultantes ser\'an mucho m\'as generales, el tratamiento din\'amico atrav\'es de las ecuaciones de Einstein para estos gases sigue la misma metodolog\'\i a que la descrita anteriormente. Sin embargo, la inclusi\'on de un campo magn\'etico introduce la existencia de una direcci\'on privilegiada, y es por lo tanto incompatible con la m\'etrica FLRW. Para este caso se utilizar\'an modelos Bianchi I, que son homog\'eneos, pero anisotr\'opicos.   
\subsection{Polvo}
Es evidente a partir del sistema de ecuaciones (\ref{EE_DE_GasRelat}), que expandiendo al l\'imite $R \gg R_0$,
la presi\'on y la energ\'ia decaen r\'apidamente como $O(R^{-5})$, y el resto de la masa energ\'ia como $O(R^{-3})$.
Por otro lado el rango $k_{_B}T/mc^2=1/\beta$ es despreciable ($\ll 10^{-3}$) para masas de nucleones o electrones y para un ancho rango de temperaturas asociadas con condiciones cl\'asicas ($T<10^8$ K), entonces es evidente a partir de (\ref{EE_DE_GasRelat_c}) que pr\'acticamente para toda su evoluci\'on, el gas ideal puede ser razonablemente aproximado suponiendo $p\approx 0$ y $\rho\approx mc^2n$. Fuentes de materia que satisfacen la ecuaci\'on de estado,
\begin{equation}
  p=0,\qquad \rho(R)\propto R^{-3}, \label{Ec_POLVO}
\end{equation}
son llamadas ``polvo'', y reducen a (\ref{TEM_FP}) a la forma $T_{ab}=\rho \, u_au_b$.
Aunque (\ref{Ec_POLVO}) puede ser justificado como l\'imite de temperatura cero de un gas ideal NR, dejando a $\rho=mc^2n$ (o sea toda la densidad de materia-energ\'ia ser\'ia igual a la energ\'ia en reposo). Esta ecuaci\'on posee una conveniente descripci\'on aproximada de ciertos tipos de materia c\'osmica no-relativista. Por ejemplo, materia fr\'ia no colisionada, o materia oscura fr\'ia no-bari\'onica (aunque esta \'ultima relaci\'on no es del todo segura para materia oscura).
\subsection{Radiaci\'on}
Para el gas ideal UR, de las ecuaciones de conservaci\'on (\ref{Ec_Cons_Materia}), y (\ref{Ec_Cons_Gibbs}) se obtiene que,
\begin{equation}
  n=n_0\left(\frac{R_0}{R}\right)^3, \qquad T=T_0\left(\frac{R_0}{R}\right),
\end{equation}
as\'i que,
\begin{equation}
  \rho(R)=3n_0k_{_B}T_0\left(\frac{R_0}{R}\right)^4=3p(R), \label{Dens_Energ_Radiacion}
\end{equation}
permiten la integraci\'on de (\ref{EEEB_1}) y (\ref{EEEB_2}). La ecuaci\'on (\ref{EE_GasUltraRelat}) puede describir aproximadamente un gas
de part\'iculas masivas ($m>0$) a muy alta temperatura, pero esta es exactamente la ecuaci\'on de estado para un gas de part\'iculas no
masivas, tales como fotones o neutrinos. Una relaci\'on equivalente a (\ref{Dens_Energ_Radiacion}) para fotones es la ley de
Steffan-Boltzmann, 
\begin{equation}
  \rho_\gamma=a_\gamma T^4=a_\gamma T_0^4\left(\frac{R_0}{R}\right)^4\,,
\end{equation}
donde $a_\gamma$ es la constante de Steffan-Boltzmann, asi que la conservaci\'on de la densidad del n\'umero de fotones es
$n_0=a_{\gamma}\,T_0^3/3k_{_B}$. El t\'ermino radiaci\'on es aplicado a un gas de fotones caracterizado por el espectro de energ\'ia de un
cuerpo negro. La Radiaci\'on C\'osmica de Fondo o CMB \footnote{CMB viene del ingl\'es Cosmic Microwave Background} es la reliquia de eras
primordiales en el Universo donde dominaba la radiaci\'on, es la m\'as importante manifestaci\'on de este tipo de gas en el contexto
cosmol\'ogico.

%
%
%
%
%
%
\newpage
\section[Ecuaci\'on de Raychaudhuri]{\Large{Ecuaci\'on de Raychaudhuri}}
\label{AP_ECU_RAYCHA}
%
%
Antes de deducir la ecuaci\'on de Raychaudhuri definamos primero la congruencia de geod\'esicas:
  \begin{itemize}
  \item
    Sea $M$ una variedad y sea $\mathcal{O}\subset\,M$ un abierto. Una {\it congruencia} en
    $\mathcal{O}$ es una familia de curvas tales que a trav\'es de cada $p\in\, \mathcal{O}$
    pasa precisamente una curva de esta familia.
  \item
    Una congruencia se dice que es {\it suave} si el correspondiente vector del campo de los vectores tangentes es suave.
\end{itemize}
%
 Consideremos una suave {\it congruencia de geod\'esicas temporaloideas} . Suponemos que las geod\'esica son parametrizables por el tiempo propio $\tau$, asi que el vector del campo $u^{a}$, tangente a las l\'ineas de materia, es normalizado a la unidad de la forma $u^{a}u_{a}=-1$. Entonces se puede definir cierto tensor del campo $B_{ab}$,
   \begin{equation}
      B_{ab}=\nabla_{b}u_{a}\,,
   \end{equation}
que ser\'a puramente espacial, o sea,
  \begin{equation}
    B_{ab}u^{a}=B_{ab}u^{b}=0\,.
  \end{equation}
  Consideremos una suave subfamilia  $\gamma_{s}(\tau)$ uni-param\'etrica de geod\'esicas en la congruencia.
Sea $\eta^{a}$ el vector de desviaci\'on ortogonal de $\gamma_{0}$ para esta subvariedad. Entonces $\eta$ representa el desplazamiento infinitesimal espacial desde $\gamma_{0}$ hasta una geod\'esica cercana,
  \begin{equation}
     \mathfrak{L}_{u}\eta^{a}=0,
  \end{equation}
  donde, la anterior derivada de Lie quiere decir que,
  \begin{equation}
    u^{b}\nabla_{b}\eta^{a}=\eta^{b}\nabla_{b}u^{a}\equiv B^{a}_{\,\,b}\eta^{b}.
  \end{equation}
  Entonces $B$ puede ser interpretado f\'isicamente como quien mide la desviaci\'on de $\eta^{a}$ al ser transportado
  paralelamente. Un observador sobre las geod\'esicas de la familia $\gamma_{0}$ podr\'a encontrar geod\'esicas
  cercanas a \'el rode\'andolas estiradas o rotadas por el mapa lineal $B^{a}_{\,\,b}$. Continuemos ahora con el siguiente conjunto de definiciones necesarias:\\
%
 La m\'etrica espacial puede ser definida como:
     \begin{equation}
       h_{ab}=g_{ab}+u_{a}u_{b}\,,
    \end{equation}
donde el tensor de proyecci\'on ortogonal sobre la hiper-superficie ortogonal a $u^{a}$ es dado por,
    \begin{equation}
      h^{a}_{\,\,b}=g^{ac}h_{cb}\,.
     \end{equation}
%
%
    La expansi\'on, se define como:
    \begin{equation}
      \theta=B^{ab}h_{ab}\,.
    \end{equation}
    El tensor de deformaciones (o cizallamiento) se define seg\'un:
    \begin{equation}
      \sigma=B_{(ab)}-\frac{\theta}{3}h_{ab}\,,
    \end{equation}
    y el tensor de vorticidad como:
    \begin{equation}
      \omega_{ab}=B_{[ab]}.
    \end{equation}
De esta forma $B_{ab}$ puede ser descompuesto como:
    \begin{equation}
     B_{ab}=\frac{\theta}{3}\,h_{ab}+\sigma_{ab}+\omega_{ab}\,.
    \end{equation}
%
   Todos ellos son p\'uramente espaciales ya que, $\sigma_{ab}u^{b}=\omega_{ab}u^{b}\equiv0$.
Ahora los respectivos significados f\'isicos de estas magnitudes son:
   \begin{itemize}
   \item
   $\theta$: Mide el promedio de expansi\'on de geod\'esicas circundantes e infinit\'esimamente cercanas.
   \item
   $\omega_{ab}$: Mide la rotaci\'on de una geod\'esica circundante e infinit\'esimamente cercana respecto a la otra.
   \item
   $\sigma_{ab}$: Mide la deformaci\'on. O sea, sea una esfera inicial en el espacio tangente la cual es transportada
   seg\'un Lie (transporte paralelo) a lo largo de $u^a$ y se distorsiona en un elipsoide con los ejes principales en la direcci\'on de los eigenvectores de $\sigma^{a}\,_{b}$. 
   \end{itemize}
%
 De la ecuaci\'on de la desviaci\'on de la geod\'esica es f\'acil derivar lo siguiente:
  \begin{eqnarray}
    u^{c}\nabla_{c}B_{ab}&=&u^{c}\nabla_{c}\nabla_{b}u_{a}=u^{c}(\nabla_{b}\nabla_{c}u_{a}+R_{cba}^{\,\,\,\,\,d}u_{d})\\
    &=&-B^{c}_{\,\,\,b}B_{ac}+R_{cba}^{\,\,\,\,d}u^{c}u_{d}.
  \end{eqnarray}
Tomando la traza, obtenemos,
  \begin{equation}\label{EcRay}
    u^{c}\nabla_{c}\theta=\frac{d\theta}{d\tau}=-\frac{\theta^2}{3}-\sigma_{ab}\sigma^{ab}+\omega_{ab}\omega^{ab}-R_{cd}u^{c}u^{d}.
  \end{equation}
 La ecuaci\'on (\ref{EcRay}) es la conocida ecuaci\'on de Raychaudhuri y es la piedra angular para demostrar los teoremas de singularidad. Usando las ecuaciones de Einstein's y la normalizaci\'on de $u^{b}$ en el \'ultimo t\'ermino de la ecuaci\'on de Raychaudhuri (\ref{EcRay}) tenemos: 
   \begin{equation}\label{EscRay}
     R_{ab}u^{a}u^{b}=8\pi[T_{ab}-\frac{T}{2}g_{ab}]u^{a}u^{b}=8\pi [T_{ab}u^{a}u^{b}+\frac{T}{2}]\,,
   \end{equation}
   donde $T_{ab}u^{a}u^{b}$ representa la densidad de energ\'ia de la materia medida por un observador cuya 4-velocidad es $u^{a}$ (temporaloidea). De la expresi\'on (\ref{EscRay}) se pueden definir las siguientes desigualdades que constituyen las condiciones de energ\'ia para la materia:
 \begin{eqnarray}
     \mbox{(Condici\'on de energ\'ia d\'ebil)} \ \ \ T_{ab}u^{a}u^{b}\geq 0, \\
     \mbox{(Condici\'on de energ\'ia fuerte)} \ \ \ T_{ab}u^{a}u^{b}+\frac{T}{2}\geq 0,\\
     \mbox{(Condici\'on de energ\'ia dominante)} \ \ \ -T^{a}_{b}u^{b}=\mbox{ vector temporaloideo o nulo}. \label{WeakEnergCond}
 \end{eqnarray}
\begin{itemize}
  \item
   Esta \'ultima condici\'on (\ref{WeakEnergCond}) debe ser un vector temporaloideo o nulo, \'el representa la densidad de 4-corriente del tensor energ\'ia-momentum de la materia. Medida por un observador con 4-velocidad $u^{a}$. Puede ser interpretado como la velocidad del flujo de energ\'ia de la materia con velocidad menor que $c$.
  \item
  Si $\nabla^{a}T_{ab}=0$ entonces la condici\'on de energ\'ia dominate se satisface. Es decir, se desvanece sobre un conjunto cerrado acronal $S$, entonces tambi\'en se desvanece en $D(S)$.\footnote{Esto constituye un lema, o sea el Lema 4.3.1, enunciado por Hawking--Ellis en 1973.} 
  \end{itemize}
  Note que la condici\'on de energ\'ia dominante implica la condici\'on de energ\'ia d\'ebil, pero en otro caso todas estas condiciones ser\'an matem\'aticamente independientes.
%
%
%
%
%
%
\newpage
\section[Significado del tiempo adimensional $(\tau)$]{\Large{Significado del tiempo adimensional $(\tau)$}}
\label{Apend_Time_Tau}
El tiempo adimensional $(\tau)$ se define a partir de la ecuaci\'on
(\ref{def_tau}):

\begin{equation}\label{definicion de tau}
\frac{d}{d\tau}=\frac{1}{H_0}\frac{d}{dt} \ \ \Rightarrow \ \
\tau=H_0t,
\end{equation}
en la cual podemos notar que el signo de $\tau$ depende del signo de
$t$ y del signo de $H_0$. Para comprender el sentido f\'isico del
tiempo adimensional $\tau$ observemos que de (\ref{ec_teta_K}), (\ref{def_tau}) y
(\ref{adim_var}) obtenemos:

\begin{eqnarray}
\HH
H_0&=&H=\frac{1}{3}(\frac{\dot{Q}_{1}}{Q_{1}}+\frac{\dot{Q}_{2}}{Q_{2}}+\frac{\dot{Q}_{3}}{Q_{3}})=\frac{1}{3}\frac{d}{dt}\ln(V),
\label{relacion de HH, V ,H0_1}\\
V&=&ABC\Rightarrow \frac{V}{V_0}=e^{3\int_0^t{ Hdt} }\,,   
  \label{relacion de HH, V ,H0_2}
\end{eqnarray}

y despejando $\HH$ de (\ref{DV3d}) tenemos:
\begin{equation}\label{ec de H}
\HH=\pm\frac{1}{3}\sqrt{3\Gamma_U+S_2^2+S_3^2+S_2S_3}\,. \ \ \
\end{equation}

Si fijamos $t\geqslant0$, en (\ref{relacion de HH, V ,H0_2}) tenemos
la siguiente interpretaci\'on f\'isica:


\begin{eqnarray}
I=3\int\limits_0^tHdt\Rightarrow  & = & \, \mbox{Si} \,I>0\Rightarrow V>V_0  \, \mbox{(expansi\'on)},  \\ 
       & = & \, \mbox{Si} \,I<0\Rightarrow V<V_0 \, \mbox{(colapso)}.  
\end{eqnarray}

Las posibles combinaciones de signo se dan en el cuadro
(\ref{Cuadro_Time_Tau}), en el cual aparecen resaltados los convenios que
hemos escogido en esta tesis.

\begin{table}[h] 
\begin{tabular}{@{}|l|l|l|l|}
\hline
    {\it Casos} & {\it Expansi\'on} & {\it Volumen elemental} & $ \tau $ \\
   &  &  &  \\
\hline
    $ H>0 $ & $\HH>0$ y $H_{0}>0$ & {\it Expansi\'on} &  $\tau>0$ \\
    \cline{2-2}   \cline{4-4}
            & $\HH<0$ y $H_{0}<0$ & $V>V_{0} $        &  $\tau<0$ \\
    \cline{1-4}
     $H<0$  & $\HH<0$ y $H_{0}>0$ & {\it Colapso}     & $\tau>0$ \\
    \cline{2-2}    \cline{4-4}
            & $\HH>0$ y $H_{0}<0$ & $V<V_{0}$         & $\tau<0$ \\
    \cline{1-4}
     $H=0$  & $\HH=0$             & $V=V_{0}$         &           \\
\hline
\end{tabular}
\caption{\label{Cuadro_Time_Tau}{\small Posibles combinaciones de signos entre $\HH$, $H_0$ y $\tau$ para $t>0$. }}
\end{table}

Notemos que el sistema comienza a evolucionar a partir de $t=\tau=0$, para $\tau>0$ el sistema colapsa y para $\tau<0$ el
sistema tiende hacia un estado diluido como se puede ver en la Figura \ref{S3espfase}.

%
%
%
%
\newpage
\section[Espacios de fase]{\Large{Espacios de fase}}

\label{Ap_Esp_Fase}

\FIGURE{\epsfig{file=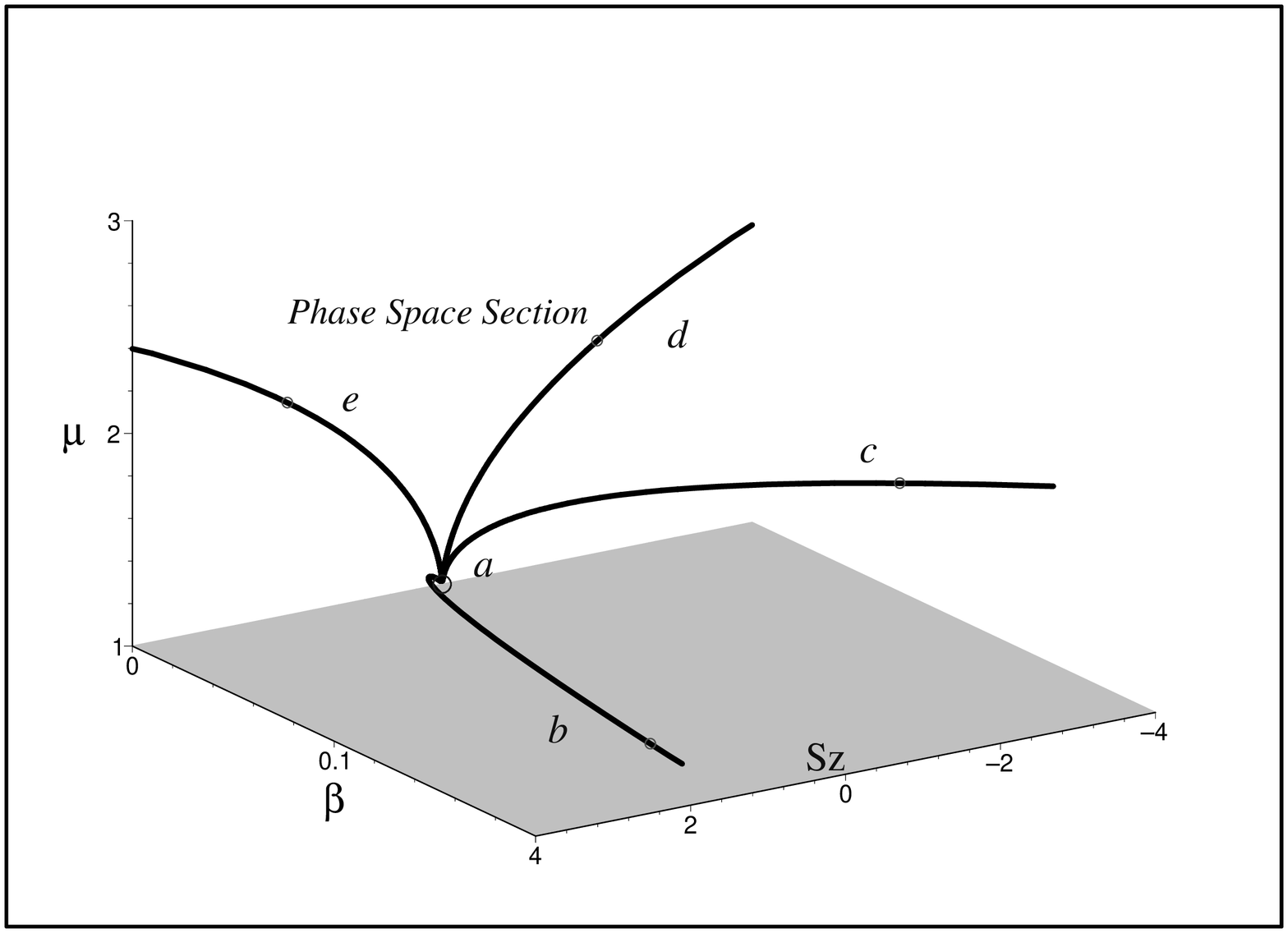,height=10cm,width=10cm,angle=270}
\caption[Ejemplo figura1]{Trayectorias en la secci\'on del espacio de fase ($S^{z},\beta,\mu$) para cuatro condiciones iniciales
diferentes de un gas magnetizado y auto-gravitante de electrones degenerados. El plano sombreado $\mu=1$, est\'a acotado por
$-4<S^{z}<4$ y $0<\beta<0.2$. El atractor ``estable'' es el punto marcado por $a$, y las curvas $b,c,d$ y $e$ son soluciones num\'ericas
sobre la secci\'on 3-dimensional del espacio de fase. Todas las trayectorias caen desde $\tau=0$ tendiendo al atractor estable $a$.
Similarmente, desde $\tau=0$ (esfera vac\'ia) todas escapan tendiendo a la singularidad.
\label{EF_Electrones}}}

\newpage 

\FIGURE{\epsfig{file=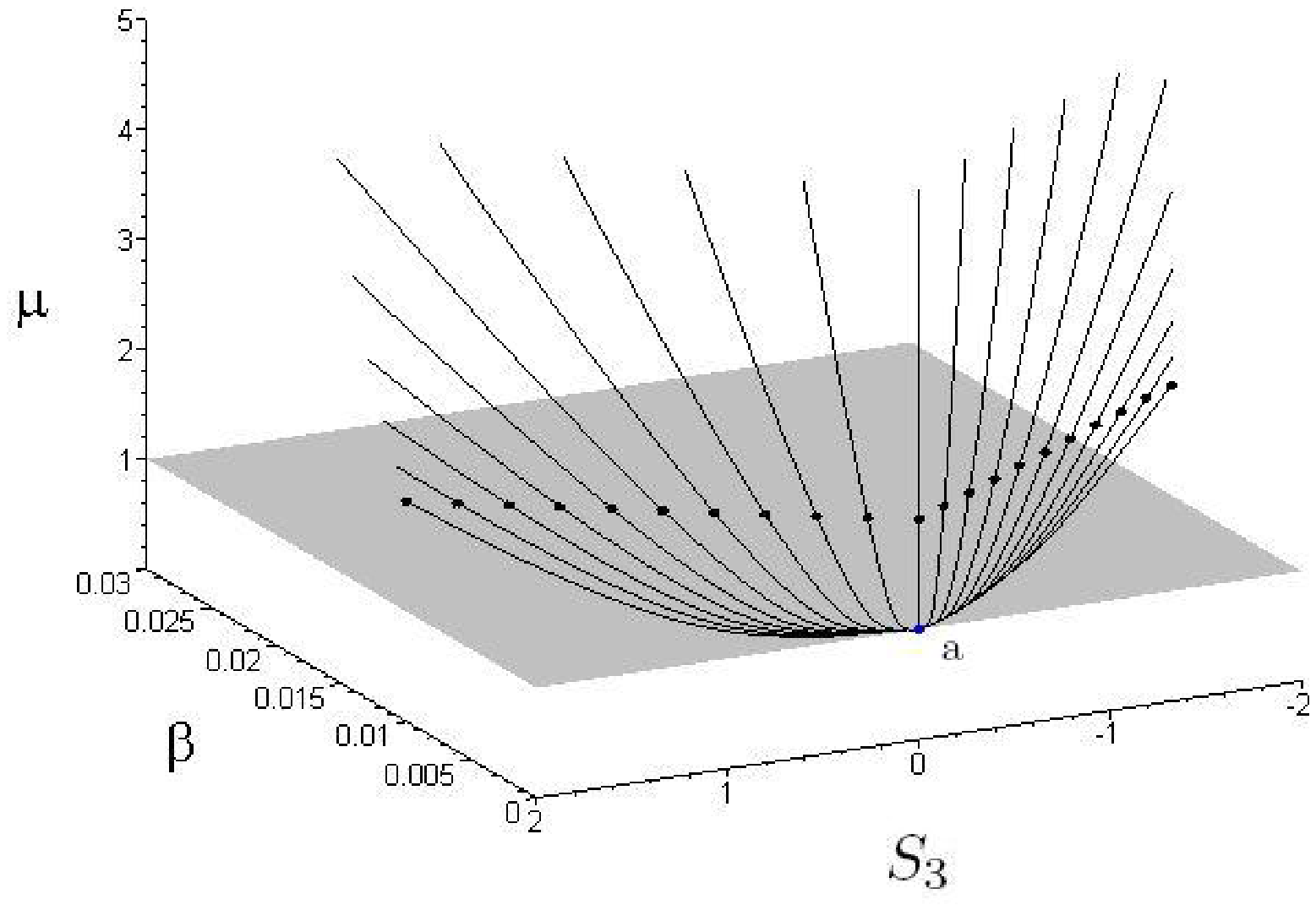,height=10cm,width=10cm,angle=0}
\caption{El gr\'afico muestra varios caminos en una subsecci\'on 3-dimensional $(S_3, \beta, \mu)$ del
espacio de fase para un gas magnetizado y auto-gravitante de neutrones degenerados. Los puntos oscuros representan las condiciones
iniciales a ``$\tau=0$''. El punto ``a'' representa un ``atractor'' con coordenadas $(S_3=0,\beta=0,\mu=1,\HH=0)$.}
\label{espafase_Neutrones}}

\newpage

\section[Invariantes de Riemann para $AdS_{2} \times S^{d-2}$]{\Large{Invariantes de Riemann para $AdS_{2} \times S^{d-2}$}}
\label{Apend_Invariantes}

\begin{table}[h] 
\begin{tabular}{@{}|l|l|l|l|}
\hline
    $\,Invariantes $ & $ \,Definiciones $ & $\,\, Soluciones $ & $ Soluciones $ \\
  $ $ & $ $ & $\,d=4$ & $\,d=5 $ \\
\hline
    $ R $ & $g^{ad}g^{bc}R_{abcd} $ & $\,2\,\gamma_{1} $ & $2\gamma_{3}$ \\
    \cline{1-4}
        $r_{1}=R_{2} $ & $\frac{1}{4}S_{a}^{\,\,\,b}S_{b}^{\,\,\,a} $ & $\frac{1}{4}\gamma^{2}_{2} $ & $\frac{3}{10}\gamma^{2}_{4}$ \\
    \cline{1-4}
        $r_{2}=R_{3} $ & $-\frac{1}{8}S_{a}^{\,\,b}S_{b}^{\,\,c}S_{c}^{\,\,a}$ & $0 $ & $\frac{3}{10^2}\gamma^{3}_{4}$\\
    \cline{1-4}
        $r_{3}=R_{4} $ & $\frac{1}{16}S_{a}^{\,\,b}S_{b}^{\,\,c}S_{c}^{\,\,d}S_{d}^{a} $ & $\frac{1}{64}\gamma^{4}_{2} $ & $\frac{21}{10^3}
        \gamma^{4}_{4} $ \\
\hline
       ${\mathfrak{Re}}(w_{1})=\mathfrak{Re}(W_{2}) $  & $\frac{1}{8}C_{abcd}C^{\,abcd} $ & $\frac{1}{6}\gamma^{2}_{1} $ & $\hspace{0.5cm}-$ \\
     \cline{1-4}
       ${\mathfrak{Re}}(w_{2})=\mathfrak{Re}(W_{3})$ & $-\frac{1}{16}C_{ab}^{\,\,cd}C_{cd}^{\,\,ef}C_{ef}^{\,\,ab} $ & $\frac{-1}{36}\gamma^{3}_{1}$ & $\hspace{0.5cm}-$ \\
     \cline{1-4}
       $\mathfrak{Re}(m_{1})=\mathfrak{Re}(M_{3})$ & $\frac{1}{8}S^{ab}S^{cd}C_{acdb} $ & $\frac{-1}{12}\gamma_{1}\gamma^{2}_{2} $ & $\hspace{0.5cm}-$ \\
   \cline{1-4}
       $\mathfrak{Re}(m_{2})=\mathfrak{Re}(M_{4})$ & $\frac{1}{16}S^{cd}S_{ef}(C_{acdb}C^{aefb}-C^{*}_{acdb}C^{*aefb}) $ & $\frac{1}{36}\gamma^{2}_{1}
       \gamma^{2}_{2} $ & $\hspace{0.5cm}-$ \\
     \cline{1-4}
       $m_{3}={M^{+}_{4}}$ & $\frac{1}{16}S^{cd}S_{ef}(C_{acdb}C^{aefb}+C^{*}_{acdb}C^{*aefb}) $ & $\frac{1}{36}\gamma^{2}_{1}
       \gamma^{2}_{2}$ & $\hspace{0.5cm}-$  \\
     \cline{1-4}
 \cline{1-4}
       $\mathfrak{Re}(m_{5})=\mathfrak{Re}(M_{5})$ & $\frac{1}{32}S^{cd}S^{ef}C^{aghb}(C_{acdb}C_{gefh}+C^{*}_{acdb}C^{*}_{gefh}) $ & $\frac{-1}{108}\gamma^{3}_{1}
       \gamma^{2}_{2}$ & $\hspace{0.5cm} -$  \\
     \cline{1-4}
      $\hspace{1.7cm}\,donde:$ & $\gamma_{1}=(v_{1}-v_{2})/v_{1}v_{2},\,\,\gamma_{2}=(v_{1}+v_{2})/v_{1}v_{2},$ & $ $ & $$ \\
      $ $ & $\,\gamma_{3}=(3v_{1}-v_{2})/v_{1}v_{2},\,\gamma_{4}=(2v_{1}+v_{2})/v_{1}v_{2} $ & $ $ & $$ \\
\hline
\end{tabular}
\caption{\label{Definicion_de_Invariantes} 
{\small Cuadro de definiciones del conjunto completo de invariantes de Riemann y sus resultados para la geometr\'ia tipo $AdS_{2}\times S^{d-2}$ cerca del horizonte de un AN en $d=4$ y $d=5$ dimensiones (el resto de los invariantes son nulos). Todas las definiciones son basadas sobre el tensor de Ricci de traza nula $S_{a\,b}$, el tensor de Weyl $C_{abcd}$  y el tensor de Riemann $R_{abcd}$. Aunque, estos tensores tambi\'en pueden ser definidos sobre una base espinorial \cite{JCarminati}. Aqu\'i, ``-'' significa que no existe una definici\'on para los invariantes complejos, para el caso de cinco dimensiones. Nosotros usamos aqu\'i letras min\'usculas para los invariantes de Carminati, pero en el Lagrangeano usamos letras may\'usculas. El s\'imbolo $\mathfrak{Re}$ significa la parte real del invariante complejo, y para re--marcar el grado del tensor (en lugar de los sub\'indices de Carminati) hemos escrito el grado del tensor como un sub\'indice ({\it ej.} $\mathfrak{Re}(W_{2})$ significa la parte real del invariante complejo de Weyl de segundo orden $w_{1}$).}}
\end{table}
\newpage 
\begin{table}[h]
\begin{tabular}{@{}|l|l|l|l|}
    \hline
         $\,d $ & $ \,Grado $ & $\,\, Invariantes $ & $Equivalencias$ \\
    \hline
         $ d=4 $ & $ 1^{\underline{st}} $ & $ R=2\gamma_{1} $ & $\,\,\gamma_{1} $ \\
    \cline{2-4}
         $ $ & $2^{\underline{nd}} $ & $R^{2}=4\gamma^{2}_{1},\qquad R_{2}=\frac{1}{4}{\gamma}^{2}_{2}, \qquad \mathfrak{Re}(W_{2})=
         \frac{1}{6}{\gamma}^{2}_{1}, $
         & $\gamma^{2}_{1},\,\gamma^{2}_{2}$ \\
    \cline{2-4}
        $ $ & $3^{\underline{rd}} $ & $R^{3}=8\gamma^{3}_{1},\qquad R\times R_{2}=\frac{1}{2}\gamma_{1}\gamma^{2}_{2},\qquad R\times \mathfrak{Re}(W_{2})=\frac{1}
        {3}\gamma^{3}_{1},\,$ & $\,\,\gamma^{3}_{1},\,
        \gamma_{1}\gamma^{2}_{2}$\\
        $ $ & $ $ & $\mathfrak{Re}(W_{3})=\frac{-1}{36}{\gamma}^{3}_{1},\qquad \mathfrak{Re}(M_{3})=\frac{-1}{12}\gamma_{1}{\gamma}^{2}_{2} $ & $ $ \\
    \cline{2-4}
        $ $ & $4^{\underline{th}} $ & $R^{4}=16\gamma^{4}_{1}, \qquad \mathfrak{Re}^2({W_{2}})=\frac{\gamma^{4}_{1}}{36}, \qquad
         R^{2}\times R_{2}=\gamma^{2}_{1} \gamma^{2}_{2},$ & $ $ \\
         $ $ & $ $ & $ \qquad R_{2}^{2}=\frac{1}{16}\gamma^{4}_{2}, \qquad R^{2}\times \mathfrak{Re}(W_{2})=\frac{2}{3}\gamma^{4}_{1},  $ & $ $ \\
        $ $ & $ $ & $ R_{2}\times \mathfrak{Re}(W_{2})=\frac{1}{24}\gamma^{2}_{1}\gamma^{2}_{2},\, \qquad \mathfrak{Re}(M_{4})=M^{+}_{4}=\frac{1}{36}
        {\gamma}^{2}_{1}{\gamma}^{2}_{2},  $ & $\,\gamma^{4}_{1},\,\gamma^{4}_{2},\,\gamma^{2}_{1}\gamma^{2}_{2}$   \\
        $ $ & $ $ & $\qquad R\times  \mathfrak{Re}(W_{3})=\frac{-\gamma^{4}_{1}}{18},
        \qquad R\times \mathfrak{Re}(M_{3})=\frac{-\gamma^{2}_{1}\gamma^{2}_{2}}{6}, $ & $ $ \\
        $ $ & $ $ & $ R_{4}=\frac{1}{64}{\gamma}^{4}_{2},$ & $ $\\
    \cline{2-4}
        $ $ & $5^{\underline{th}} $ & $R^5={{32}}\gamma^{5}_{1}, \qquad R^3\times R_{2}=2\gamma^{3}_{1}\gamma^{2}_{2}, \qquad R^3\times
        \mathfrak{Re}(W_{2})=\frac{4}{3}\gamma^{5}_{1}, $ & $ $ \\
        $ $ & $ $ & $\qquad R^2\times \mathfrak{Re}(W_{3})=-\frac{\gamma^{5}_{1}}{9}, \qquad R^{2}_{2}\times{R}=\frac{\gamma_{1}\gamma^{4}_{2}}{8},$ & $ $ \\
        $ $ & $ $ &  $ R^2\times \mathfrak{Re}(M_{3})=-\frac{\gamma^{3}_{1}\gamma^{2}_{2}}{3},\qquad R\times \mathfrak{Re}^{2}(W_{2})=
        \frac{\gamma^{5}_{1}}{18},$ & $\gamma^{5}_{1},\gamma_{1}\gamma^{4}_{2},\gamma^{3}_{1}\gamma^{2}_{2}$ \\
        $ $ & $ $ & $\qquad \,R\times R_{2}\times \mathfrak{Re}(W_{2})=
        \frac{\gamma^{3}_{1}\gamma^{2}_{2}}{12}, $ & $ $ \\
        $ $ & $ $ &  $R_{2}\times \mathfrak{Re}(W_{3})=\frac{-\gamma^{3}_{1}\gamma^{2}_{2}}{144}, \qquad R_{2}\times
        \mathfrak{Re}(M_{3})=\frac{-\gamma_{1}\gamma^{4}_{2}}{48}, \qquad $ & $ $ \\
        $ $ & $ $ & $ \mathfrak{Re}(W_{2})\times \mathfrak{Re}(M_{3})=\frac{-\gamma^{3}_{1}\gamma^{2}_{2}}{72},\qquad
         \mathfrak{Re}(W_{2})\times \mathfrak{Re}(W_{3})=\frac{-\gamma^{5}_{1}}{216}, $ & $ $ \\
        $ $ & $ $ & $ R\times R_{4}=\frac{\gamma_{1}\gamma^{4}_{2}}{32}, \qquad R\times \mathfrak{Re}(M_{4})=R\times M^{+}_{4}\equiv\frac{\gamma^{3}_{1}\gamma^{2}_{2}}{18}$ & $ $ \\
\hline
       $ d=5 $  & $1^{\underline{st}} $ & $R=2\gamma_{3} $ & $ \, \gamma_{3}$ \\
     \cline{2-4}
       $ $ & $2^{\underline{nd}} $ & $R^{2}=4\gamma^{2}_{3}, \qquad R_{2}=\frac{3}{10}\gamma^{2}_{4}\qquad$ & $\gamma^{2}_{3},\,\gamma^{2}_{4}$ \\
     \cline{2-4}
       $ $ & $3^{\underline{rd}} $ & $R^{3}=8\gamma^{3}_{3},\qquad R\times\,R_{2}=\frac{3}{5}\gamma_{3}\gamma^{2}_{4},\qquad R_{3}=\frac{3}{100}\gamma^{3}_{4} $
       & $\gamma^{3}_{3},\,\gamma_{3}\gamma^{2}_{4},\,\gamma^{3}_{4}$ \\
     \cline{2-4}
       $ $ & $4^{\underline{th}} $ & $R^{4}=16\gamma^{4}_{3},\qquad R^{2}_{2}=\frac{9}{100}\gamma^{4}_{4},\qquad R^{2}\times\,
       R_{2}=\frac{6}{5}\gamma^{2}_{3}\gamma^{2}_{4}, \qquad
        $ & $\,\gamma^{4}_{3},\,\gamma^{4}_{4},\,\gamma^{2}_{3}\gamma^{2}_{4},$  \\
        $ $ & $ $ & $\qquad R\times R_{3}=\frac{3}{50}\gamma_{3}\gamma^{3}_{4}, \qquad R_{4}=\frac{21}{10^3}\gamma^{4}_{4}$ & $\gamma_{3}\gamma^{3}_{4} $ \\
     \cline{2-4}
       $ $ & $ $ & $donde:\,\,\,\gamma_{1}=(v_{1}-v_{2})/v_{1}v_{2},\,\gamma_{2}=(v_{1}+v_{2})/v_{1}v_{2}, $ & $$ \\
       $ $ & $ $ & $\hspace{1.3cm}\gamma_{3}=(3v_{1}-v_{2})/v_{1}v_{2},\,\gamma_{4}=(2v_{1}+v_{2})/v_{1}v_{2} $ & $$ \\
\hline
\end{tabular}
\caption{\label{Tabla_de_Invariantes} 
Conjunto completo de invariantes no nulos, ellos han sido organizados por $grados$ para $d=4$ y $d=5$, bas\'andonos en la geometr\'ia $AdS_{2}\times S^{d-2}$.}
\end{table}

\end{appendices}

\newpage

\section*{Acknowledgements}
{This work has been supported by \emph{Ministerio de Ciencia, Tecnolog\'{\i}a y Medio Ambiente} under the grant CB0407
and the ICTP Office of External Activities through NET-35. A.U.R also acknowledges the Program of ICTP-CLAF fellowship as
well as the hospitality of ICN, UNAM and the support from the research grant \emph{DGAPA--UNAM PAPIIT--IN119309}.}
%
%
%
%
%

%
\end{document}